\documentclass[usenatbib]{mnras}
\usepackage{multirow}
\usepackage{rotating}
\usepackage{colortbl}
\usepackage{color}
\usepackage[fleqn]{amsmath}
\usepackage{amssymb}
\usepackage{amsfonts}
\usepackage{verbatim}
\usepackage{scalefnt}
\usepackage[percent]{overpic}

\newcommand{\comments}[1]{} 
\newcommand\T{\rule{0pt}{2.6ex}}       
\newcommand\B{\rule[-1.2ex]{0pt}{0pt}} 

\title[Dual AGN: frequency and properties]{A survey of dual active galactic nuclei in simulations of galaxy mergers: frequency and properties}

\author[P.~R. Capelo et al.]{Pedro~R. Capelo,$^{1}$\thanks{E-mail: pcapelo@physik.uzh.ch} Massimo Dotti,$^{2,3}$ Marta Volonteri,$^{4}$ \newauthor Lucio Mayer,$^{1}$ Jillian~M. Bellovary$^{5}$ and Sijing Shen$^{6}$\\ 
$^1$Center for Theoretical Astrophysics and Cosmology, Institute for Computational Science, University of Zurich,\\
Winterthurerstrasse 190, CH-8057 Z$\ddot{u}$rich, Switzerland\\
$^2$Dipartimento di Fisica G. Occhialini, Universit$\grave{a}$ degli Studi di Milano--Bicocca, Piazza della Scienza 3, I-20126 Milano, Italy\\
$^3$INFN, Sezione Milano--Bicocca, Piazza della Scienza 3, I-20126 Milano, Italy\\
$^4$Institut d'Astrophysique de Paris, Sorbonne Universit\'{e}s, UPMC Univ. Paris 06 et CNRS, UMR 7095,\\ 98bis boulevard Arago, F-75014 Paris, France\\
$^5$Department of Physics, Queensborough Community College, 222-05 56th Avenue, Bayside, NY 11364, USA\\
$^6$Institute of Astronomy, University of Cambridge, Madingley Road, Cambridge CB3 0HA, UK
}

\pubyear{2017}

\begin{document}

\date{Accepted 2017 April 28. Received 2017 April 21; in original form 2016 November 28}

\label{firstpage}


\maketitle

\begin{abstract}
We investigate the simultaneous triggering of active galactic nuclei (AGN) in merging galaxies, using a large suite of high-resolution hydrodynamical simulations. We compute dual-AGN observability time-scales using bolometric, X-ray, and Eddington-ratio thresholds, confirming that dual activity from supermassive black holes (BHs) is generally higher at late pericentric passages, before a merger remnant has formed, especially at high luminosities. For typical minor and major mergers, dual activity lasts $\sim$20--70 and $\sim$100--160~Myr, respectively. We also explore the effects of X-ray obscuration from gas, finding that the dual-AGN time decreases at most by a factor of $\sim$2, and of contamination from star formation. Using projected separations and velocity differences rather than three-dimensional quantities can decrease the dual-AGN time-scales by up to $\sim$4, and we apply filters which mimic current observational-resolution limitations. In agreement with observations, we find that, for a sample of major and minor mergers hosting at least one AGN, the fraction harbouring dual AGN is $\sim$20--30 and $\sim$1--10 per cent, respectively. We quantify the effects of merger mass ratio (0.1 to 1), geometry (coplanar, prograde, retrograde, and inclined), disc gas fraction, and BH properties, finding that the mass ratio is the most important factor, with the difference between minor and major mergers varying between factors of a few to orders of magnitude, depending on the luminosity and filter used. We also find that a shallow imaging survey will require very high angular resolution, whereas a deep imaging survey will be less resolution-dependent.
\end{abstract}

\begin{keywords}
galaxies: active -- galaxies: interactions -- galaxies: nuclei.
\end{keywords}

\section{Introduction}\label{dualagnpaper:sec:Introduction}

A direct consequence of the hierarchical paradigm of structure formation \citep[e.g.][]{Blumenthal84} are mergers between galaxies. Since supermassive black holes (BHs) are believed to exist at the centre of most massive galaxies \citep[e.g.][]{KormendyRichstone95,Ferrarese2005}, merging systems with two BHs are expected to be common. Moreover, as galaxy mergers can trigger active galactic nuclei (AGN; e.g. \citealt{DiMatteo2005,hopkins2006,Younger2008,Johansson2009,HQ2010,Hayward_et_al_2014,Capelo_et_al_2015}), dual AGN -- defined in this paper as systems with two AGN with a separation $0.1 \lesssim d \lesssim 100$~kpc -- should also be quite frequent.

Both observations and simulations show that AGN fraction/activity increases with decreasing galaxy separation \citep[e.g.][]{Ellison2011,Silverman2011,Koss2012,Ellison_et_al_2013,Capelo_et_al_2015}, implying that the same should happen with dual AGN activity. However, even though several dual AGN with separations $\gtrsim$10~kpc have been detected \citep[e.g.][]{Myers_et_al_2008,Hennawi_et_al_2010,Liu2011,Koss2012}, observed dual AGN with separations $\lesssim$10~kpc are very rare. One reason for this paucity is the inherent difficulty in spatially resolving two AGN at such small separations, especially at redshift $z \gtrsim 0.1$. One of the most promising diagnostics for the existence of dual AGN is the presence of double-peaked narrow-line (NL) emission with line-of-sight velocity splitting of $\gtrsim$100~km~s$^{-1}$. Unfortunately, this splitting may be due to a variety of other phenomena, including outflows and rotating discs (see, e.g. \citealt{Komossa_Zensus_2016}), giving raise to many negative outcomes \citep[e.g.][]{Tingay_Wayth_2011,Gabanyi_et_al_2016} and necessitating follow-up observations to confirm the candidates. Hence, only a handful of systems have been so far confirmed \citep[e.g.][]{Fu_et_al_2011b,Liu_et_al_2013,Comerford_et_al_2015,Muller-Sanchez_et_al_2015}.

On the numerical side, there are two main avenues of research. Cosmological simulations \citep[e.g.][]{Steinborn_et_al_2016,Volonteri_et_al_2016,Tremmel_et_al_2016} can produce relatively large numbers of galaxy mergers and systems of multiple BHs and AGN, ideally providing the theoretical counterpart to observations yielding the fraction of dual AGN over the total number of AGN \citep[e.g.][]{Shen2011}. These simulations, however, usually lack the required resolution to reliably follow the dynamics and accretion of BHs. Idealised merger simulations, on the other hand, have the advantage of resolving $<$kpc scales, which are extremely important to follow the dynamics of the two BHs \citep[e.g.][]{Van_Wassenhove_et_al_2014}, and of letting one study the effect of merger, galactic, and BH parameters in a controlled environment.

In the context of isolated merger simulations, \citet{Van_Wassenhove_et_al_2012} follow the dynamics and accretion onto the BHs of merging galaxies in three idealised merger simulations (1:2 spiral--spiral, 1:2 elliptical--spiral, and 1:10 spiral--spiral), finding that strong dual AGN activity occurs during the late stages of the encounters, at separations $<$1--10~kpc, and that much of the AGN activity in mergers is not simultaneous (see also \citealt{Blecha2013}, who simulate the connection between dual AGN activity and double-peaked NLs).

We build upon the work by \citet{Van_Wassenhove_et_al_2012} by analysing a much larger suite of 12 mergers, studying more initial mass ratios (1:1, 1:2, 1:4, 1:6, and 1:10), disc gas fractions (30 and 60 per cent), and geometries (coplanar, prograde--prograde, retrograde--prograde, prograde--retrograde, and inclined), therefore being able to assess the importance of merger and galactic properties. We also include control runs, in which we study the effects of varying BH properties such as mass and feedback efficiency. In order to better mimic observational limitations, we additionally `observe' our mergers from 100 random lines of sight, study AGN activity with different thresholds (bolometric and X-ray luminosity, and Eddington ratio), and explore the effects of gas obscuration and star formation contamination on X-ray dual-AGN time-scales.

In Sections~\ref{dualagnpaper:sec:Numerical_setup} and \ref{dualagnpaper:sec:Analysis}, we describe the numerical setup of our simulations and the methods of our analysis, respectively. In Section~\ref{dualagnpaper:sec:Results}, we discuss in detail the results of one of our runs (Section~\ref{dualagnpaper:sec:Default_merger}); explore the effects of obscuration and contamination from star formation (Section~\ref{dualagnpaper:sec:X-ray_obscuration_and_dilution}); assess the importance of merger, galactic, and BH parameters (Section~\ref{dualagnpaper:sec:Dependence_on_merger_properties}); and compare the results to other theoretical and observational data (Section~\ref{dualagnpaper:sec:Comparison}). We summarize and conclude in Section~\ref{dualagnpaper:sec:Conclusions}.

\section{Numerical Setup}\label{dualagnpaper:sec:Numerical_setup}

In the following, we summarize the setup of the 12 simulated mergers of our suite. We refer the reader to \citet{Capelo_et_al_2015} for more details on 11 of these mergers (Runs 01--10 and C3 in their Table~1) and to \citet{Capelo_Dotti_2017} for an additional merger -- which here we name C4 for consistency of notation -- where we increased the initial mass of the BHs (see below). The main parameters of each merger are shown in Table~\ref{dualagnpaper:tab:merger_params}. This suite is a follow-up of a similar suite of mergers \citep[][]{Callegari2009,Callegari2011,Van_Wassenhove_et_al_2012,Van_Wassenhove_et_al_2014}, initially constructed to study the pairing time-scales of BHs in unequal-mass galaxy mergers \citep[see also][]{Pfister_et_al_2017}.

We simulate isolated mergers of disc galaxies at $z = 3$, initially at a distance equal to the sum of their virial radii, set on parabolic orbits \citep[][]{Benson05} with a first pericentric distance equal to 20 per cent of the virial radius of the larger (`primary') galaxy \citep[][]{Khochfar2006}, which is the same for all encounters. Whereas the global orbital parameters are identical for all the mergers in the suite,\footnote{Ignoring the fact that the sum of the virial radii changes with the initial mass ratio, since the smaller (`secondary') galaxy varies.} the internal orbital parameters are not: we include coplanar, prograde--prograde, retrograde--prograde, and prograde--retrograde encounters, and inclined mergers, by varying the angle (0, $\pi/4$, or $\pi$ radians) between the initial individual galactic angular momentum vector of each galaxy and the global angular momentum vector.

All galaxies are composed of a dark matter halo, a stellar and gaseous disc, a stellar bulge, and a central BH. The dark matter halo is described by a Navarro--Frenk--White \citep[NFW,][]{NFW1996} profile with spin and concentration parameters 0.04 and 3, respectively, up to the virial radius, and by an exponentially decaying NFW profile for larger radii \citep[][]{Springel_White_1999}. The disc is described by an exponential disc with an isothermal sheet \citep[][]{Spitzer_1942,Camm_1950}, with its mass, equal to 4 per cent of the virial mass of the galaxy, divided into gas (30 or 60 per cent, depending on the merger) and stars. The initial value of the disc scale radius, $r_{\rm disc}$, comes from imposing conservation of specific angular momentum of the material that forms the disc, whereas the initial disc scale height, $z_{\rm disc}$, is set equal to 10 per cent of $r_{\rm disc}$. The bulge is described by a spherical \citet{Hernquist1990} profile with a mass ($M_{\rm bulge}$) and scale radius equal to 0.8 per cent of the virial mass of the galaxy and 20 per cent of $r_{\rm disc}$, respectively. The primary galaxy in each merger has an initial virial mass of $2.21 \times 10^{11}$~M$_{\odot}$, whereas the secondary galaxy has a mass that is a fraction (1/10, 1/6, and 1/4 in the case of `minor' mergers; 1/2 and 1 in the case of `major' mergers) of that of the larger galaxy. The initial disc scale radius varies from 0.53~kpc (for the secondary galaxy in the 1:10 merger) to 1.13~kpc (for the primary galaxy in all mergers). 

The central BH has a mass $M_{\rm BH}$ proportional to that of the stellar bulge \citep[][]{MarconiHunt2003}, where the constant of proportionality is $2 \times 10^{-3}$ (`standard') for all mergers except for one \citep[Run~C4;][]{Capelo_Dotti_2017}, in which we instead use $5 \times 10^{-3}$ (`large'), as it was recently shown that the BH--bulge relation might evolve with redshift \citep[][]{Merloni2010}. BHs are sink particles which accrete preferentially nearby, cold, and dense gas \citep[][]{Bellovary10}, according to a Bondi--Lyttleton--Hoyle \citep[hereafter, Bondi;][]{Bondi1952,BondiHoyle1944,Hoyle_Lyttleton_1939} formula:\footnote{At each time-step $dt$, in case of an accretion event, the code takes a quantity $dm$ from the gas, according to the Bondi formula, and adds the same amount to the mass of the BH, therefore neglecting the radiative efficiency and introducing a 10-per-cent error.}

\begin{equation}\label{dualagnpaper:eq:bondi}
\dot{M}_{\rm BH} = \frac{4 \pi \alpha M^2_{\rm BH} \rho}{(c^2_{\rm s} + v^2)^{3/2}},
\end{equation}

\noindent where $\rho$ is the local gas density, $c_{\rm s}$ is the local sound speed, $v$ is the relative velocity between gas and BH, $G$ is the gravitational constant, and $\alpha$ is a boost factor.\footnote{Our standard boost factor is $\alpha = 3$ as originally reported in \citet{Capelo_et_al_2015}, but we recently found that, in some segments of some simulations, the boost factor is doubled \citep[][]{Gabor_et_al_2016}. Since BH growth is self-regulated by AGN feedback in these simulations, our results are insensitive to the exact value of the boost factor. We partially re-ran one merger simulation with different values of the boost factor (varying it between 1 and 6) and found no significant differences in the BH accretion levels.} Accretion is then capped at mildly super-Eddington values, by limiting $f_{\rm Edd} \equiv \dot{M}_{\rm BH}/\dot{M}_{\rm BH,\,Edd}$ at $\alpha$, where

\begin{equation}\label{dualagnpaper:eq:eddington}
\dot{M}_{\rm BH,\,Edd} = \frac{4 \pi G M_{\rm BH} m_{\rm p}}{\epsilon_{\rm r} \sigma_{\rm T} c},
\end{equation}

\noindent where $m_{\rm p}$ is the proton mass, $\sigma_{\rm T}$ is the Thomson cross-section, $c$ is the speed of light in vacuum, and $\epsilon_{\rm r} = 0.1$ \citep{ShakuraSunyaev} is the radiative efficiency. The latter gives the fraction of BH accretion energy rate ($\dot{M}_{\rm BH} c^2$) that is emitted as radiation: $L_{\rm bol} = \epsilon_{\rm r} \dot{M}_{\rm BH} c^2$. Part of this radiation is then coupled to the gas, in the form of thermal energy \citep[][]{Bellovary10}. The amount of coupling is set by the BH feedback efficiency parameter $\epsilon_{\rm f}$, which is equal to 0.001 (`standard') in all runs except for one of the control runs (Run~C3), in which $\epsilon_{\rm f} = 0.005$ (`high').

All 12 simulations were performed with the $N$-body smoothed particle hydrodynamics (SPH) code {\scshape gasoline} \citep[][]{gasoline} -- an extension of {\scshape pkdgrav} \citep[][]{stadel01} -- which has realistic implementations of line cooling for atomic hydrogen and helium, and metals \citep[][]{Shen_et_al_2010}; star formation (stars can stochastically form with a star formation efficiency of 0.015 from gas colder than 6000~K and denser than 100~a.m.u.~cm$^{-3}$), supernova feedback, and stellar winds \citep[][]{Stinson2006}; and the above mentioned recipe for BH accretion and feedback \citep[][]{Bellovary10}. No additional corrections \citep[such as an over-pressurized equation of state; e.g.][]{Springel_Hernquist_2003} to the gas thermodynamics were adopted. The star formation density threshold was chosen to be near the maximum density at which the Jeans mass ($M_{\rm Jeans}$) of gas at the temperature floor of the simulations ($T_{\rm floor}$) is resolved by $N$ particles:

\begin{equation}\label{dualagnpaper:eq:jeans1}
n_{\rm max} = C \left(\frac{T_{\rm floor}}{200~{\rm K}}\right)^3 \left(\frac{N}{64}\right)^{-2} \left(\frac{m_{\rm gas}}{10^3~{\rm M}_{\odot}}\right)^{-2}~{\rm cm}^{-3},
\end{equation}

\noindent where $m_{\rm gas}$ is the gas particle mass and $C \simeq 179$.\footnote{This value was obtained using the following definition: $M_{\rm Jeans} = (\pi^{5/2}/6) [\gamma k_{\rm B} T/(\mu \, m_{\rm H} G)]^{3/2} \rho^{-1/2}$, where $k_{\rm B}$ is the Boltzmann constant, $m_{\rm H}$ is the hydrogen mass, and $\rho$, $T$, $\gamma = 5/3$, and $\mu \simeq 0.6$ are the mass density, temperature, adiabatic index, and mean molecular weight of the gas, respectively. Using an alternative definition, $M_{\rm Jeans} = [10 k_{\rm B} T/(3 (\gamma-1) \mu \, m_{\rm H} G)]^{3/2}(4 \pi \rho/3)^{-1/2}$, we obtain $C \simeq 136$.} In our simulations, $T_{\rm floor} = 500$~K, and we require $N = 64$.

The particle mass of the stellar, gas, and dark matter particles is $3.3 \times 10^3$, $4.6 \times 10^3$, and 5--$11 \times 10^4$~M$_{\odot}$, respectively. The gravitational softening, for the same particle types, is 10, 20, and 23--30~pc, respectively (and that of BHs is 5~pc). The range in the dark matter quantities is due to the fact that we want to ensure that the dark matter particles have a mass smaller than 15 per cent of that of the smaller BH in each merger, thus successfully limiting excursions of BHs from the centre of each galaxy. For this reason, we do not need to artificially model unresolved dynamical friction (see, e.g. \citealt{Tremmel_et_al_2015} and references therein). All the parameters described here are summarized in Tables~1 and 2 of \citet{Capelo_et_al_2015}, which the reader should refer to for more details.

\begin{table} \centering
\vspace{-3.5pt}
\caption[Merger parameters]{Main (initial) simulation parameters for our six major mergers (Runs~01--06), four minor mergers (Runs~07--10), and two control runs (Runs~C3--C4). For consistency, we use the same notation as in \citet{Capelo_et_al_2015}, where their Runs~C1--C2 had zero BH accretion and therefore are not relevant for the current study. (1) Run number. (2) Mass ratio $q_{\textnormal{\tiny \textsc{G}}}$ between the merging galaxies. (3) Angle $\theta_1$ between the primary galaxy's angular momentum vector and the overall orbital angular momentum vector, in radians. (4) Angle $\theta_2$ between the secondary galaxy's angular momentum vector and the overall orbital angular momentum vector, in radians. (5) Gas fraction in the galactic disc. (6) BH mass (in units of $2 \times 10^{-3} M_{\rm bulge}$). (7) BH feedback efficiency $\epsilon_f$.
\label{dualagnpaper:tab:merger_params}}
\vspace{10pt}
{\small
\begin{tabular*}{0.475\textwidth}{m{22pt}m{22pt}m{22pt}m{22pt}m{22pt}m{22pt}m{25pt}}
\hline
Run & $q_{\textnormal{\tiny \textsc{G}}}$ & $\theta_1$ & $\theta_2$ & gas & $M_{\rm BH}$ & $\epsilon_{\rm f}$ \T \B \\
\hline
01 & 1:1 & 0 & 0 & 0.3 & 1 & 0.001 \T \B \\
02 & 1:2 & 0 & 0 & 0.3 & 1 & 0.001 \B \\
03 & 1:2 & $\pi/4$ & 0 & 0.3 & 1 & 0.001 \B \\
04 & 1:2 & $\pi$ & 0 & 0.3 & 1 & 0.001 \B \\
05 & 1:2 & 0 & $\pi$ & 0.3 & 1 & 0.001 \B \\
06 & 1:2 & 0 & 0 & 0.6 & 1 & 0.001 \B \\
\hline
07 & 1:4 & 0 & 0 & 0.3 & 1 & 0.001 \T \B \\
08 & 1:4 & $\pi/4$ & 0 & 0.3 & 1 & 0.001 \B \\
09 & 1:6 & 0 & 0 & 0.3 & 1 & 0.001 \B \\
10 & 1:10 & 0 & 0 & 0.3 & 1 & 0.001 \B \\
\hline
C3 & 1:2 & 0 & 0 & 0.3 & 1 & 0.005 \T \B \\
C4 & 1:2 & 0 & 0 & 0.3 & 2.5 & 0.001 \B \\
\hline
\end{tabular*}
\vspace{5pt}
}
\end{table}


\section{Analysis}\label{dualagnpaper:sec:Analysis}

Consistently with previous work \citep{Capelo_et_al_2015}, we divide the encounter in three distinct stages: the `stochastic stage', from the beginning of the simulation to the second pericentric passage; the `merger stage', which ends when the specific angular momentum of the central gas ceases its dramatic oscillations caused by the encounter; and the `remnant stage', which ends when the BH separation (averaged over 5~Myr) is below twice the gravitational softening of the BHs (i.e. the stellar gravitational softening: 10~pc).\footnote{In one case in our suite (1:1 merger -- Run~01), the 10-pc threshold is reached slightly before the end of the merger stage (as opposed to after, as in all other mergers). For consistency with the other mergers, in this case we decided to not change the length of the merger stage, since the difference in time is minimal. In another case (1:4 inclined-primary merger -- Run~08), we stopped the simulation when the remnant stage was as long as the stochastic stage, since the 10-pc threshold had not been reached yet. We note that we have also computed dual-activity observability time-scales considering a much longer remnant stage (i.e. always as long as the stochastic stage, regardless of the 10-pc limit, as in \citealt{Capelo_et_al_2015}) and found that, for typical imaging and spectroscopy resolutions, there is no significant change.} After this point, we cannot study reliably the dynamics of the two BHs, due to resolution limitations. Depending on the merger, there are cases when the last stage is fairly short (e.g. Run 02: 0.1~Gyr; see Section~\ref{dualagnpaper:sec:Default_merger}) and cases when it is long (e.g. Run~05, the 1:2 coplanar, retrograde--prograde merger, and Run~08, the 1:4 inclined-primary merger: 0.5 and 1~Gyr, respectively).

Since we are interested in the observability of dual activity, we `observe' the same encounter from several random lines of sight and translate three-dimensional (3D) quantities into projected (onto the plane perpendicular to the line of sight) BH separations ($r$) and projected (onto the line of sight) BH velocity differences ($v$). More specifically, the code returns the mass, bolometric luminosity, 3D separation, and 3D velocity difference of the BHs every $\Delta t = 0.1$~Myr, which is comparable to the characteristic dynamical times in proximity of the BHs (given the BH masses and gravitational softenings involved). We then project the 3D properties onto $N$ random and isotropically distributed lines of sight, where $N = 10$, 100, or 1000. The fraction of projections that results into an either spatial or spectroscopic AGN pair corresponds to the probability $p$ of detecting a dual AGN, and the expected dual-activity time within $\Delta t$ is estimated as $\Delta t \times p$. For the remainder of this paper, we use $N = 100$ (see Section~\ref{Effects_of_projection} for more details).

Throughout the paper, we adopt different thresholds for bolometric luminosity (from $10^{42}$ to  $10^{44}$ erg~s$^{-1}$) and for X-ray luminosity in the 2--10~keV band (from $10^{41.3}$ to  $10^{43.3}$ erg~s$^{-1}$). Note that the secondary BHs in the stochastic stages of the minor mergers (mass ratios 1:4, 1:6, and 1:10) cannot reach the highest X-ray threshold we consider ($10^{43.3}$~erg~s$^{-1}$). We also assume typical thresholds of 1 and 10~kpc (for imaging), and 150~km~s$^{-1}$, as the minimum velocity difference needed to separate two AGN in spectroscopic observations.


\section{Results}\label{dualagnpaper:sec:Results}

We first study the dual-activity observability time-scales for one individual merger (Section~\ref{dualagnpaper:sec:Default_merger}) and discuss the effects of obscuration caused by intervening gas in the host galaxy (Section~\ref{dualagnpaper:sec:X-ray_obscuration_and_dilution}). We then compare the results of all encounters (Section~\ref{dualagnpaper:sec:Dependence_on_merger_properties}) to assess the effects of merger (mass ratio, geometry) or galactic parameters (gas fraction), and of BH-related quantities (BH mass, feedback efficiency). Finally, we compare our results to those from observational surveys and theoretical studies (Section~\ref{dualagnpaper:sec:Comparison}).

\begin{figure}
\centering
\vspace{2.0pt}
\includegraphics[width=0.99\columnwidth,angle=0]{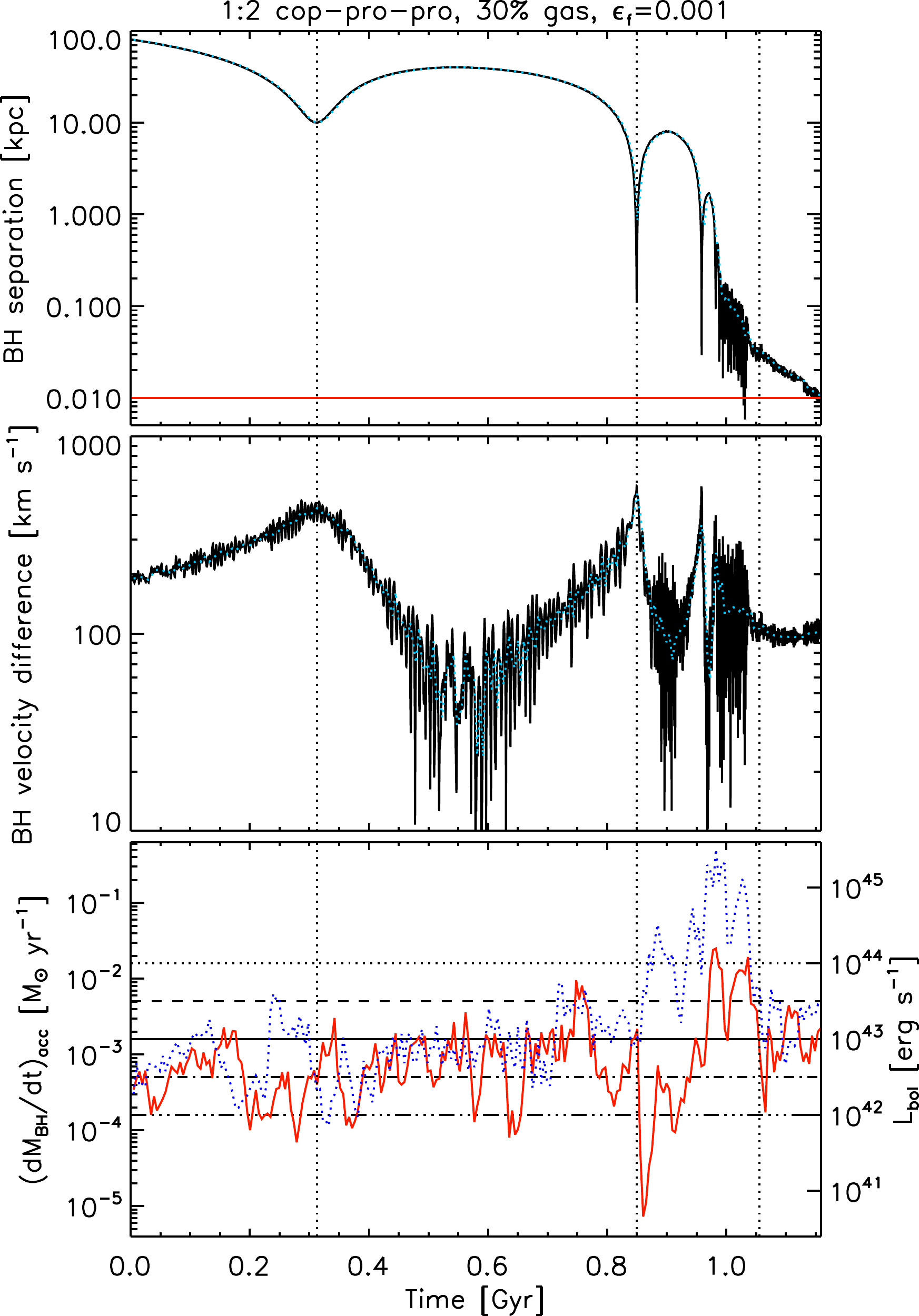}
\vspace{-5.0pt}
\caption[]{Temporal evolution of the 1:2 coplanar, prograde--prograde merger with 30 per cent gas fraction and standard BH mass and feedback efficiency (the default merger; total encounter time 1.158~Gyr), from the beginning of the stochastic stage to the end of the remnant stage. In all panels, the vertical, dotted, black lines show, from left to right: (i) the first pericentric passage, (ii) the end of the stochastic stage, and (iii) the end of the merger stage. Top panel: 3D separation between the two BHs, shown every 0.1~Myr (black, solid), together with its average every 5~Myr (cyan, dot). The horizontal, solid, red line shows the stellar gravitational softening (10~pc). Middle panel: 3D velocity difference between the two BHs, shown every 0.1~Myr (black, solid), together with its average every 5~Myr (cyan, dot). Bottom panel: BH accretion rate (and bolometric luminosity) of the primary (blue, dot) and secondary (red, solid) BH, averaged every 5~Myr. The horizontal lines show different levels of bolometric luminosity, between $10^{42}$ and $10^{44}$~erg~s$^{-1}$ [the line styles match those of Figs~\ref{dualagnpaper:fig:m2_hr_gf0_3_BHeff0_001_phi000000_dual_agn_Mdot_deltat_proj_no3b} (left panels), \ref{dualagnpaper:fig:dual_agn_Bolom_Mdot_deltat_proj_norm2_no3b_coppropro_onlyr_3thr}, \ref{dualagnpaper:fig:m2_hr_gf0_3_BHeff0_001_phi000000_dual_agn_morelos_no3b}, \ref{dualagnpaper:fig:dual_agn_Bolom_Mdot_deltat_proj_norm1_no3b_coppropro_onlyr_3thr}--\ref{dualagnpaper:fig:dual_agn_Bolom_Mdot_deltat_proj_norm3_no3b_coppropro_onlyr_3thr}, and \ref{dualagnpaper:fig:dual_agn_Bolom_Mdot_deltat_proj_no3b_coppropro_onlyr_3thr}--\ref{dualagnpaper:fig:dual_agn_Bolom_Mdot_deltat_proj_norm2_no3b_bhphysics_onlyr_3thr}].}
\label{dualagnpaper:fig:m2_hr_gf0_3_BHeff0_001_phi000000_three_panels_dual}
\end{figure}


\subsection{The default merger}\label{dualagnpaper:sec:Default_merger}

The `default merger' (Run~02 in Tables~\ref{dualagnpaper:tab:merger_params}--\ref{dualagnpaper:tab:comparison}) is the 1:2 coplanar, prograde--prograde merger with 30 per cent gas fraction and standard BH mass and feedback efficiency. We chose this encounter because it is at the intersection of most of our parameter-space studies (see Section~\ref{dualagnpaper:sec:Dependence_on_merger_properties}).


\subsubsection{General behaviour}\label{dualagnpaper:sec:General_behaviour}

In Fig.~\ref{dualagnpaper:fig:m2_hr_gf0_3_BHeff0_001_phi000000_three_panels_dual}, we show the detailed history of the default merger. In the top two panels, we show the (3D) separation and velocity difference between the two BHs, as a function of time. In the remainder of this paper, unless otherwise stated, we use projected quantities, as explained in Section~\ref{dualagnpaper:sec:Analysis}. In the bottom panel, we show the BH accretion rate (and bolometric luminosity) of the two BHs. This figure allows one to appreciate that, based on 3D quantities, spectroscopic searches for large velocity differences should identify for the most part dual AGN at the first pericentre, when their separation is 10--20~kpc, where BHs spend a large amount of time with sufficiently large velocity difference, and also at subsequent pericentres, but for much shorter-lived episodes at higher velocities. Conversely, searches for physically distinct nuclei would be more successful at finding AGN at apocentres. As shown in the bottom panel, this has to be, however, convolved with the AGN luminosities and re-assessed taking projection effects into account.  Projecting a vector onto a random line of sight (what we do for velocity differences) or onto the plane perpendicular to such line of sight (what we do for separations) has vastly different results. The probability distribution of projected velocity differences is flat between 0 and the 3D value. On the other hand, the probability distribution of projected separations peaks at the 3D separation, with the probability of having small projected separations being very small (see Fig.~\ref{dualagnpaper:fig:projvectordistr}).

In order to understand better the dual-activity results presented later, we first show in Fig.~\ref{dualagnpaper:fig:m2_hr_gf0_3_BHeff0_001_phi000000_dist_vel_Mdot_deltat_proj_no3b_1a1btogether} the orbital history of the encounter in a complementary way, by computing the  time spent {\it above} a given $r$ or $v$. We also consider the results on a stage-by-stage basis, in order to more easily interpret the importance of each phase for the dual-activity time-scales. The stochastic stage (total time 0.849~Gyr) is clearly the dominant stage for all $r$ and all $v$, except at very high $v$, when the time spent during the merger stage (total time 0.206~Gyr) is comparable. The remnant stage (total time 0.103~Gyr), on the other hand, is never visible in the $r$-panel (because the largest separation during such stage is $< 0.1$~kpc) and negligible in the $v$-panel (especially for $v \gtrsim 100$~km~s$^{-1}$, the typical range for spectroscopy surveys).


\subsubsection{Dual activity}\label{dualagnpaper:sec:Dual_activity}

In Fig.~\ref{dualagnpaper:fig:m2_hr_gf0_3_BHeff0_001_phi000000_dual_agn_Mdot_deltat_proj_no3b}, we show the amount of  time the two BHs spend {\it above} a given $r$ or $v$ when they accrete above a given threshold (bolometric -- $L_{\rm bol}$ -- or hard-X-ray -- $L_{\rm 2-10\,keV}$ -- luminosity, or Eddington ratio -- $f_{\rm Edd}$). We show it in a `cumulative' way (time spent {\it above} $r$ or $v$, rather than time spent {\it at} $r$ or $v$) to highlight the dependence on any {\it minimum} projected separation or velocity difference, to mimic observational limitations. To calculate the X-ray time-scales, we applied a simple inverse bolometric correction \citep[][]{Hopkins2007}: $L_{\rm 2-10\,keV} = L_{\rm bol} [c_1(L_{\rm bol}/10^{10}\,$L$_{\odot})^{k_1}+c_2(L_{\rm bol}/10^{10}\,$L$_{\odot})^{k_2}]^{-1}$, where $c_1 = 10.83$, $k_1 = 0.28$, $c_2 = 6.08$, and $k_2 = -0.02$. The solid line in the bolometric and X-ray panels represents what is typically used as the threshold for AGN activity: $10^{43}$ \citep[i.e. $2.6 \times 10^9$~L$_{\odot}$; e.g.][]{Steinborn_et_al_2016} and $10^{42.3}$~erg~s$^{-1}$ \citep[e.g.][]{Silverman2011}, respectively.

\begin{figure}
\centering
\vspace{3.0pt}
\includegraphics[width=0.99\columnwidth,angle=0]{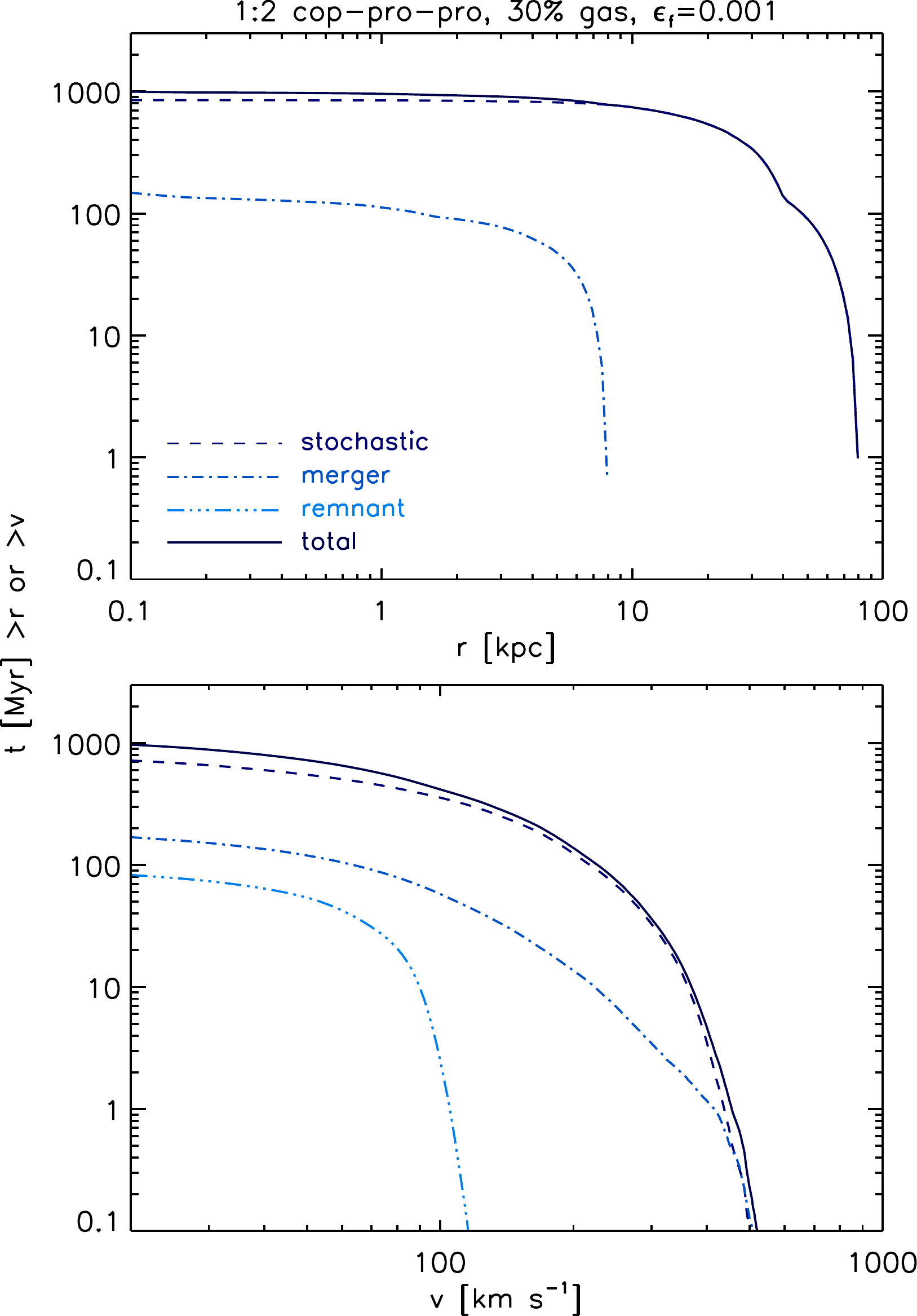}
\vspace{-5.0pt}
\caption[]{For the default merger: time spent {\it above} a given projected separation (top panel) and projected velocity difference (bottom panel) between the two BHs, during the stochastic stage (dash), the merger stage (dash-dot), the remnant stage (dash-triple-dot), and the entire encounter (solid). The remnant stage is actually never visible in the $r$-panel, because the BH projected separation during such stage is always lower than 0.1~kpc, and is negligible in the $v$-panel.}
\label{dualagnpaper:fig:m2_hr_gf0_3_BHeff0_001_phi000000_dist_vel_Mdot_deltat_proj_no3b_1a1btogether}
\end{figure}

\begin{figure*}
\centering
\vspace{2.0pt}
\includegraphics[width=1.70\columnwidth,angle=90]{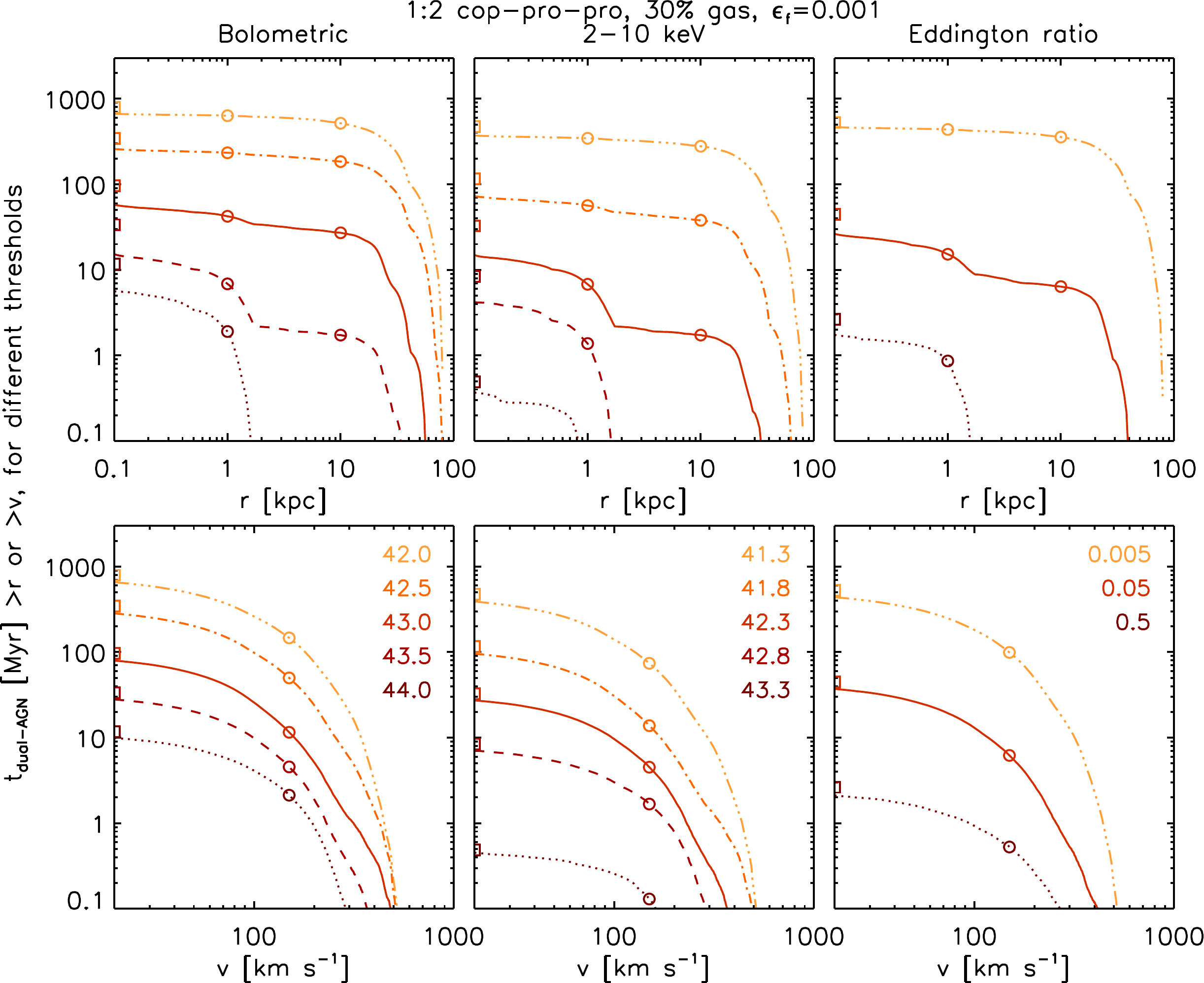}
\vspace{5.0pt}
\caption[]{For the default merger: dual-activity time {\it above} a given projected separation (top panels) and projected velocity difference (bottom panels) between the two BHs, for several bolometric [left panels; $10^{42}$ (dash-triple-dot), $10^{42.5}$ (dash-dot), $10^{43}$ (solid), $10^{43.5}$ (dash), and $10^{44}$ (dot) erg~s$^{-1}$], 2--10~keV [middle panels; $10^{41.3}$ (dash-triple-dot), $10^{41.8}$ (dash-dot), $10^{42.3}$ (solid), $10^{42.8}$ (dash), and $10^{43.3}$ (dot) erg~s$^{-1}$], and Eddington-ratio [right panels; 0.005 (dash-triple-dot), 0.05 (solid), and 0.5 (dot)] thresholds. The squares indicate the dual-activity time regardless of any $r$ or $v$ filter (i.e. $r = v = 0$). The circles refer to the following projected separations and velocity differences: 1 and 10~kpc, and 150~km~s$^{-1}$ (see Table~\ref{dualagnpaper:tab:m2_hr_gf0_3_BHeff0_001_phi000000}). The breaks visible in the upper panels occur when the BHs are at apocentre, i.e. when the BHs spend a relatively large amount of time (e.g. the break clearly visible at $\sim$2~kpc is due to the third apocentric passage at $\sim$0.96~Gyr; see Fig.~\ref{dualagnpaper:fig:m2_hr_gf0_3_BHeff0_001_phi000000_three_panels_dual}). Similar figures, for all other mergers, can be found in the online-only material.}
\label{dualagnpaper:fig:m2_hr_gf0_3_BHeff0_001_phi000000_dual_agn_Mdot_deltat_proj_no3b}
\end{figure*}

The dual-activity curves at low luminosity thresholds (e.g. $L_{\rm bol} = 10^{42}$~erg~s$^{-1}$ or $L_{\rm 2-10\, keV} = 10^{41.3}$~erg~s$^{-1}$) follow the orbital history of the encounter (compare with Fig.~\ref{dualagnpaper:fig:m2_hr_gf0_3_BHeff0_001_phi000000_dist_vel_Mdot_deltat_proj_no3b_1a1btogether}). This is because, at low thresholds, the BHs are both active for a significant fraction of the encounter. When the threshold increases, the dual-activity time obviously decreases, but also ceases to follow the orbital history. Activity (and especially simultaneous activity) is high only during the merger stage, when gas gets efficiently funnelled into the central regions of the galaxies due to tidal torques and/or ram-pressure shocks \citep[][]{Capelo_Dotti_2017}. During the initial part of the merger stage (i.e. between the second and third pericentric passages), only the primary BH is active (see Fig.~\ref{dualagnpaper:fig:m2_hr_gf0_3_BHeff0_001_phi000000_three_panels_dual}). This is the reason why, at e.g. $L_{\rm bol} = 10^{43.5}$~erg~s$^{-1}$ or $L_{\rm 2-10\, keV} = 10^{42.3}$~erg~s$^{-1}$, we see a break in the dual-activity curves at $r \sim 2$~kpc, which is the separation at third apocentre, rather than at $\sim$8~kpc (second apocentre).

If we consider a typical bolometric threshold for AGN activity ($10^{43}$~erg~s$^{-1}$) and typical thresholds for resolving dual AGN in imaging (1 and 10~kpc) and spectroscopy (150~km~s$^{-1}$), we obtain a dual-activity observability time-scale equal to 3.7, 2.3, and 1.0 per cent, respectively, of the total encounter time (see Run~02 in Table~\ref{dualagnpaper:tab:comparison}). These fractions increase to 4.4, 3.7, and 4.5 per cent when we divide the dual-activity time by the time spent above those separations and velocity differences (which can be compared to theoretical results from cosmological simulations which provide the number of dual AGN out of the total number of BH pairs -- see Section~\ref{dualagnpaper:sec:Comparison_with_simulations}). The numbers increase further, to 12.5, 12.7, and 14.1 per cent, when we consider the dual-activity time divided by the  activity time, defined as the time when at least one of the BHs is active, spent above those $r$ and $v$: the proper comparison, in this case, is to observational results which provide the number of dual AGN out of the total number of pairs in which there is at least one AGN (see Section~\ref{dualagnpaper:sec:Comparison_with_observations}).

\begin{figure*}
\centering
\vspace{-1.0pt}
\begin{overpic}[width=1.55\columnwidth,angle=0]{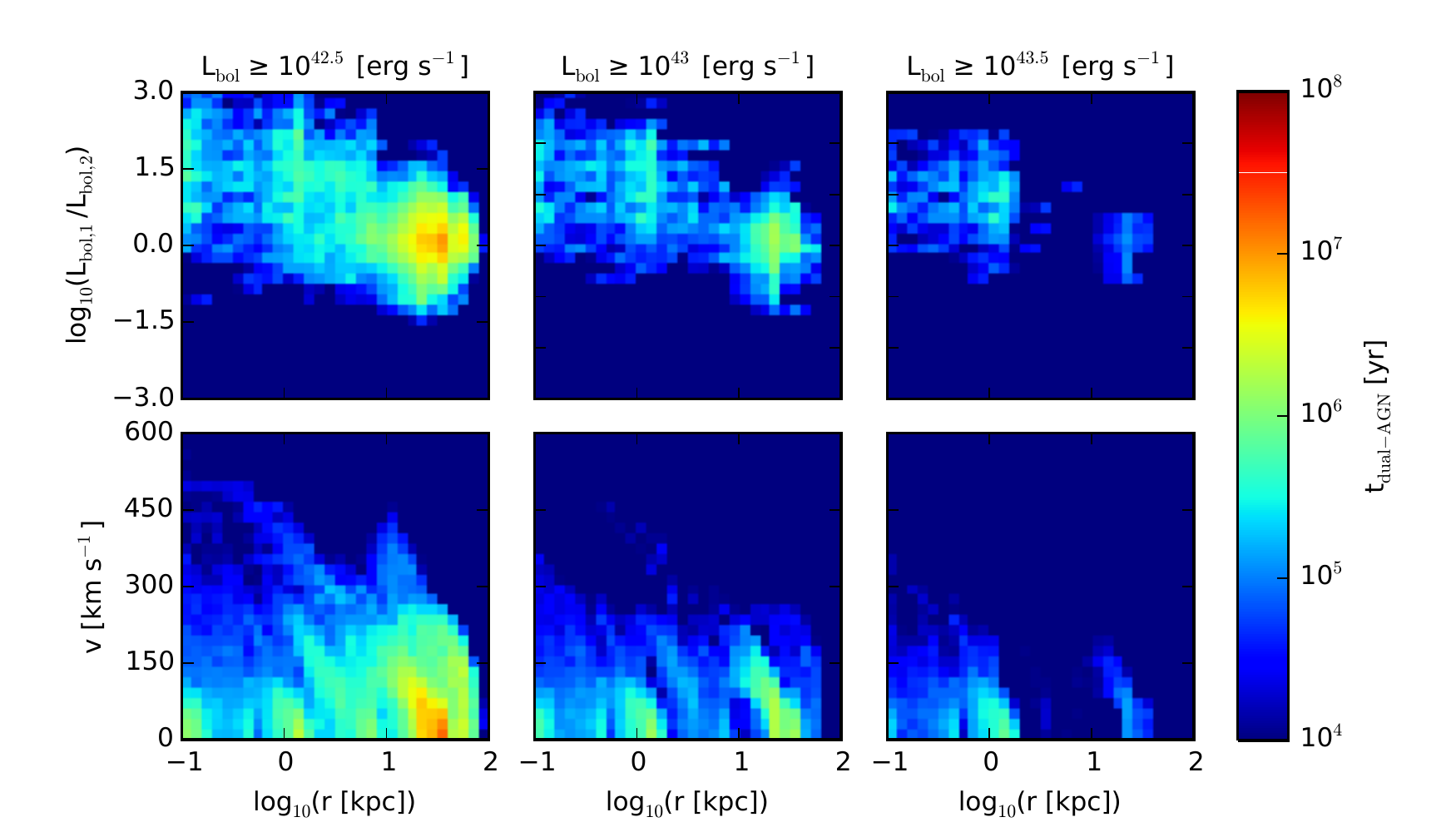}
\put (29,54.5) {\textcolor{black}{1:2 cop-pro-pro, 30\% gas, $\epsilon_{\rm f}=0.001$}}
\end{overpic}\\
\vspace{-4.0pt}
\begin{overpic}[width=1.55\columnwidth,angle=0]{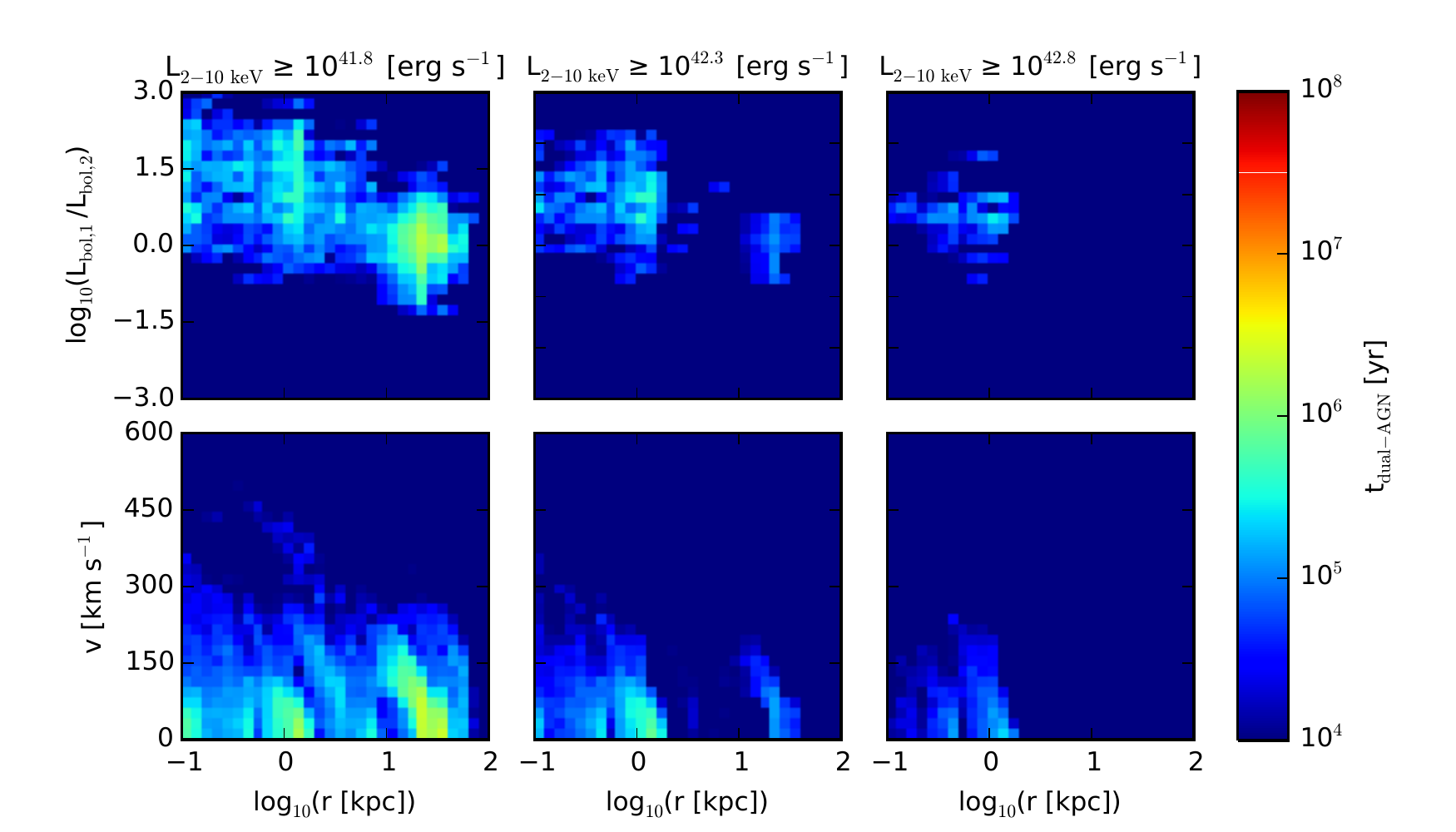}
\end{overpic}\\
\vspace{-4.0pt}
\begin{overpic}[width=1.55\columnwidth,angle=0]{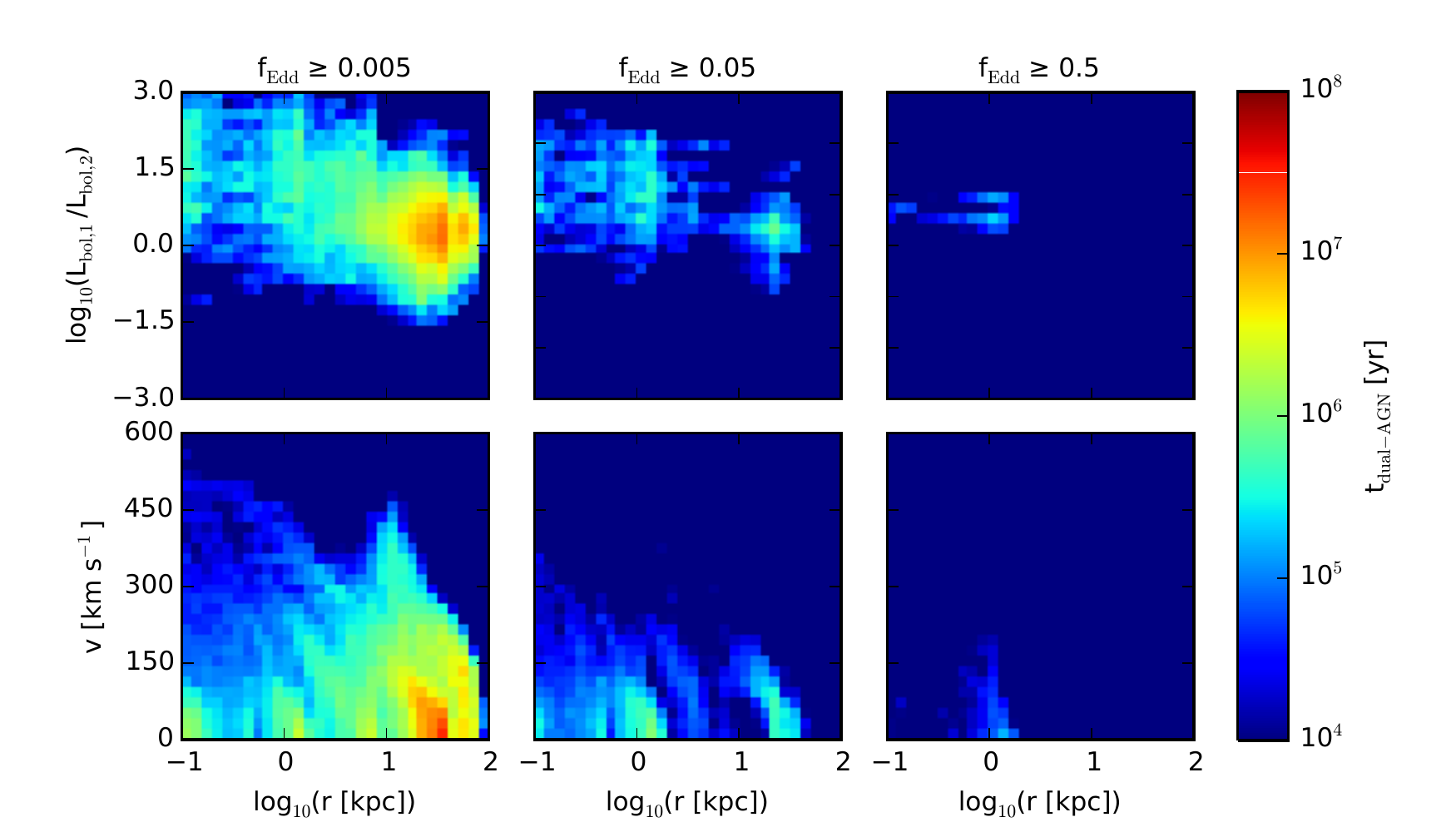}
\end{overpic}
\caption[]{For the default merger: dual-activity time as a function of projected separation and velocity difference, and bolometric-luminosity ratio between the two BHs, for three bolometric [top six panels; $10^{42.5}$ (left), $10^{43}$ (centre), and $10^{43.5}$ (right) erg~s$^{-1}$], 2--10~keV [middle six panels; $10^{41.8}$ (left), $10^{42.3}$ (centre), and $10^{42.8}$ (right) erg~s$^{-1}$], and Eddington-ratio [bottom six panels; 0.005 (left), 0.05 (centre), and 0.5 (right)] thresholds. Each side is divided in 30 bins ($\Delta \log_{10} (r \, [{\rm kpc}])= 0.1$, $\Delta v = 20$~km~s$^{-1}$, and $\Delta \log_{10} (L_{\rm bol,1}/L_{\rm bol,2}) = 0.2$). Similar figures, for all other mergers, can be found in the online-only material.}
\label{dualagnpaper:fig:m2_hr_gf0_3_BHeff0_001_phi000000_dual_agn_map_proj_100los_stageall_loglog_no3b}
\end{figure*}

In Fig.~\ref{dualagnpaper:fig:m2_hr_gf0_3_BHeff0_001_phi000000_dual_agn_map_proj_100los_stageall_loglog_no3b}, the dual-activity time is shown as a function of two variables, chosen from $r$, $v$, and $L_{\rm bol,1}/L_{\rm bol,2}$ (we note that {\it only} in this figure we show the dual-activity time {\it at} $r$ and/or $v$, rather than {\it above} $r$ and/or $v$). 
The top three panels (in each group of six panels) show the bolometric-luminosity ratio between the two active BHs, as a function of projected separation. This ratio can be as high as $10^3$, but only at small separations ($r \lesssim 2$~kpc), when the two galaxies are likely interacting. At larger separations ($r \gtrsim 8$~kpc), when the two galaxies are mostly isolated, activity is much lower. This means that, in the case both BHs are active, they are likely to be barely active (i.e. their luminosity is just above the given threshold), which is the reason why the bolometric-luminosity ratio is close to unity (and approaches unity with increasing $r$). Overall, the ratio is more often above unity than below: the primary, more massive BH is on average more luminous than the secondary (see \citealt{Capelo_et_al_2015} for more discussion on this). The `cloud' of dual activity around $\sim$20--40~kpc is simply due to the fact that the BHs spend a large amount of time at that distance during the first apocentre.

The BHs are active mostly when their projected velocity difference is $v \lesssim 150$~km~s$^{-1}$. These results are given for projected quantities, as opposed to 3D quantities. The time-scales do in fact change quite significantly, especially when filtering by $v$, and we quantify it in Section~\ref{Effects_of_projection} (see Figs~\ref{dualagnpaper:fig:m2_hr_gf0_3_BHeff0_001_phi000000_dual_agn_morelos_no3b}--\ref{dualagnpaper:fig:m2_hr_gf0_3_BHeff0_001_phi000000_dual_agn_map_proj_1los_stageall_loglog_no3b}).

\subsection{X-ray obscuration and dilution}\label{dualagnpaper:sec:X-ray_obscuration_and_dilution}

\begin{figure}
\centering
\vspace{3.0pt}
\includegraphics[width=0.99\columnwidth,angle=0]{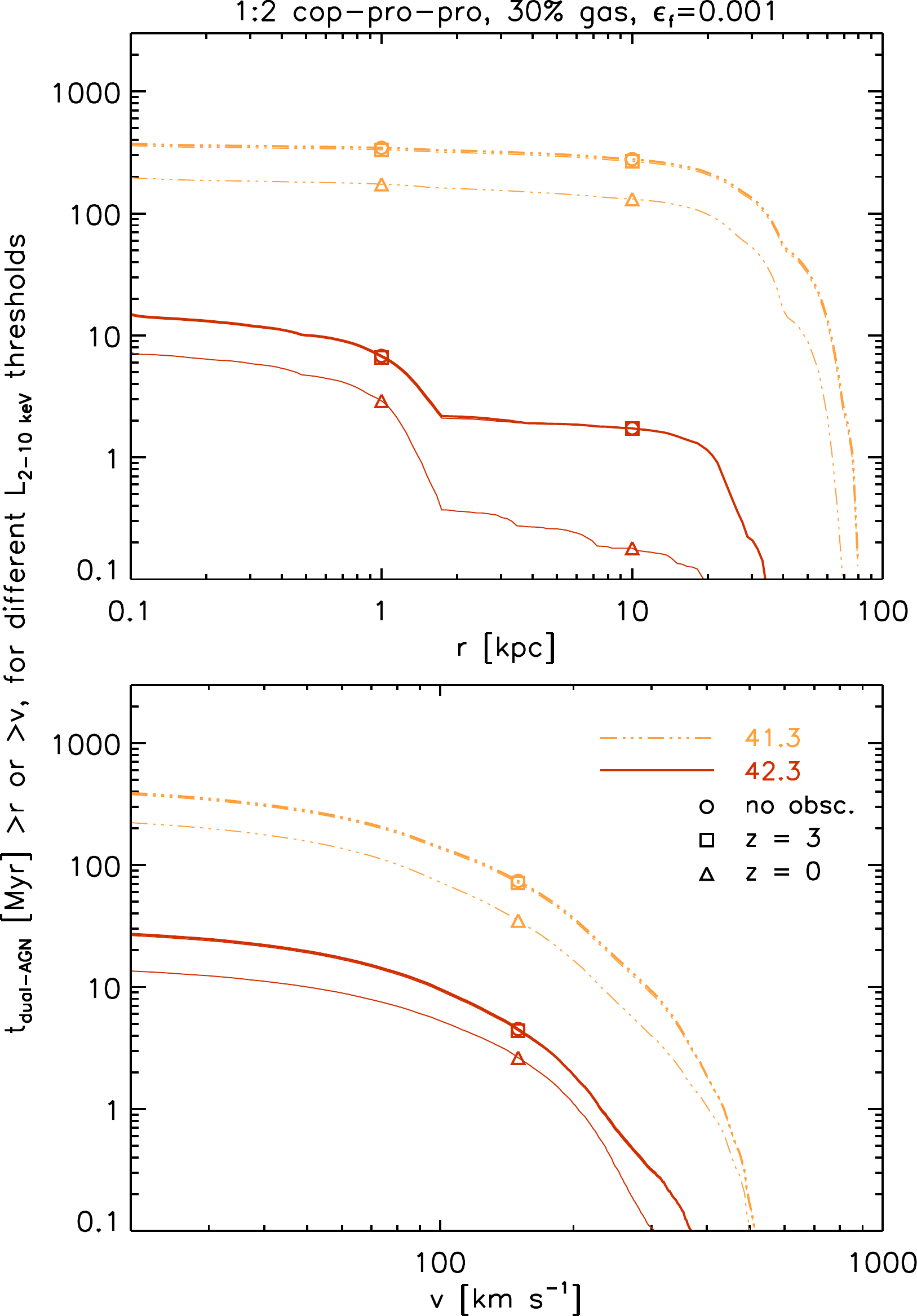}
\vspace{-5.0pt}
\caption[]{For the default merger: dual-activity time above a given projected separation (top panel) and projected velocity difference (bottom panel) between the two BHs, for two 2--10~keV thresholds [$10^{41.3}$ (dash-triple-dot) and $10^{42.3}$ (solid) erg~s$^{-1}$]. The symbols refer to the following projected separations and velocity differences: 1 and 10~kpc, and 150~km~s$^{-1}$. For each luminosity threshold, we show three cases: the system is not obscured (thick lines and circles; same as the middle panels of Fig.~\ref{dualagnpaper:fig:m2_hr_gf0_3_BHeff0_001_phi000000_dual_agn_Mdot_deltat_proj_no3b}); the system is obscured and assumed to be at $z = 3$ (case A; thin lines and squares); and the system is obscured and assumed to be at $z = 0$ (case B; thin lines and triangles). Case A is almost not visible, as it is almost indistinguishable from the case with no obscuration.}
\label{dualagnpaper:fig:m2_hr_gf0_3_BHeff0_001_phi000000_dual_agn_no3b_obsc}
\end{figure}

The results described in the previous section assume that the intrinsic luminosity is affected for the observer only by distance dimming. In reality, observations at optical or X-ray wavelengths are also affected by dimming caused by intervening dust or gas, as well as contamination from stellar sources,  stars or X-ray binaries in optical or X-ray, respectively. In this paper, we only discuss X-ray luminosities, therefore we will consider ($i$) the obscuration produced by the gas surrounding the BH and ($ii$) the contamination (or dilution) from X-ray binaries.

Here we neglect the contribution of the Milky Way's interstellar medium and of the intergalactic medium. We also neglect obscuration in the vicinity of the BH (torus or broad-line region), below our resolution, as the correction is statistical, i.e. it depends on the inclination with respect to the line of sight and is not related to properties of the merger.  We focus instead on the role of the gas belonging to the BHs' hosts, which is expected to screen the AGN differently during the merger. 

The column density, $N_{\rm H}$, can obviously vary wildly during the encounter and is also dependent on the line of sight: the central BH of an isolated galaxy viewed face-on will be less obscured than that in a merging galaxy viewed edge-on. Since here we are only interested in order-of-magnitude estimates, we consider one typical column density per BH per time-step (i.e. every 5~Myr, the frequency of the main outputs). To obtain these values, we first compute the average face-on column density by calculating the total gas mass inside a cylinder of radius $R_{\rm cyl} = 0.1$~kpc (five times the gas gravitational softening), whose axis is parallel to the $z$-axis and includes the position of the BH, between the BH itself and the observer (at infinity). Thus, we have two masses (one per direction: $+z$ and $-z$), which we average. We then divide this average mass by $\mu \,m_{\rm H} S$, where $S = \pi R_{\rm cyl}^2$, to obtain the face-on column density. We do the same for the edge-on column density, this time averaging four different gas masses, using cylinders with their axes parallel to the $x$ and $y$ axes. At the beginning of the stochastic stage, we obtain a column density of $\sim$$3 \times 10^{22}$ and $\sim$$4 \times 10^{23}$~cm$^{-2}$, for the primary BH's face-on and edge-on case, respectively, which is very close to what is expected analytically from the initial conditions (the factor of $\sim$10 difference comes from the factor of 10 between $r_{\rm disc}$ and $z_{\rm disc}$). We computed the same quantities in the case when BHs do not accrete \citep[Run~C1 in][]{Capelo_et_al_2015}, to check for dependency on BH accretion and feedback, and obtained similar values. Finally, we average the two column densities (face-on and edge-on) to obtain a typical column density $N_{\rm H}$ for each BH every 5~Myr. We find that $N_{\rm H}$ is always very close to $10^{23}$~cm$^{-2}$ during the stochastic and remnant stages and within less than one order of magnitude from such value during the merger stage. We computed the same quantities also for the merger with 60 per cent gas fraction (Run~06 in Tables~\ref{dualagnpaper:tab:merger_params}--\ref{dualagnpaper:tab:comparison}) and found similar values. The values for the secondary BH are very similar to those of the primary BH, as expected, since the (analytical) column density ratio is proportional to the cube root of the mass ratio. We believe that the little variation in column density, compared, e.g. to that found in \citet{Hopkins_et_al_2012}, is due to the relatively coarser resolution of our simulations.

Using the online tool {\scshape webpimms},\footnote{\url{https://heasarc.gsfc.nasa.gov/Tools/w3pimms.html} (powered by {\scshape pimms} version 4.8d; \citealt{Mukai_1993}). The settings used are the following: Convert from `Unabsorbed Flux' into `Flux'; Input Energy Range = Output Energy Range = 2--10~keV; Galactic $N_{\rm H} = 0.0$; Power Law with Photon Index = 1.8 \citep[the same used in our inverse bolometric correction;][]{Hopkins2007}; redshift = 0 or 3; Intrinsic $N_{\rm H}$ was computed from the outputs. For $N_{\rm H} = 10^{23}$~cm$^{-2}$: $f_{\rm through} = 0.559$ and 0.965, for $z = 0$ and 3, respectively.} we calculated the detection factor, $f_{\rm through}$, defined as the fraction of detected over unabsorbed flux. We considered two redshifts:  $z = 3$, since our galaxies are constructed as typical of that redshift (case A), and $z = 0$,  in which our galaxies are representative of relatively small, gas-rich local galaxies (case B). The obscuration for case A is almost negligible, whereas for case B it produces a change in the dual-activity time-scales of a factor $\sim$2, more prominent at high thresholds (Fig.~\ref{dualagnpaper:fig:m2_hr_gf0_3_BHeff0_001_phi000000_dual_agn_no3b_obsc}). We note that these calculations were done for the 2--10~keV band. Observations with harder X-rays (i.e. higher energy) would be less obscured, and vice versa \citep[see, e.g.][]{Koss2016}.

Amongst the many other galactic sources of X-rays, the dominant source is the population of high-mass X-ray binaries (HMXB). Since high-mass stars are short-lived, X-ray binaries follow closely the star formation rate (SFR): $L_{\rm HMXB} = 6.7 \times 10^{39} \times$~SFR~[M$_{\odot}$~yr$^{-1}$]~erg~s$^{-1}$ \citep{Grimm_et_al_2003}. Since the global SFR in our systems never exceeds $\sim$20~M$_{\odot}$~yr$^{-1}$ \citep{Capelo_et_al_2015,Volonteri_et_al_2015a,Volonteri_et_al_2015b}, the X-ray luminosity from the stellar component is always $<1.3 \times 10^{41}$~erg~s$^{-1}$ and reaches such values only for brief bursts during the merger stage. This is one order of magnitude lower than the typical X-ray threshold ($10^{42.3}$~erg~s$^{-1}$) used to define AGN activity. Hence, contamination from X-ray binaries is negligible.


\pagebreak
\subsection{Dependence on merger properties}\label{dualagnpaper:sec:Dependence_on_merger_properties}

\begin{figure}
\centering
\vspace{3.0pt}
\includegraphics[width=0.99\columnwidth,angle=0]{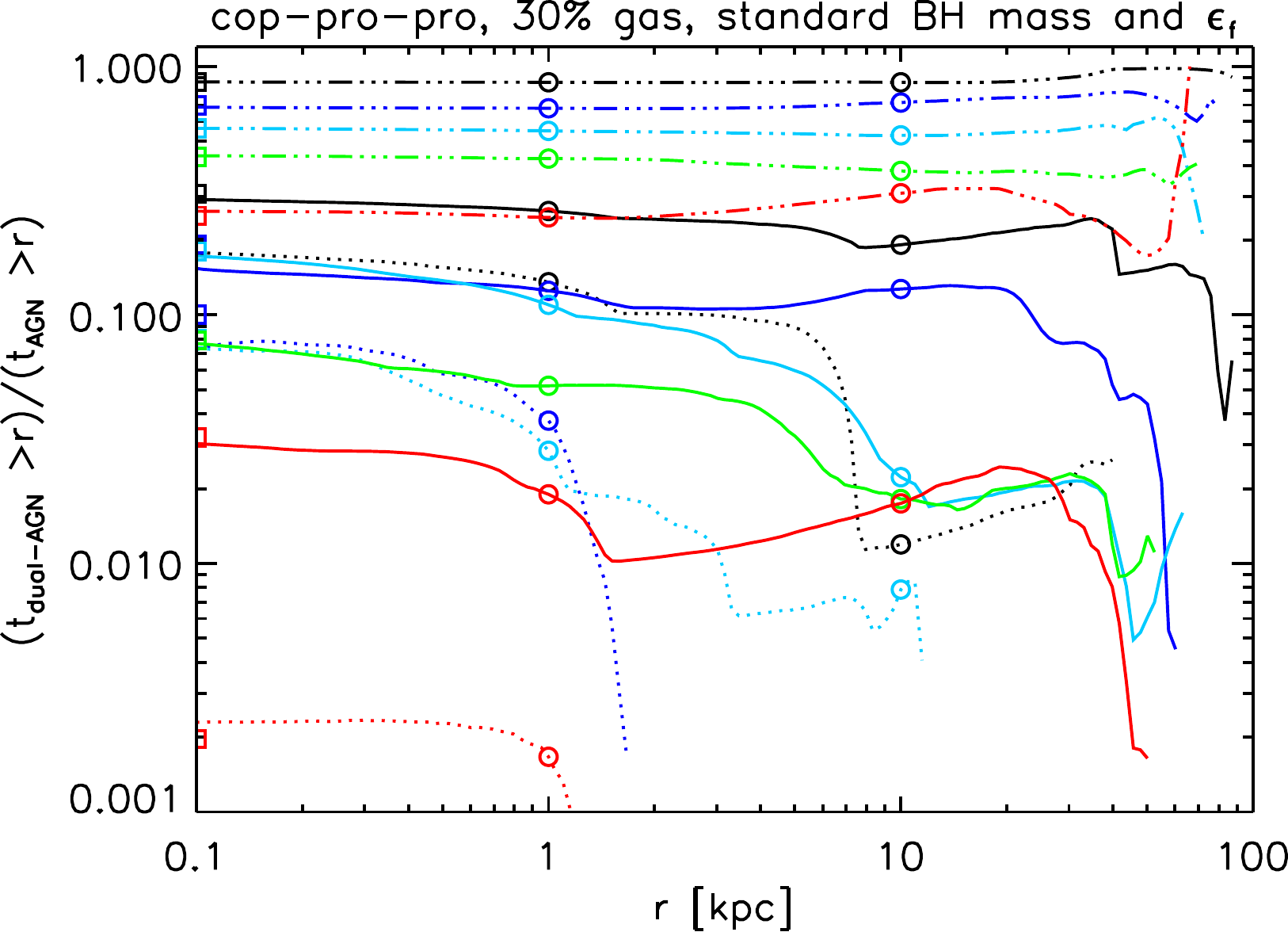}
\vspace{-5.0pt}
\caption[]{Dual-activity time above a given projected separation between the two BHs, divided by the activity time above the same $r$, using three bolometric thresholds [$10^{42}$ (dash-triple-dot), $10^{43}$ (solid), and $10^{44}$ (dot) erg~s$^{-1}$], for the coplanar, prograde--prograde mergers with 30 per cent gas fraction, standard BH mass and feedback efficiency, and different mass ratios (1:1: black, total encounter time 0.978~Gyr; 1:2: blue, 1.158~Gyr; 1:4: cyan, 1.271~Gyr; 1:6: green, 1.682~Gyr; and 1:10: red, 3.305~Gyr). The squares and circles refer to the following projected separations: 0, 1, and 10~kpc. The seemingly random behaviour of the curves at high $r$ is due to statistical fluctuations, as the time spent at those high projected separations is relatively low (see Fig.~\ref{dualagnpaper:fig:m2_hr_gf0_3_BHeff0_001_phi000000_dist_vel_Mdot_deltat_proj_no3b_1a1btogether}). At $r = 1$~kpc, the normalized dual-activity time is highest for the 1:1 merger, followed by the 1:2, 1:4, 1:6, and 1:10 mergers, regardless of the luminosity threshold shown (except for $L_{\rm thr} = 10^{44}$~erg~s$^{-1}$, when the 1:6 merger is not visible).}
\label{dualagnpaper:fig:dual_agn_Bolom_Mdot_deltat_proj_norm2_no3b_coppropro_onlyr_3thr}
\end{figure}

\begin{table*} \centering
\vspace{-3.5pt}
\caption[]{Dual-activity observability time-scales for all mergers (numbered as in Table~\ref{dualagnpaper:tab:merger_params}), assuming a threshold $L_{\rm bol} = 10^{43}$~erg~s$^{-1}$ and using different imaging (1 and 10~kpc) and spectroscopy (150~km~s$^{-1}$) filters, normalized in three different ways: (1) to the total encounter time (see also Fig.~\ref{dualagnpaper:fig:dual_agn_Bolom_Mdot_deltat_proj_norm1_no3b_coppropro_onlyr_3thr}); (2) to the time spent above the same filter (see also Figs~\ref{dualagnpaper:fig:dual_agn_Bolom_Mdot_deltat_proj_norm3_no3b_coppropro_onlyr_3thr} and \ref{dualagnpaper:fig:dual_agn_time_norm5_3thr}); and (3) to the activity time spent above the same filter (see also Fig.~\ref{dualagnpaper:fig:dual_agn_time_norm4_3thr}). All numbers are fractions. 
\label{dualagnpaper:tab:comparison}}
\vspace{5pt}
\begin{tabular*}{0.97\textwidth}{m{39pt}|m{38pt}m{38pt}m{38pt}|m{38pt}m{38pt}m{38pt}|m{38pt}m{38pt}m{38pt}}
Run    & \multicolumn{3}{c|}{1~kpc} & \multicolumn{3}{c|}{10~kpc} & \multicolumn{3}{c}{150~km~s$^{-1}$} \B \\
                      & (1) & (2) & (3) & (1) & (2) & (3) & (1) & (2) & (3) \B \\
\hline
01 &  0.126 &  0.137 &  0.262 &  0.062 &  0.086 &  0.192 &  0.024 &  0.083 &  0.187 \T \B \\
02 &  0.037 &  0.044 &  0.125 &  0.023 &  0.037 &  0.127 &  0.010 &  0.045 &  0.141 \T \B \\
03 &  0.089 &  0.097 &  0.265 &  0.017 &  0.024 &  0.091 &  0.009 &  0.038 &  0.116 \T \B \\
04 &  0.051 &  0.080 &  0.269 &  0.005 &  0.011 &  0.050 &  0.007 &  0.032 &  0.105 \T \B \\
05 &  0.072 &  0.096 &  0.313 &  0.006 &  0.010 &  0.050 &  0.017 &  0.077 &  0.227 \T \B \\
06 &  0.080 &  0.087 &  0.184 &  0.052 &  0.071 &  0.155 &  0.040 &  0.152 &  0.268 \T \B \\
07 &  0.027 &  0.029 &  0.110 &  0.003 &  0.004 &  0.022 &  0.010 &  0.051 &  0.169 \T \B \\
08 &  0.012 &  0.023 &  0.087 &  0.006 &  0.014 &  0.081 &  0.003 &  0.020 &  0.074 \T \B \\
09 &  0.007 &  0.008 &  0.052 &  0.001 &  0.001 &  0.018 &  0.002 &  0.015 &  0.068 \T \B \\
10 &  0.004 &  0.005 &  0.019 &  0.002 &  0.003 &  0.017 &  0.001 &  0.007 &  0.024 \T \B \\
C3 &  0.004 &  0.006 &  0.060 &  0.000 &  0.000 &  0.011 &  0.005 &  0.020 &  0.140 \T \B \\
C4 &  0.085 &  0.097 &  0.201 &  0.042 &  0.062 &  0.156 &  0.032 &  0.134 &  0.275 \T \B \\
\end{tabular*}
\vspace{5pt}
\end{table*}

In this section, we study the dependence of the dual-activity observability time on several parameters of both galaxies and mergers, namely the initial mass ratio of the merger, the initial geometry, the initial gas fraction, and, mostly for control reasons, the BH mass and feedback efficiency. We focus on the relative differences, to highlight the effects of the above mentioned parameters, rather than on the absolute results, which are instead the focus of Section~\ref{dualagnpaper:sec:Comparison}, where we compare our data to other theoretical and observational work. The conclusions in this section should not be affected by the specifics of the initial conditions nor by the sub-grid recipes (e.g. disc and bulge mass fraction; SF recipe), since they are kept the same in this parameter study.


\subsubsection{Initial mass ratio}\label{dualagnpaper:sec:Initial_mass_ratio}

The mass ratio is likely the most important factor in determining BH activity. As already shown in \citet{Capelo_et_al_2015}, the smaller (secondary) galaxy always responds strongly to the interaction (mostly during the merger stage), regardless of the initial mass ratio. On the other hand, the larger (primary) galaxy is affected in different ways depending on the mass of the secondary. In minor mergers, the primary BH shows almost no response, because the secondary galaxy is too little to produce any significant effects (in other words, the primary galaxy is almost `unaware' of the interaction). In major mergers, instead, the companion is large enough to significantly affect the primary and produce higher levels of primary BH activity.

In Fig.~\ref{dualagnpaper:fig:dual_agn_Bolom_Mdot_deltat_proj_norm2_no3b_coppropro_onlyr_3thr}, we show the dual-activity time above a given projected separation, divided by the  activity time above the same $r$, for all five coplanar, prograde--prograde mergers with 30 per cent gas fraction and standard BH mass and feedback efficiency, which have initial mass ratios ranging from 1 to 0.1 (Runs~01, 02, 07, 09, and 10, in Tables~\ref{dualagnpaper:tab:merger_params}--\ref{dualagnpaper:tab:comparison}). The trend with mass ratio is as expected: major mergers are more active than minor mergers.\footnote{The trend is the same also in the $v$ case (not shown in the paper). In case we do not normalize, the longer duration of the minor mergers is not sufficient to offset their intrinsically low levels of activity, except at low thresholds (see Fig.~\ref{dualagnpaper:fig:dual_agn_Bolom_Mdot_deltat_proj_no3b_coppropro_onlyr_3thr}).} We can quantify the ratio of normalized dual-activity time between major and minor mergers, which ranges from factors of a few at $10^{42}$~erg~s$^{-1}$ to $\sim$10 at $10^{43}$~erg~s$^{-1}$, to $\gg 10$ at $10^{44}$~erg~s$^{-1}$.

Fig.~\ref{dualagnpaper:fig:dual_agn_Bolom_Mdot_deltat_proj_norm2_no3b_coppropro_onlyr_3thr} highlights an important result. At low luminosity thresholds ($<10^{43}$~erg~s$^{-1}$) the curves are mostly flat, i.e. independent of $r$. Conversely, at high luminosity thresholds ($10^{44}$~erg~s$^{-1}$) dual activity is captured only at small $r$, when BHs are most active, albeit for short times. The prediction is that, in order to identify dual AGN, shallow surveys will be more affected by an increase of angular resolution, with respect to deeper surveys.

In Fig.~\ref{dualagnpaper:fig:dual_agn_time_norm4_3thr}, we show again the normalized time (for $r = 1$ and 10~kpc and for $v = 150$~km~s$^{-1}$) for the five mergers of this section, together with all the other simulated encounters. The same numbers, together with other normalizations, are summarized in Table~\ref{dualagnpaper:tab:comparison}.


\subsubsection{Initial geometry, gas fraction, and black holes}\label{dualagnpaper:sec:Other_parameters}

The initial geometry, i.e. the angle between the galactic and global angular momentum vectors, is potentially important, because merger-induced tidal torques and ram-pressure shocks are most effective in coplanar encounters. To quantify these effects, we compare four 1:2 mergers: coplanar, prograde--prograde; retrograde--prograde; prograde--retrograde; and inclined-primary  (Runs~02, 03, 04, and 05). We also separately compare two 1:4 mergers: coplanar, prograde--prograde and inclined-primary (Runs~07 and 08). Overall, the effect of geometry, once all other parameters are fixed, is not very significant, as the normalized dual-activity time-scales vary only by a factor of $\sim$2--2.5, with no clear trend: amongst the 1:2 mergers, the coplanar, prograde--prograde merger is the least active above 1~kpc and the most active above 10~kpc (see Fig.~\ref{dualagnpaper:fig:dual_agn_Bolom_Mdot_deltat_proj_norm2_no3b_1to2moregeometries_onlyr_3thr} for more details); the opposite happens for the 1:4 mergers (see Fig.~\ref{dualagnpaper:fig:dual_agn_Bolom_Mdot_deltat_proj_norm2_no3b_1to4moregeometries_onlyr_3thr}).

The initial amount of gas is obviously very important, especially in isolated simulation like ours, where there is no replenishment from the cosmic web. The more gas is available, the more can potentially be funnelled towards the central BHs. In runs~02 and 06, which differ only by the initial gas fraction in the disc (30 versus 60 per cent), overall, activity increases with gas availability, but not significantly: by a factor of $\sim$1.2--1.9, depending on the filter used (see Fig.~\ref{dualagnpaper:fig:dual_agn_Bolom_Mdot_deltat_proj_norm2_no3b_lowvshighgasfrac_onlyr_3thr} for more details).

BH parameters have obviously a direct effect on the dual-activity observability time-scales. Increasing the initial BH mass potentially increases luminosities, as BH accretion in our simulations scales as $M_{\rm BH}^2$ [see Equation~\eqref{dualagnpaper:eq:bondi}], but at the same time the additional power would increase AGN feedback, which is proportional to the luminosity.  If we instead increase the BH feedback efficiency, i.e. the constant of proportionality with luminosity, the gas surrounding the BH is less cold and dense, causing BH accretion to decrease. To quantify these effects on the dual-activity observability time-scales, we compare Runs~02, C3, and C4, which differ only by the initial BH mass or feedback efficiency. Increasing the BH mass (by a factor of 2.5) increases the dual-activity time, but only by a factor $\lesssim$2. Increasing the BH feedback efficiency (by a factor of 5) tends to decrease the normalized dual-activity time, although the factor widely depends on the filter one chooses (see Fig.~\ref{dualagnpaper:fig:dual_agn_Bolom_Mdot_deltat_proj_norm2_no3b_bhphysics_onlyr_3thr} for more details). We note that different prescriptions for BH accretion (e.g. taking into account the angular momentum of the accreting gas; \citealt{Debuhr2010,Hopkins_Quataert_2011,AnglesAlcazar_et_al_2013,AnglesAlcazar_et_al_2015,AnglesAlcazar_et_al_2017}) or AGN feedback \citep[e.g.][]{Wurster_Thacker_2013,Newton_Kay_2013} may also affect the results.


\subsection{Comparison with other work}\label{dualagnpaper:sec:Comparison}

In this section, we compare our dual-activity time-scales to a few select studies, both observational and theoretical, to validate our results.


\subsubsection{Comparison with observations}\label{dualagnpaper:sec:Comparison_with_observations}

Most surveys provide the total number of AGN pairs out of the total number of AGN \citep[e.g.][]{Shen2011}. These results are difficult to compare to ours since, by construction, we do not simulate isolated galaxies which could be harbouring a single AGN.

\citet{Koss2012} study the fraction of dual AGN in the all-sky Swift Burst Alert Telescope (BAT) survey. For each of the 167 BAT AGN from the survey, they search for apparent companions, finding that 77 of them have at least a companion\footnote{For this and the following numbers, we refer to the online-only version of Table~1 of \citet{Koss2012}. When a BAT AGN has more than one companion, we consider only the pair with the largest $M/r^2$, where $M$ is the mass of the BAT AGN's companion and $r$ is the projected separation.} within 1--100~kpc. Of these 77 companions, 15 have an AGN, making the fraction of dual AGN systems out of interacting systems (in which at least one system has a AGN) equal to 19.5 per cent (15/77). If we divide their sample in major and minor pairs (defined as pairs with mass ratios $>$ and $\leq 0.25$, respectively), the fraction changes to 37.1 per cent (13/35) for the major pairs and to 4.8 per cent (2/42) for the minor pairs. Since in this case only interacting systems are considered, we can compare the number from \citet{Koss2012} to our results. To do that, we simply consider, for each of our mergers, the dual-activity time when the BHs are at $r > 1$~kpc (our galaxies, by construction, are never above $r \sim 90$~kpc) and divide it by the  activity time at the same projected separations.\footnote{We do not consider the additional velocity-difference filter $v < 300$~km~s$^{-1}$ performed by \citet{Koss2012} -- done to avoid contamination from the Hubble flow -- because the time spent above 300~km~s$^{-1}$ is negligible (see Figs~\ref{dualagnpaper:fig:m2_hr_gf0_3_BHeff0_001_phi000000_dual_agn_Mdot_deltat_proj_no3b} and \ref{dualagnpaper:fig:m2_hr_gf0_3_BHeff0_001_phi000000_dual_agn_map_proj_100los_stageall_loglog_no3b}).}

\begin{figure}
\centering
\vspace{3.0pt}
\includegraphics[width=0.99\columnwidth,angle=0]{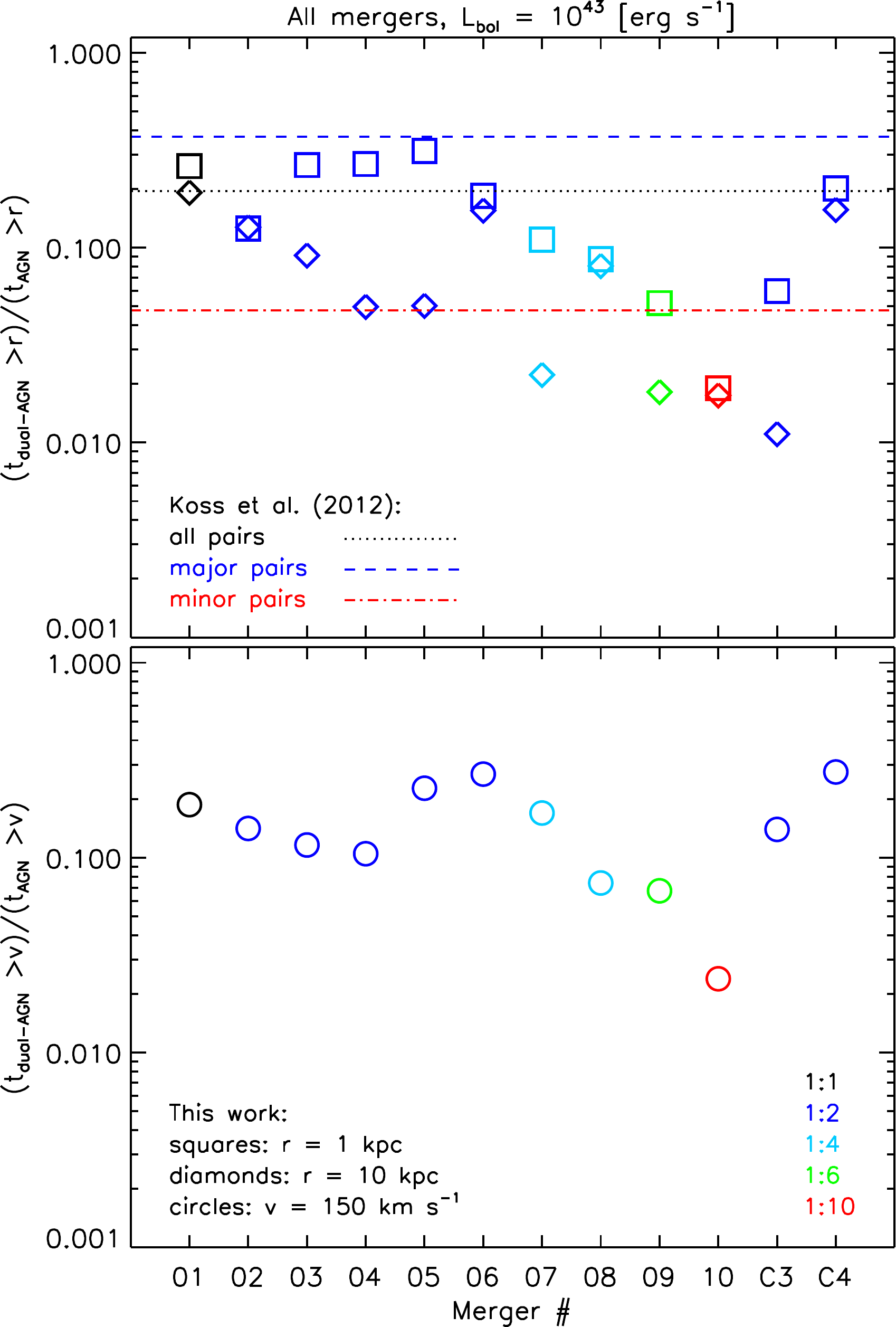}
\vspace{-5.0pt}
\caption[]{Dual-activity time above a given $r$ (top panel: 1~kpc -- squares -- and 10~kpc -- diamonds) and $v$ (bottom panel: 150~km~s$^{-1}$ -- circles), divided by the activity time above the same threshold, assuming a bolometric-luminosity threshold of $10^{43}$~erg~s$^{-1}$, for all the mergers in our suite. The merger number in the $x$-axis is the same as in Tables~\ref{dualagnpaper:tab:merger_params}--\ref{dualagnpaper:tab:comparison}. The colours refer to the initial mass ratio: 1:1 -- black, 1:2 -- blue, 1:4 -- cyan, 1:6 -- green, and 1:10 -- red. The horizontal lines in the top panel show the observed fraction of dual AGN systems out of interacting systems (in which at least one system has a AGN) by \citet{Koss2012}, for their full sample of BAT AGN (black, dot), their major pairs (blue, dash), and their minor pairs (red, dash-dot). Our major mergers with standard BH efficiency (squares of Runs 01--06 and C4) should be compared to the blue, dashed line, whereas our minor mergers (squares of Runs 07--10) should be compared to the red, dash-dotted line.}
\label{dualagnpaper:fig:dual_agn_time_norm4_3thr}
\end{figure}

In Fig.~\ref{dualagnpaper:fig:dual_agn_time_norm4_3thr}, we show the normalized dual-activity time above 1~kpc, 10~kpc, and 150~km~s$^{-1}$, for all our simulated mergers. We also show the results of \citet{Koss2012} for their full sample of BAT AGN and their subsamples of major and minor pairs. If we neglect the 1:2 merger with high BH feedback efficiency, which obviously has a much lower accretion rate \citep[see Section~\ref{dualagnpaper:sec:Other_parameters} and][]{Capelo_et_al_2015}, our results are within a factor of $\sim$2 from the result of \citet{Koss2012}, for both major and minor mergers. We caution the reader that our definitions of activity are very different (we impose a simple luminosity threshold, whereas they use a variety of diagnostics). We also recover the trends found in \citeauthor{Koss2012} (\citeyear{Koss2012}; see their Fig.~2), when we split our results by distance or by mass ratio: the fraction of dual AGN increases with decreasing separation and increasing mass ratio (if we assume that galaxy and BH mass ratio are correlated also in observational samples).

\citet{Comerford_et_al_2015}, in their study of optically selected dual AGN candidates, find that all their dual (and dual/offset) AGN systems have $f_{\rm Edd,1}/f_{\rm Edd,2} < 1$, that is the AGN in the less luminous stellar bulge (assuming that BH masses trace their host's stellar bulge luminosity) has the higher specific accretion rate. This is equivalent to stating ${\rm d}q/{\rm d}t > 0$, where $q \equiv M_{\rm BH,2}/M_{\rm BH,1}$. We have already studied the dependence of $q$ with time \citep[][]{Capelo_et_al_2015} and found that, during the merger stage, which is when dual activity is more frequent, (typically) ${\rm d}q/{\rm d}t < 0$ for major mergers and $> 0$ for minor mergers, indicating that not all our dual AGN have $f_{\rm Edd,1}/f_{\rm Edd,2} < 1$. We note, however, that the only clear dual AGN in \citet{Comerford_et_al_2015} is a minor merger.

\subsubsection{Comparison with simulations}\label{dualagnpaper:sec:Comparison_with_simulations}

We begin comparing our results to those of \citet{Van_Wassenhove_et_al_2012}, since our setup is very similar to theirs. Taking the typical bolometric-luminosity threshold of $10^{43}$~erg~s$^{-1}$, we obtain a dual-activity time equal to 16.8 per cent of the activity time, compared with their 16.3 per cent (we only compare the {\it total} numbers, i.e. when $r = v = 0$, since we use projected quantities and \citealt{Van_Wassenhove_et_al_2012} use 3D quantities).

We then compare our results to recent cosmological studies which, although lacking in resolution and the ability to control parameters, have the advantage of modelling relatively large areas and therefore increasing the number of galaxies one can study.

\citet{Steinborn_et_al_2016} employ a cosmological simulation with a large volume of (182~Mpc)$^3$ from the set of the Magneticum Pathfinder Simulations, using the TreePM-SPH code {\scshape gadget3} \citep{Springel_2005} and running it up to $z = 2$. If we assume that our encounters start at $z = 3$, then our major mergers (which last $\sim$1~Gyr) are thought to be at the same final redshift considered in \citet{Steinborn_et_al_2016}. From the simulation data, they count 34 BH pairs (defined as two BHs with a 3D separation in the range 2--10~kpc), of which 9 are dual AGN (with an AGN defined as a BH with bolometric luminosity higher than $10^{43}$~erg~s$^{-1}$). Since the pairs are not chosen, as it was in the case of \citet{Koss2012}, starting from one AGN and then looking for a companion, we need to compare their results to a different normalization: the dual-activity time divided by the merger time, for projected separations of 2--10~kpc.

\begin{figure}
\centering
\vspace{3.0pt}
\includegraphics[width=0.99\columnwidth,angle=0]{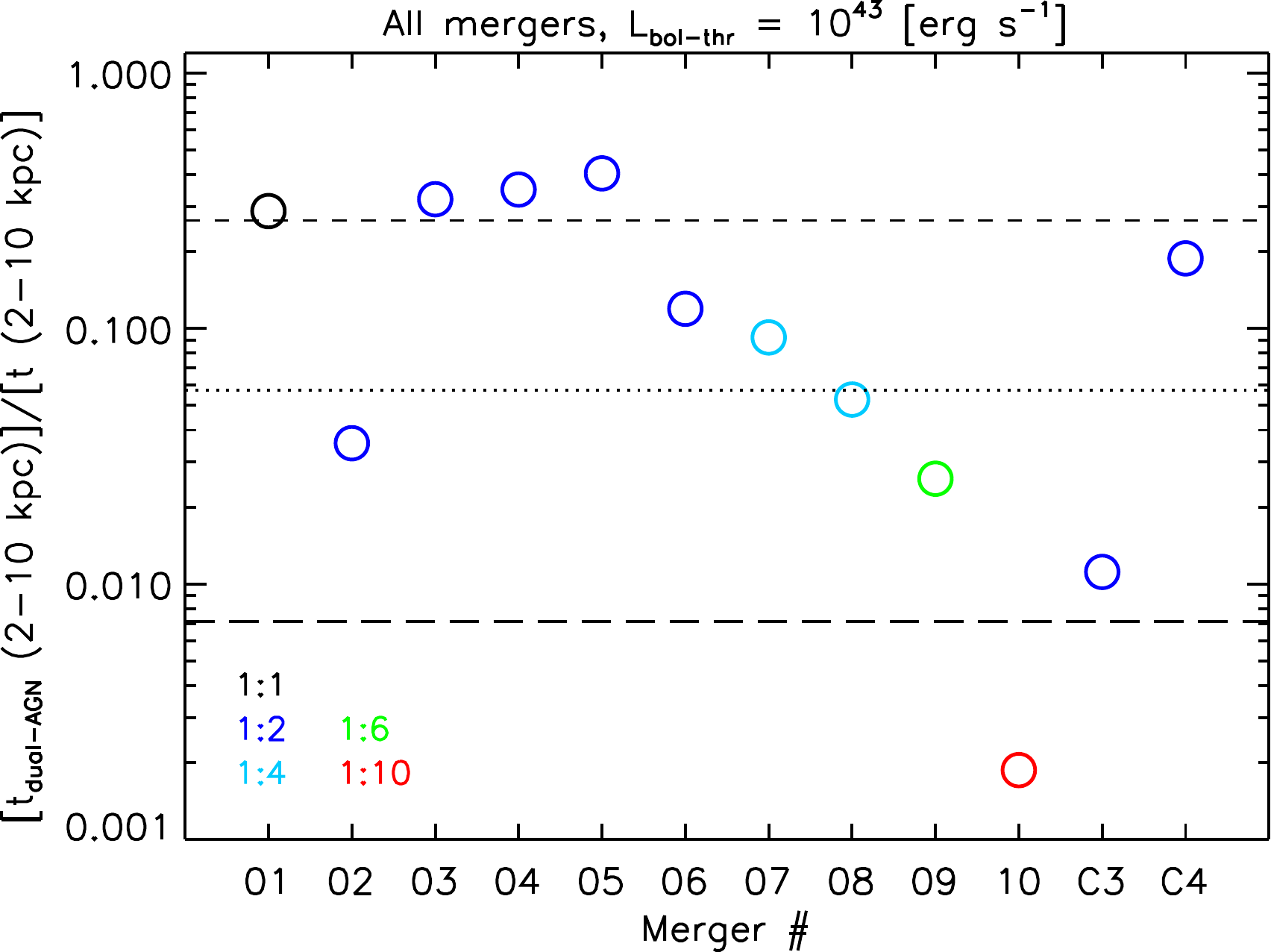}
\vspace{-5.0pt}
\caption[]{Same as the top panel of Fig.~\ref{dualagnpaper:fig:dual_agn_time_norm4_3thr}, but for the  ratio of dual-activity time to merger time, in the projected-separation range 2--10~kpc. The horizontal, dashed line shows the result of \citet{Steinborn_et_al_2016}, whereas the horizontal, long-dashed and dotted lines show the results of \citet{Volonteri_et_al_2016} for galaxies with galaxy masses greater than $10^{10}$~M$_{\odot}$ at redshifts 0 and 2, respectively.}
\label{dualagnpaper:fig:dual_agn_time_norm6_3thr}
\end{figure}

The results are shown in Fig.~\ref{dualagnpaper:fig:dual_agn_time_norm6_3thr}. Each symbol represents one of the simulated encounters, whereas the horizontal, dashed line shows the result by \citet{Steinborn_et_al_2016}: 26.5 per cent dual-AGN fraction. The dual AGN in \citet{Steinborn_et_al_2016} are mostly major (seven out of nine have a mass ratio greater than 0.25) and all our major mergers with standard BH feedback efficiency (except for one) are within a factor of $\sim$2 from their result. We caution that the galaxy and BH masses considered in \citet{Steinborn_et_al_2016} are higher than ours. Nevertheless, we are consistent with their result. For further comparison, we also show the results for galaxies with galaxy mass $> 10^{10}$~M$_{\odot}$ at redshifts 0 and 2 from the Horizon-AGN simulation \citep[][]{Volonteri_et_al_2016}. We caution the reader that, even though they use the same bolometric-luminosity threshold ($10^{43}$~erg~s$^{-1}$) as in \citet{Steinborn_et_al_2016}, they do not use a fixed separation threshold.

We also checked the trends found in \citeauthor{Steinborn_et_al_2016} (\citeyear{Steinborn_et_al_2016}; see their Fig.~3), when we split our results by (cumulative) distance: they find a flat distribution of dual-AGN fraction for separations between 5 and 10~kpc and an increase for lower separations. Overall, we also find an increase with decreasing separation but also see a somewhat flatter distribution.

\section{Conclusions}\label{dualagnpaper:sec:Conclusions}

We performed a large suite of high-resolution hydrodynamical simulations of galaxy mergers, where we vary the initial mass ratio, geometry, gas fraction, and BH parameters, to investigate the triggering of dual AGN. Using different thresholds (bolometric and hard-X-ray luminosity, and Eddington ratio) and, in the case of X-rays, accounting for obscuration from gas and contamination from star formation, we computed the dual-activity observability time-scales for each of our 12 mergers, as a function of projected BH separation and velocity difference. We also assessed the importance of each phase in the history of the mergers and applied thresholds (both to projected separations and velocity differences) to mimic observational-resolution limitations. Finally, we studied the effects of changing merger, galactic, and BH parameters and compared our results to both observational and theoretical works.

We itemize our findings below.

\begin{enumerate}

\item Dual BH activity is generally low during the stochastic and remnant stages and increases during the merger stage, especially at or right after late pericentric passages (Fig.~\ref{dualagnpaper:fig:m2_hr_gf0_3_BHeff0_001_phi000000_three_panels_dual} and Table~\ref{dualagnpaper:tab:m2_hr_gf0_3_BHeff0_001_phi000000}). At low luminosity thresholds, dual activity follows the history of the entire encounter. At high thresholds, it roughly follows the history of the merger phase. During the remnant stage, the separations and velocity differences between the two BHs are too small for detection (Figs~\ref{dualagnpaper:fig:m2_hr_gf0_3_BHeff0_001_phi000000_dist_vel_Mdot_deltat_proj_no3b_1a1btogether}--\ref{dualagnpaper:fig:m2_hr_gf0_3_BHeff0_001_phi000000_dual_agn_map_proj_100los_stageall_loglog_no3b}).

\item When accounting for gas obscuration in the case of hard-X-ray (2--10 keV) observations, we find that the decrease in dual-activity time is negligible when we assume the galaxies to be at high redshift ($z \sim 3$) and moderate (a factor of $\sim$2) if our simulated galaxies are representative of relatively small, gas-rich local galaxies (Fig.~\ref{dualagnpaper:fig:m2_hr_gf0_3_BHeff0_001_phi000000_dual_agn_no3b_obsc}). Contamination from star formation is always irrelevant.

\item Using projected BH separations ($r$) and velocity differences ($v$) rather than 3D quantities produces very different results, especially when one filters by velocity difference (mimicking spectroscopic observations): the decrease in dual-activity time can be as high as a factor of $\sim$4 (Figs~\ref{dualagnpaper:fig:m2_hr_gf0_3_BHeff0_001_phi000000_dual_agn_morelos_no3b}--\ref{dualagnpaper:fig:m2_hr_gf0_3_BHeff0_001_phi000000_dual_agn_map_proj_1los_stageall_loglog_no3b}).

\item The normalized dual-activity time -- defined as the dual-activity time divided by the time spent above the same $r$ or $v$ filter -- for the default merger (of mass ratio 1:2) is 12.5--14.1 per cent, depending on the filter.

\item The normalized dual-activity time increases with increasing mass ratio (with the factor increasing with the threshold, ranging from factors of a few to orders of magnitude), is independent of $r$ at low luminosities, and is non-negligible only at small $r$ at high luminosities: in order to identify dual AGN, shallow surveys will be more affected by an increase of angular resolution, with respect to deeper surveys (Fig.~\ref{dualagnpaper:fig:dual_agn_Bolom_Mdot_deltat_proj_norm2_no3b_coppropro_onlyr_3thr} and Table~\ref{dualagnpaper:tab:comparison}).

\item The normalized dual-activity time does not heavily depend on the geometry of the encounters, with the difference between mergers being only a factor of $\sim$2--2.5 and with no clear trend. Doubling the initial amount of gas increases the time, but only by a factor of $\sim$1.2--1.9, depending on the filter used. Doubling the initial BH mass increases the time by a factor of $\lesssim$2, whereas quintupling the BH feedback efficiency parameter decreases it, although the factor depends widely on the filter used (Figs~\ref{dualagnpaper:fig:dual_agn_time_norm4_3thr} and \ref{dualagnpaper:fig:dual_agn_Bolom_Mdot_deltat_proj_norm2_no3b_1to2moregeometries_onlyr_3thr}--\ref{dualagnpaper:fig:dual_agn_Bolom_Mdot_deltat_proj_norm2_no3b_bhphysics_onlyr_3thr}, and Table~\ref{dualagnpaper:tab:comparison}).

\item The results from our idealised merger simulations are consistent with observations, with the normalized dual-activity time of our mergers with standard BH feedback efficiency being within a factor of $\sim$2 from that of comparable systems observed in \citeauthor{Koss2012} (\citeyear{Koss2012}; Fig.~\ref{dualagnpaper:fig:dual_agn_time_norm4_3thr}), for both major and minor encounters. They are also consistent with cosmological simulations, with all our major mergers with standard BH feedback efficiency (except for one) within a factor of $\sim$2 from those simulated in \citeauthor{Steinborn_et_al_2016} (\citeyear{Steinborn_et_al_2016}; Fig.~\ref{dualagnpaper:fig:dual_agn_time_norm6_3thr}).

\end{enumerate}

We caution that all our estimates are upper limits, as we have not included the effects of dust obscuration from the torus/broad line region in the vicinity of the BH, which cannot be resolved in our suite of simulations. Moreover, given our temporal (0.1~Myr) and spatial ($\sim$10~pc) resolution, we cannot account for variability on smaller time-scales and spatial scales than those, possibly overestimating the dual-AGN time, as more accretion becomes non-simultaneous \citep{HQ2010,Levine2010}. Future numerical studies with improved resolution should partially alleviate these issues. We also note that our results may be sensitive to different treatments of BH accretion \citep[e.g.][]{Debuhr2010,Hopkins_Quataert_2011,AnglesAlcazar_et_al_2013,AnglesAlcazar_et_al_2015,AnglesAlcazar_et_al_2017}, AGN feedback \citep[e.g.][]{Wurster_Thacker_2013,Newton_Kay_2013}, or the numerical method employed \citep[e.g.][]{Hayward_et_al_2014}. A thorough analysis of the dependence of the results on all these aspects is not within the scope of the current work and will be addressed in a future work.

Another aspect that warrants future investigation is the dependence of our results on galaxy mass. More massive merging galaxies may be able to form circumnuclear discs (CNDs), as suggested by both observations \citep[][]{Downes_Solomon_1998,Medling_et_al_2014} and simulations \citep[][]{Mayer_et_al_2008,Roskar_et_al_2015}. The presence of the CND would affect the gas flow around the BHs, hence their accretion history, and could also play a role in obscuration originating at scales larger than the scale of the tori \citep[][]{Hopkins_et_al_2012}. These effects would have an impact on the dual-AGN time-scales. Yet, based on the results of pc-scale merger simulations \citep[][]{Roskar_et_al_2015}, we argue that CNDs would probably form at the end of the remnant phase studied in this work. Therefore, our results, which concern the previous evolutionary phase of the massive BH pair, should not be affected.

Another avenue of research is to study the activity of BH pairs in massive (stellar mass greater than $10^{10}$~M$_{\odot}$) galaxies at $z > 1$ which, being more dense and clumpy, might be a favourable environment for dual AGN activity. In such galaxies, the orbital decay of massive BH pairs has been shown to be inefficient at separations of a few hundred pc to a kpc, due to the perturbations of the clumpy interstellar medium onto the BH pair \citep[][]{Tamburello_et_al_2017}. The latter effect could lengthen the duration of the dual-activity phase after the galaxy merger has been completed. We note, however, that the galaxy mass range we investigate here is statistically more relevant across cosmic epochs, since more massive galaxies are rare at $z > 1$. Moreover, at low redshift, disc galaxies exhibit relatively smooth discs even at the largest mass scales.

\section*{Acknowledgements}
We thank the reviewer for the useful comments that greatly improved this work. We also thank Andrea Comastri, Davide Fiacconi, Carmen Montuori, Thomas~R. Quinn, and Sandor Van Wassenhove for their help and useful discussions. MV acknowledges funding support from NASA, through award ATP NNX10AC84G, from SAO, through award TM1-12007X, from NSF, through award AST 1107675, and from  the European Research Council under the European Community's Seventh Framework Programme (FP7/2007-2013 Grant Agreement no.\ 614199, project ``BLACK''). Resources supporting this work were provided by the NASA High-End Computing (HEC) Program through the NASA Advanced Supercomputing (NAS) Division at Ames Research Center, and by TGCC, under the allocations 2013-t2013046955 and 2014-x2014046955 made by GENCI. This research was supported in part by the National Science Foundation under grant no. NSF PHY11-25915, through the Kavli Institute for Theoretical Physics and its program `A Universe of Black Holes'. PRC thanks the Institut d'Astrophysique de Paris for hosting him during his visits and acknowledges support by the Tomalla foundation.

\scalefont{0.94}
\setlength{\bibhang}{1.6em}
\setlength\labelwidth{0.0em}
\bibliographystyle{mnras}
\bibliography{dualagnpaper}
\normalsize


\clearpage

\appendix

\section{Effects of projection and resolution}\label{Effects_of_projection}

\begin{figure}
\centering
\vspace{3.0pt}
\includegraphics[width=0.99\columnwidth,angle=0]{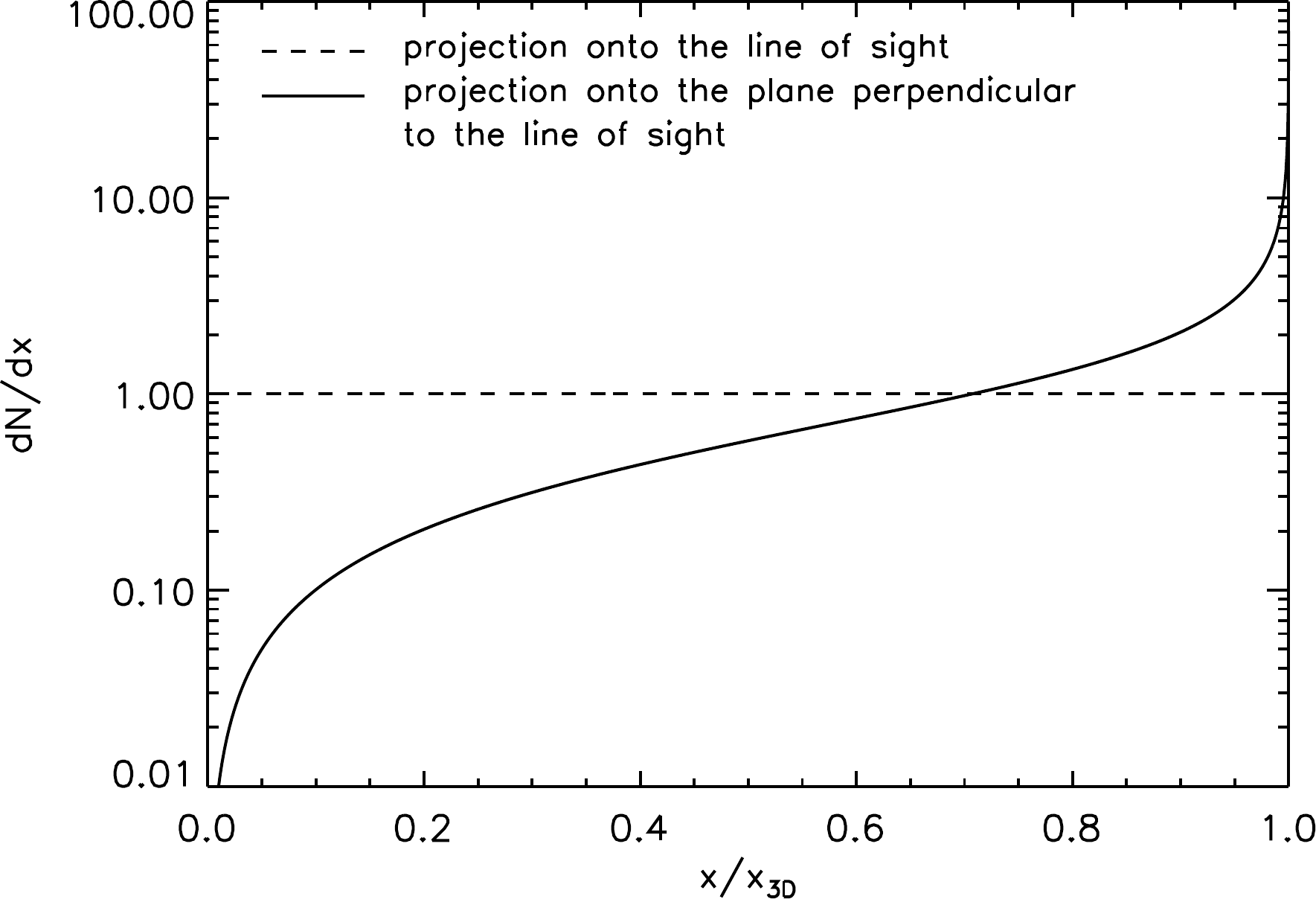}
\vspace{-5.0pt}
\caption[]{Normalized probability distributions of the magnitude $x$ of projected vectors, for two different projections of a given 3D vector of magnitude $x_{\rm 3D}$. (a) Projection onto the line of sight (in this work: velocity difference; dashed line): $dN/dx = 1/x_{\rm 3D}$; (b) Projection onto the plane perpendicular to the line of sight (in this work: separation; solid line): $dN/dx = (x/x_{\rm 3D}^2)/\sqrt{1-(x/x_{\rm 3D})^2}$.}
\label{dualagnpaper:fig:projvectordistr}
\end{figure}

\begin{figure}
\centering
\vspace{3.0pt}
\includegraphics[width=0.99\columnwidth,angle=0]{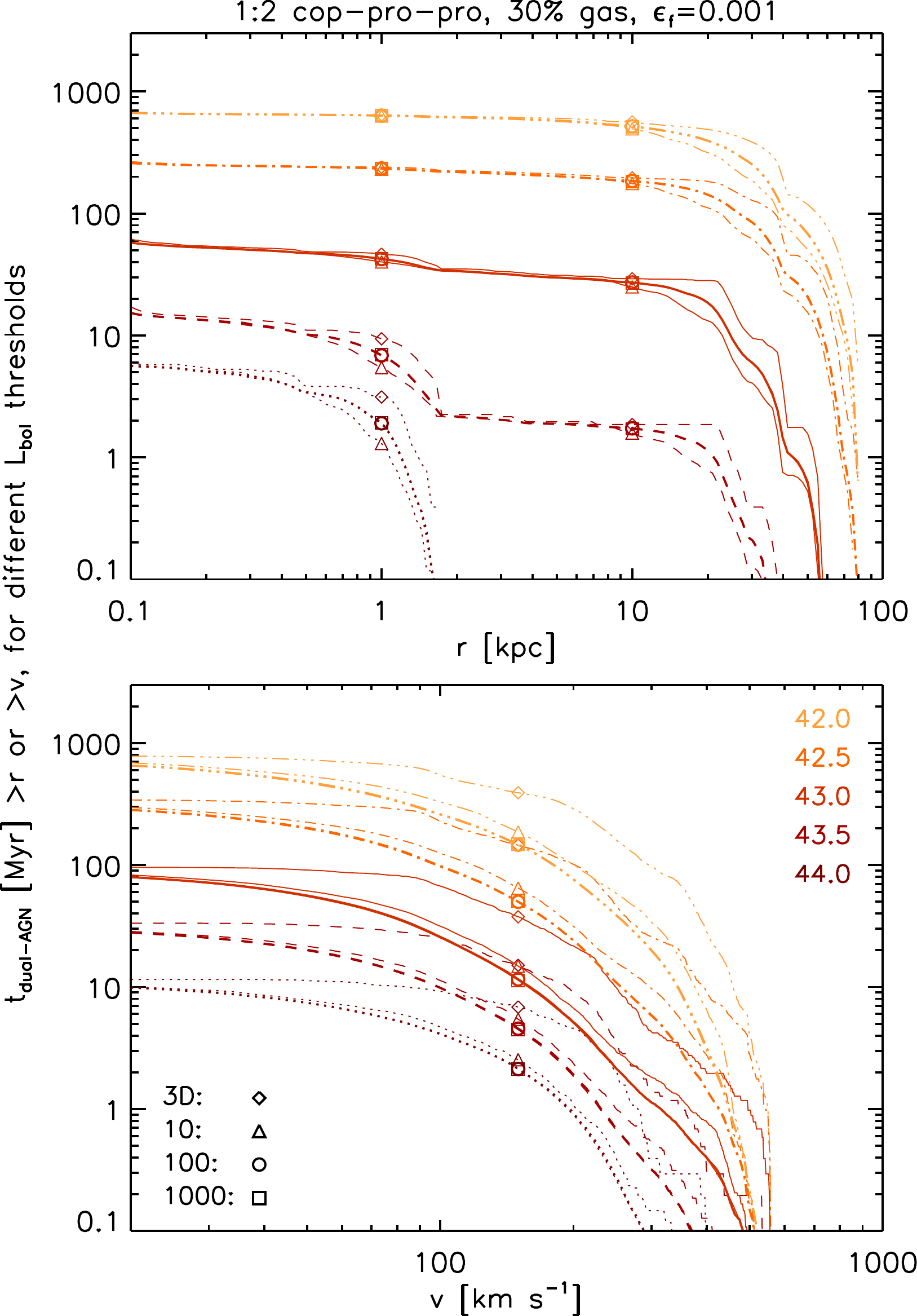}
\vspace{-5.0pt}
\caption[]{Same as the left panels of Fig.~\ref{dualagnpaper:fig:m2_hr_gf0_3_BHeff0_001_phi000000_dual_agn_Mdot_deltat_proj_no3b}, but using 3D quantities and different numbers of lines of sight. The symbols refer to the following projected separations and velocity differences: 1 and 10~kpc, and 150~km~s$^{-1}$. For each luminosity threshold, we show four lines: 1000 (thin lines and squares), 100 (thick lines and circles), and 10 lines of sight (thin lines and triangles), and 3D quantities (thin lines and diamonds). As expected, the 3D time is always higher than the projected times, although the projection does not change significantly the $r$-plots. On the other hand, the time-decrease factor in the $v$-plots can be as much as 3--4. The difference between using 100 and 1000 lines of sight is negligible.}
\label{dualagnpaper:fig:m2_hr_gf0_3_BHeff0_001_phi000000_dual_agn_morelos_no3b}
\end{figure}

\begin{figure*}
\centering
\vspace{-1.0pt}
\begin{overpic}[width=1.55\columnwidth,angle=0]{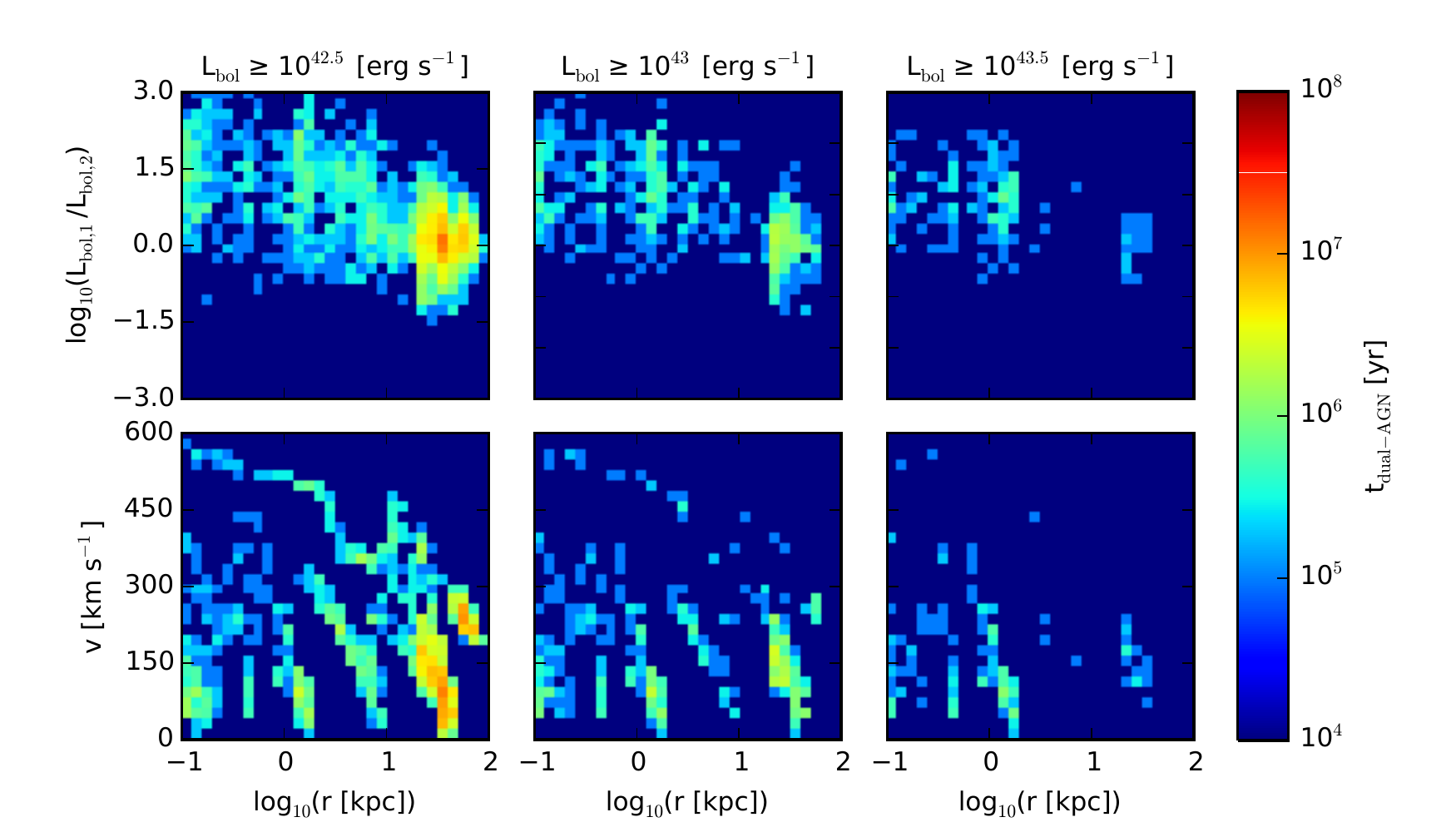}
\put (29,54.5) {\textcolor{black}{1:2 cop-pro-pro, 30\% gas, $\epsilon_{\rm f}=0.001$}}
\end{overpic}\\
\vspace{-4.0pt}
\begin{overpic}[width=1.55\columnwidth,angle=0]{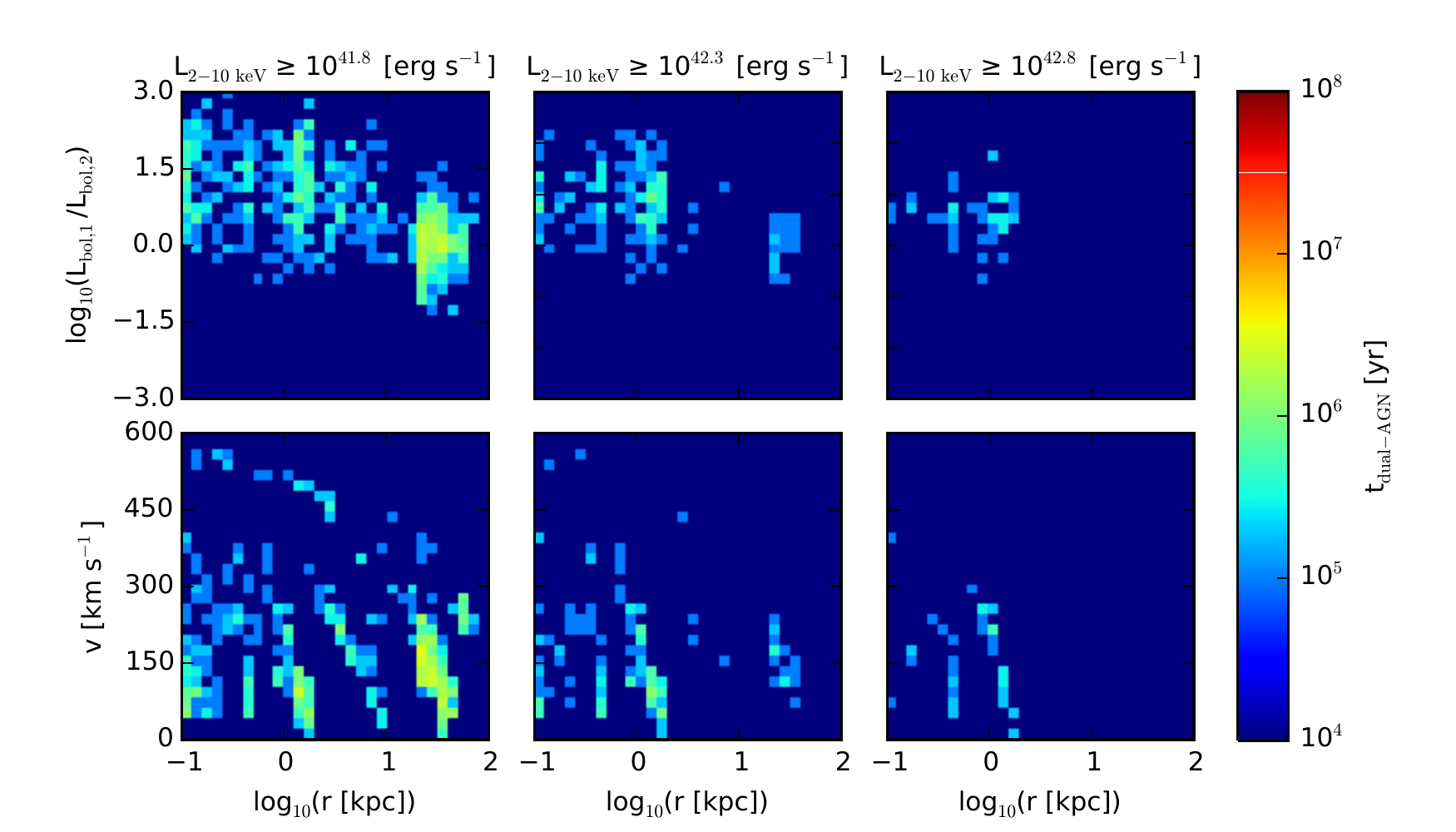}
\end{overpic}\\
\vspace{-4.0pt}
\begin{overpic}[width=1.55\columnwidth,angle=0]{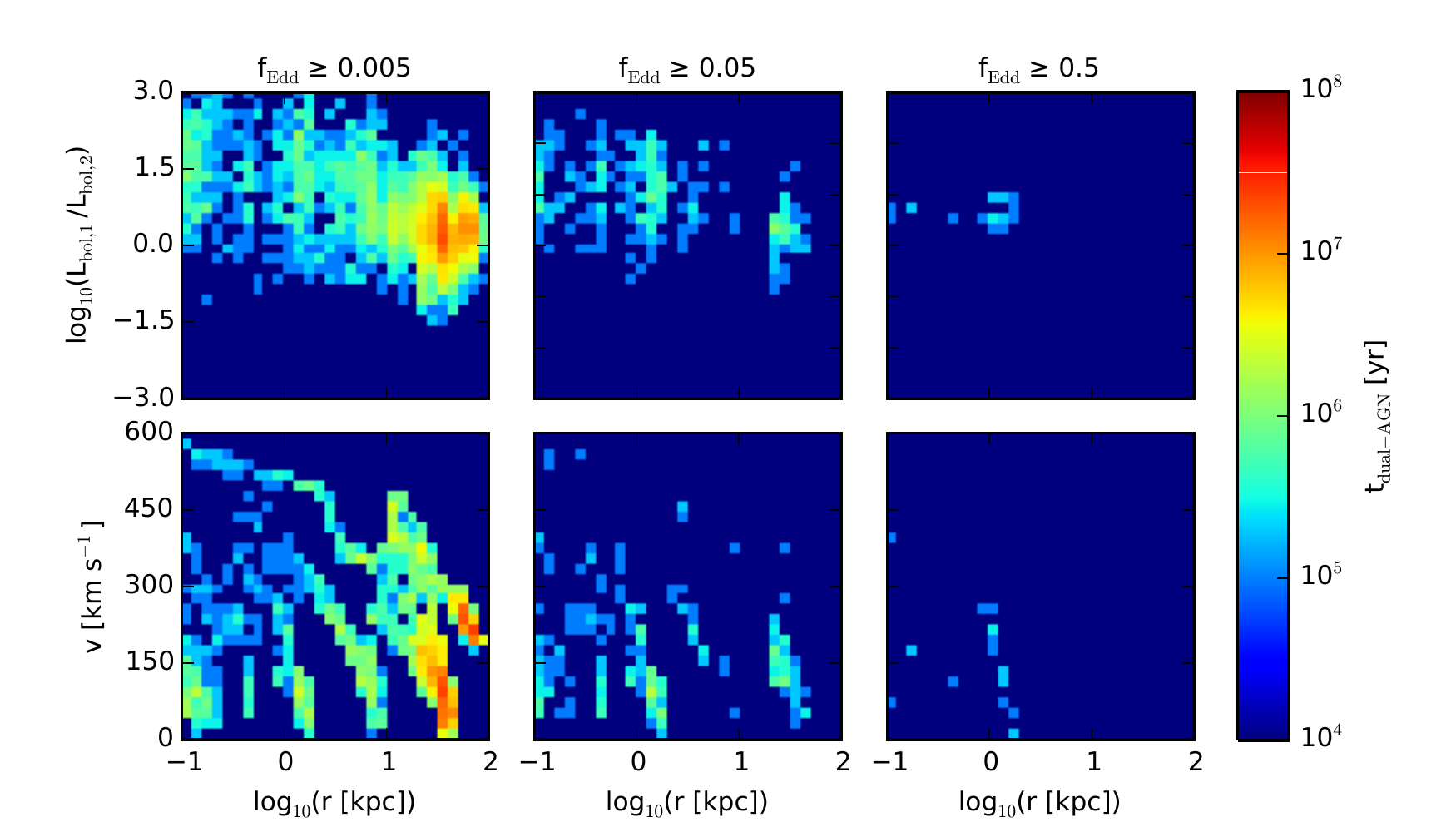}
\end{overpic}
\caption[]{Same as Fig.~\ref{dualagnpaper:fig:m2_hr_gf0_3_BHeff0_001_phi000000_dual_agn_map_proj_100los_stageall_loglog_no3b}, using 3D quantities.}
\label{dualagnpaper:fig:m2_hr_gf0_3_BHeff0_001_phi000000_dual_agn_map_proj_1los_stageall_loglog_no3b}
\end{figure*}

In this section, we show the effects of using projected lines of sight as opposed to 3D quantities.

In Fig.~\ref{dualagnpaper:fig:projvectordistr}, we show the normalized distributions of the magnitude of projected vectors, for the two cases when we project a vector onto a random line of sight (in this work: velocity difference) and when we project a vector onto the plane perpendicular to a random line of sight (in this work: separation). In the former case, the probability distribution is flat. In the latter, the probability distribution is skewed towards large values, i.e. the magnitude of the 3D vector. As an example, the probability that the magnitude of the projected vector is less than 80 per cent of that of the 3D vector is 80 per cent in the $v$ case and 40 per cent in the $r$ case.

In Fig.~\ref{dualagnpaper:fig:m2_hr_gf0_3_BHeff0_001_phi000000_dual_agn_morelos_no3b}, we quantify the effects of projection, by showing the dual-activity observability time-scales in the default merger using 3D quantities and projected quantities, based on 10, 100, and 1000 random lines of sight. The results with 100 and 1000 lines of sight are indistinguishable, which justifies our use of 100 lines of sight throughout the analysis. On the other hand, using a lower number of random lines of sight (e.g. 10) or even 3D quantities can give results that are different by up to a factor of 3--4 in the $v$-plots and are similar in the $r$-plots (except for very high luminosity thresholds).

In Fig.~\ref{dualagnpaper:fig:m2_hr_gf0_3_BHeff0_001_phi000000_dual_agn_map_proj_1los_stageall_loglog_no3b}, we show the dual-activity observability time-scales as a function of $r$ and $v$, and of $r$ and $L_{\rm bol,\,1}/L_{\rm bol,\,2}$, in the same way shown in Fig.~\ref{dualagnpaper:fig:m2_hr_gf0_3_BHeff0_001_phi000000_dual_agn_map_proj_100los_stageall_loglog_no3b}, this time using 3D quantities \citep[as used in][]{Van_Wassenhove_et_al_2012}. The difference in the `$v$ versus $r$' plots is striking. For example, from looking at the $L_{\rm bol} = 10^{42.5}$ results of Fig.~\ref{dualagnpaper:fig:m2_hr_gf0_3_BHeff0_001_phi000000_dual_agn_map_proj_1los_stageall_loglog_no3b}, one would expect to find a relatively large population of systems at large separations ($\gtrsim 40$~kpc; before the first pericentric passage) with high velocity differences (around 180--300~km~s$^{-1}$), an ideal case for current spectroscopic surveys. However, if we consider the effects of projection (see Fig.~\ref{dualagnpaper:fig:m2_hr_gf0_3_BHeff0_001_phi000000_dual_agn_map_proj_100los_stageall_loglog_no3b}), the result changes dramatically, with the dual-activity time in the same region of velocity--separation phase-space basically going to zero.

We note that our results change when we degrade the resolution, as already shown in \citet{Volonteri_et_al_2015a}. At low resolution, we cannot resolve small-scale gravitational torques and overdensities. This leads, in the case of the default merger, to a BH accretion rate generally higher than that of the high-resolution run, where the region near the BHs is better resolved, causing an increase in the dual-activity time-scales. For more details on the effects of resolution, we refer the reader to the Appendix of \citet{Volonteri_et_al_2015a}.

\clearpage

\section{Additional figures and table}\label{Additional_figures}

\begin{figure}
\centering
\vspace{3.0pt}
\includegraphics[width=1.00\columnwidth,angle=0]{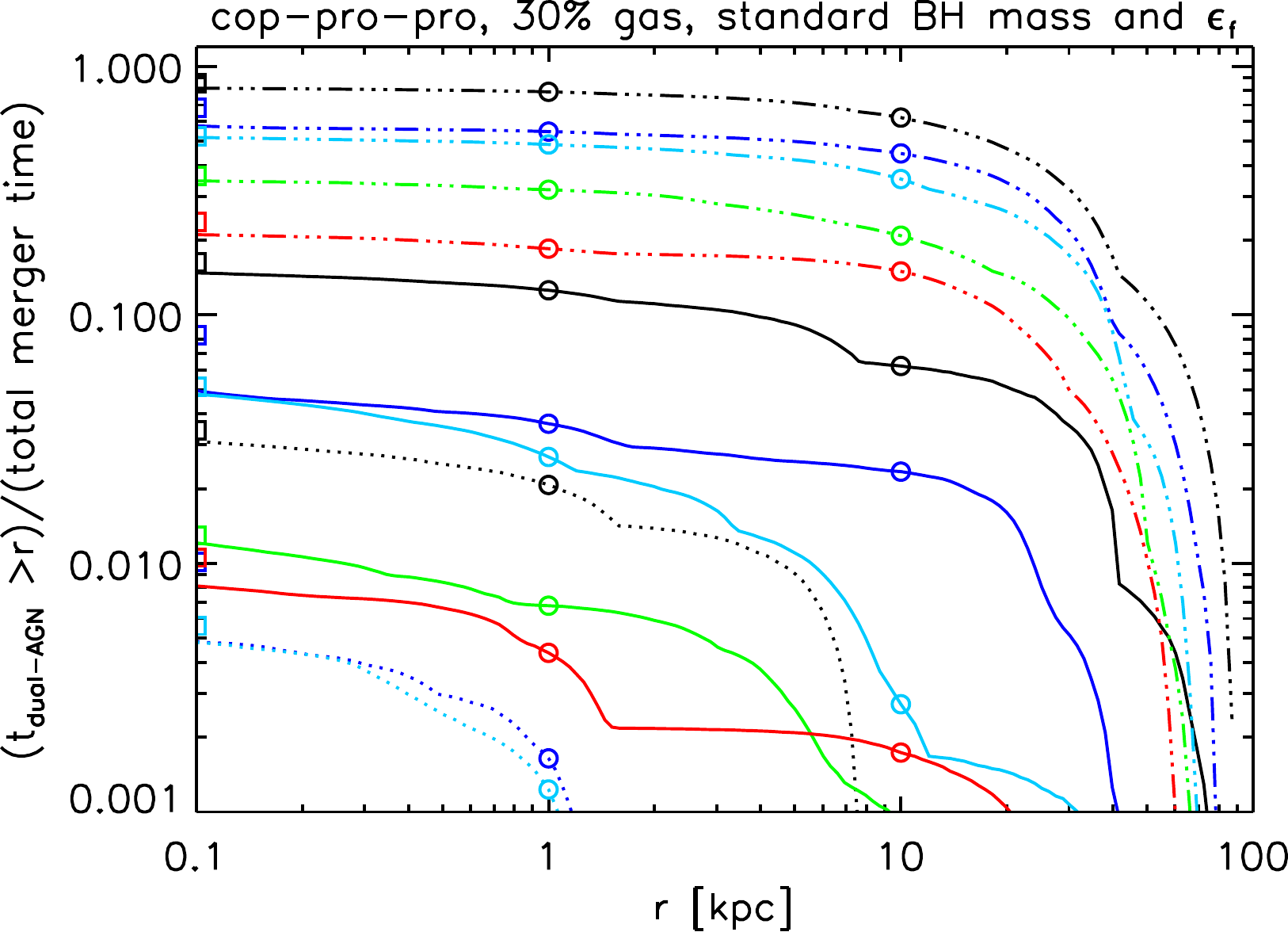}
\vspace{-5.0pt}
\caption[]{Same as Fig.~\ref{dualagnpaper:fig:dual_agn_Bolom_Mdot_deltat_proj_norm2_no3b_coppropro_onlyr_3thr}, except that we divide the dual-activity time above a given BH projected separation by the total merger time. At $r = 1$~kpc, the normalized dual-activity time is highest for the 1:1 merger, followed by the 1:2, 1:4, 1:6, and 1:10 mergers, regardless of the luminosity threshold shown.}
\label{dualagnpaper:fig:dual_agn_Bolom_Mdot_deltat_proj_norm1_no3b_coppropro_onlyr_3thr}
\end{figure}

\begin{figure}
\centering
\vspace{3.0pt}
\includegraphics[width=1.00\columnwidth,angle=0]{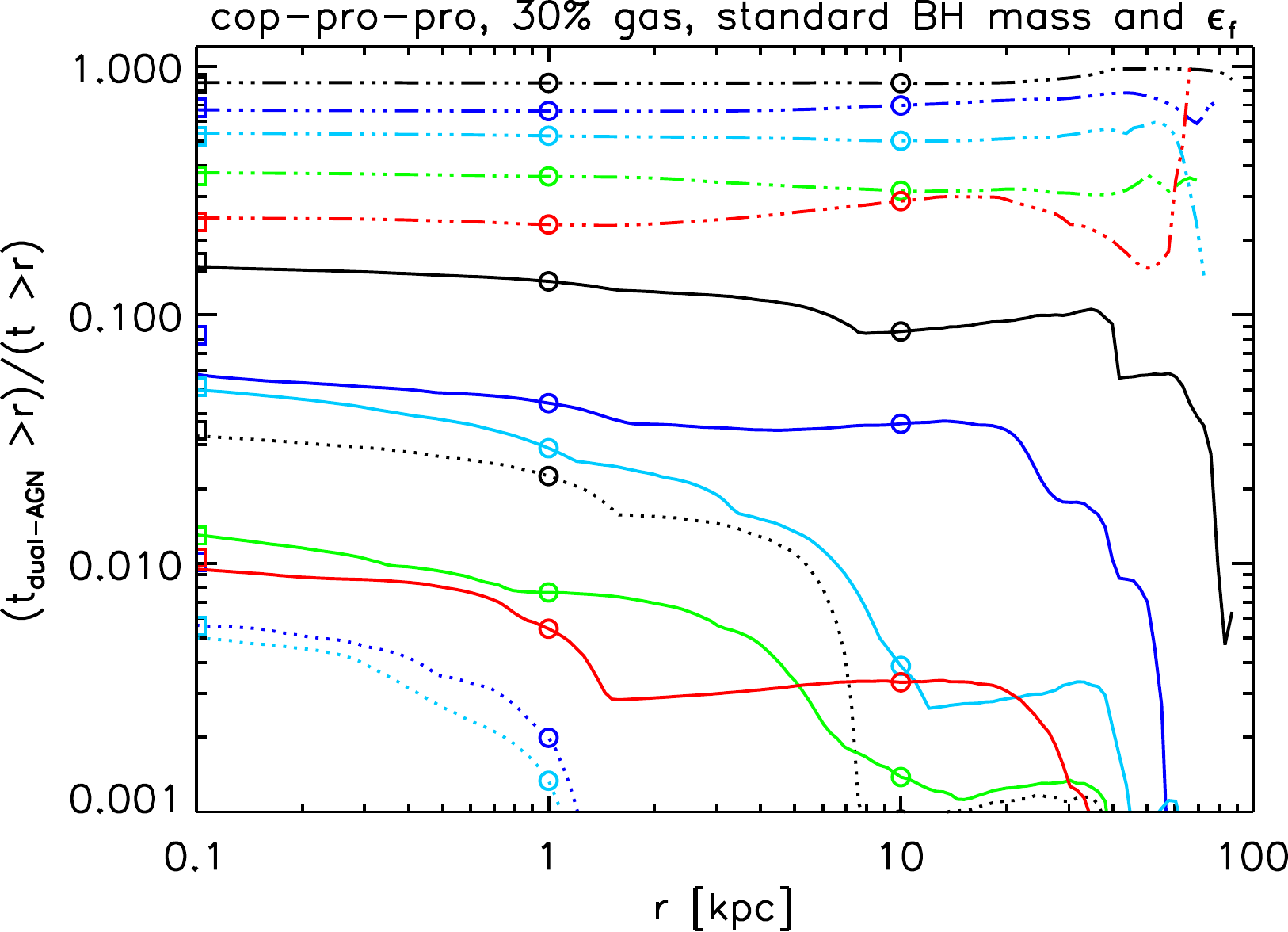}
\vspace{-5.0pt}
\caption[]{Same as Fig.~\ref{dualagnpaper:fig:dual_agn_Bolom_Mdot_deltat_proj_norm2_no3b_coppropro_onlyr_3thr}, except that we divide the dual-activity time above a given BH projected separation by the  time spent above the same $r$. At $r = 1$~kpc, the normalized dual-activity time is highest for the 1:1 merger, followed by the 1:2, 1:4, 1:6, and 1:10 mergers, regardless of the luminosity threshold shown.}
\label{dualagnpaper:fig:dual_agn_Bolom_Mdot_deltat_proj_norm3_no3b_coppropro_onlyr_3thr}
\end{figure}

\begin{figure}
\centering
\vspace{1.8pt}
\includegraphics[width=0.98\columnwidth,angle=0]{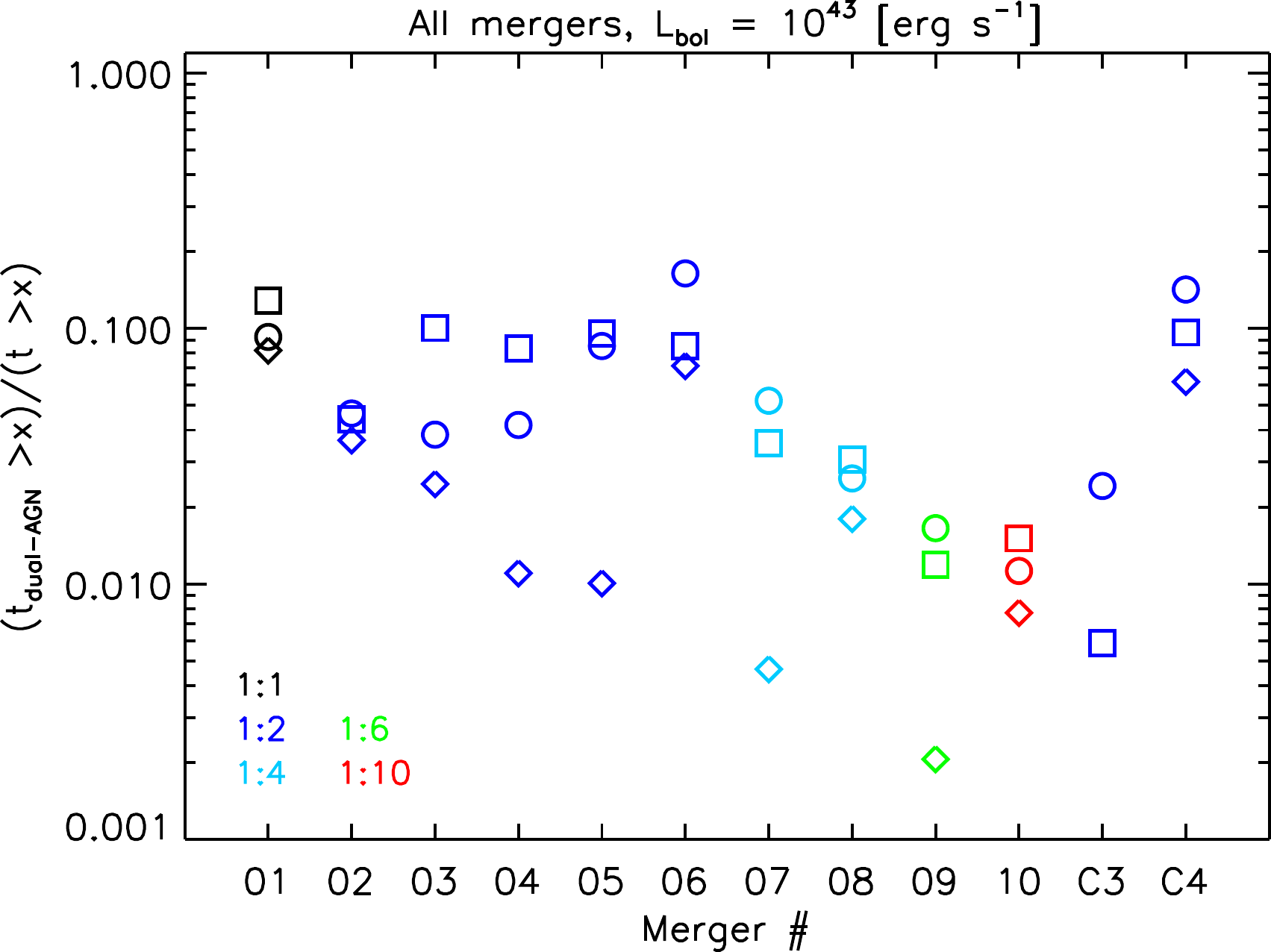}
\vspace{-0.0pt}
\caption[]{Same as Fig.~\ref{dualagnpaper:fig:dual_agn_time_norm4_3thr}, except that we divide the dual-activity time above a given $r$ and $v$ by the  time above the same threshold (and use one panel).\\\\\\}
\label{dualagnpaper:fig:dual_agn_time_norm5_3thr}
\end{figure}

\begin{figure}
\centering
\vspace{-2.0pt}
\includegraphics[width=0.99\columnwidth,angle=0]{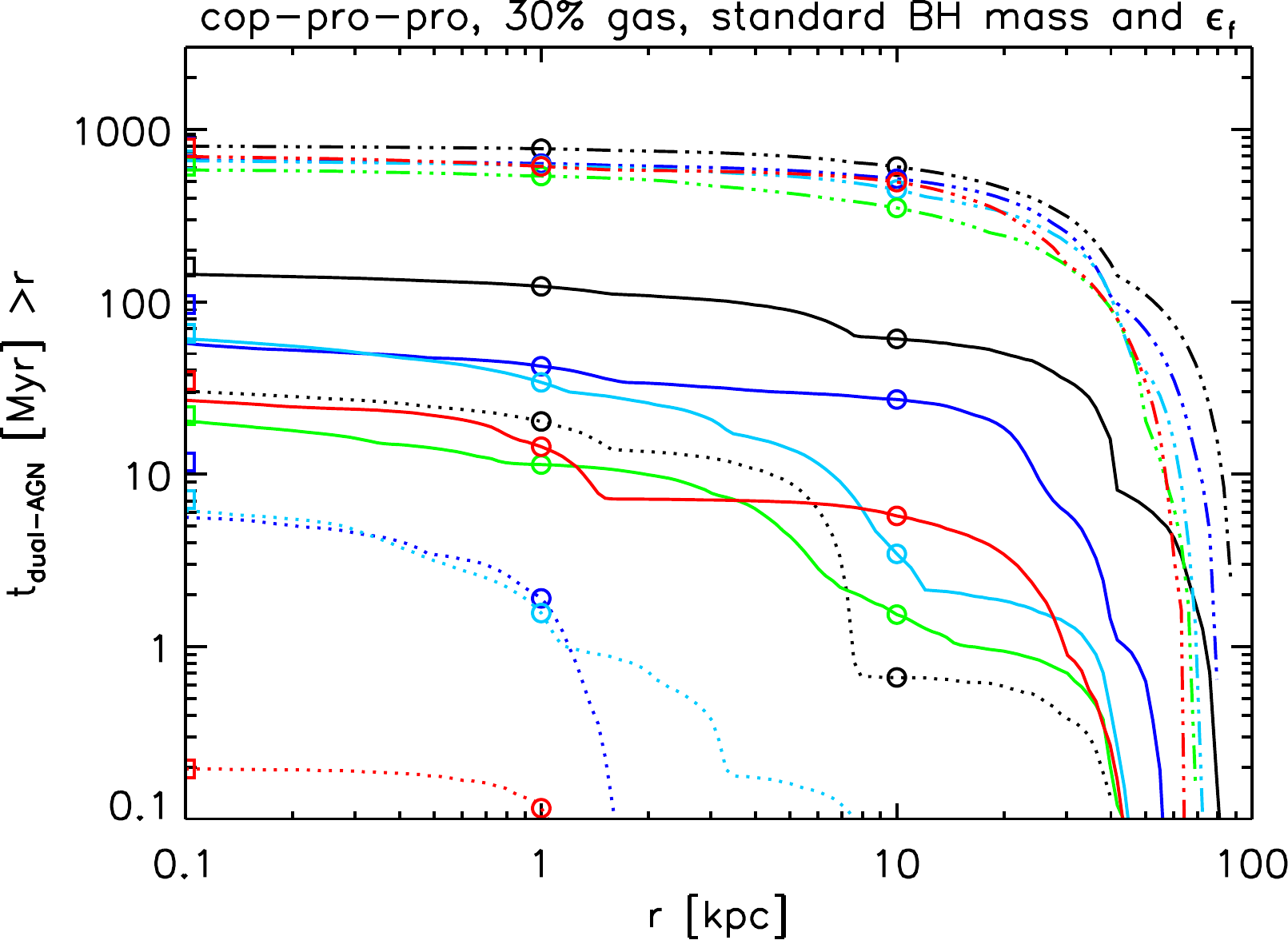}
\vspace{-5.0pt}
\caption[]{Same as Fig.~\ref{dualagnpaper:fig:dual_agn_Bolom_Mdot_deltat_proj_norm2_no3b_coppropro_onlyr_3thr}, except that we do not normalize the time. At $r = 1$~kpc, the normalized dual-activity time is highest for the 1:1 merger, followed by the 1:2, 1:4, 1:10, and 1:6 mergers, regardless of the luminosity threshold shown.}
\label{dualagnpaper:fig:dual_agn_Bolom_Mdot_deltat_proj_no3b_coppropro_onlyr_3thr}
\end{figure}

\begin{figure}
\centering
\vspace{3.0pt}
\includegraphics[width=1.00\columnwidth,angle=0]{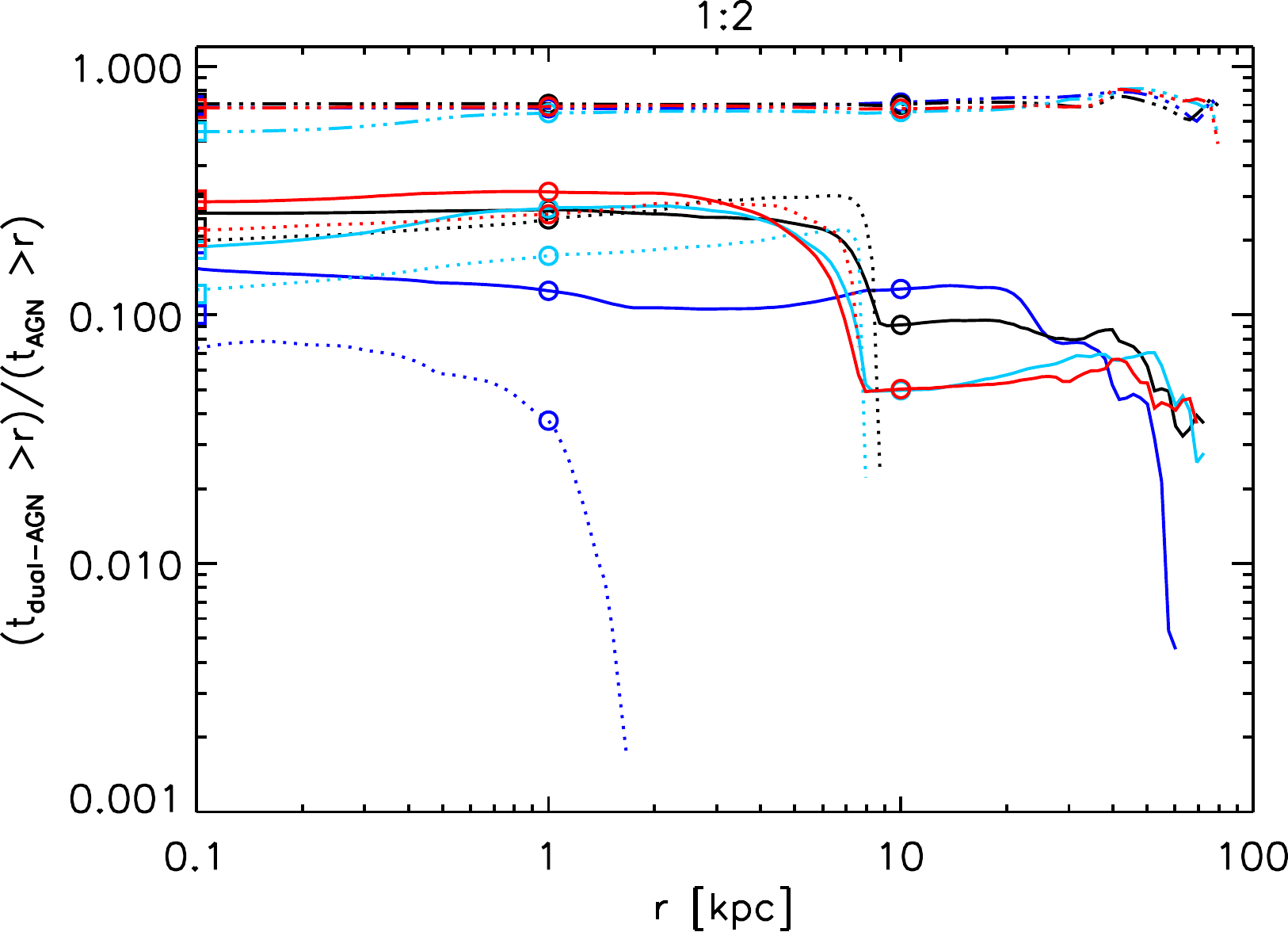}
\vspace{-5.0pt}
\caption[]{Same as Fig.~\ref{dualagnpaper:fig:dual_agn_Bolom_Mdot_deltat_proj_norm2_no3b_coppropro_onlyr_3thr}, but for the 1:2 coplanar, prograde--prograde mergers with 30 per cent gas fraction, standard BH mass and feedback efficiency, and different geometries (coplanar, prograde--prograde: blue, total encounter time 1.158~Gyr; inclined-primary: black, 1.085~Gyr; coplanar, retrograde--prograde: cyan, 1.579~Gyr; and coplanar, prograde--retrograde: red, 1.266~Gyr). At $r = 1$~kpc, the normalized dual-activity time for $L_{\rm thr} = 10^{42}$~erg~s$^{-1}$ is highest for the inclined-primary merger, followed by the prograde--retrograde, prograde--prograde, and retrograde--prograde mergers. For $L_{\rm thr} = 10^{43}$~erg~s$^{-1}$, it is highest for the prograde--retrograde merger, followed by the retrograde--prograde, inclined-primary, and prograde--prograde mergers. For $L_{\rm thr} = 10^{44}$~erg~s$^{-1}$, it is highest for the prograde--retrograde merger, followed by the inclined-primary, retrograde--prograde, and prograde--prograde mergers.}
\label{dualagnpaper:fig:dual_agn_Bolom_Mdot_deltat_proj_norm2_no3b_1to2moregeometries_onlyr_3thr}
\end{figure}

\begin{figure}
\centering
\vspace{3.0pt}
\includegraphics[width=1.00\columnwidth,angle=0]{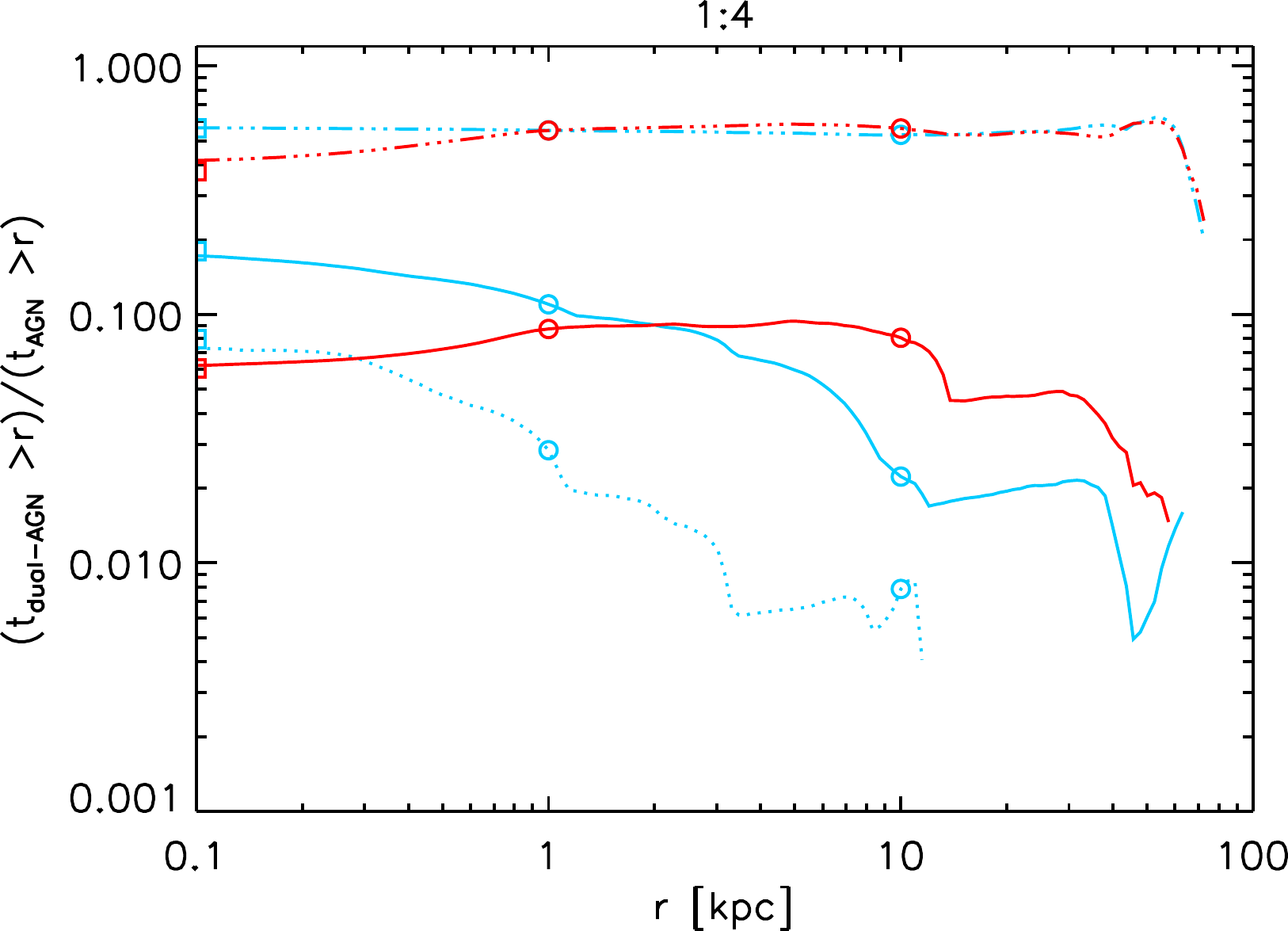}
\vspace{-5.0pt}
\caption[]{Same as Fig.~\ref{dualagnpaper:fig:dual_agn_Bolom_Mdot_deltat_proj_norm2_no3b_coppropro_onlyr_3thr}, but for the 1:4 coplanar, prograde--prograde mergers with 30 per cent gas fraction, standard BH mass and feedback efficiency, and different geometries (coplanar, prograde--prograde: cyan, total encounter time 1.271~Gyr; and inclined-primary: red, 2.357~Gyr). At $r = 0.1$~kpc, the normalized dual-activity time is highest for the coplanar merger, regardless of the luminosity threshold shown.}
\label{dualagnpaper:fig:dual_agn_Bolom_Mdot_deltat_proj_norm2_no3b_1to4moregeometries_onlyr_3thr}
\end{figure}

\begin{figure}
\centering
\vspace{3.0pt}
\includegraphics[width=1.00\columnwidth,angle=0]{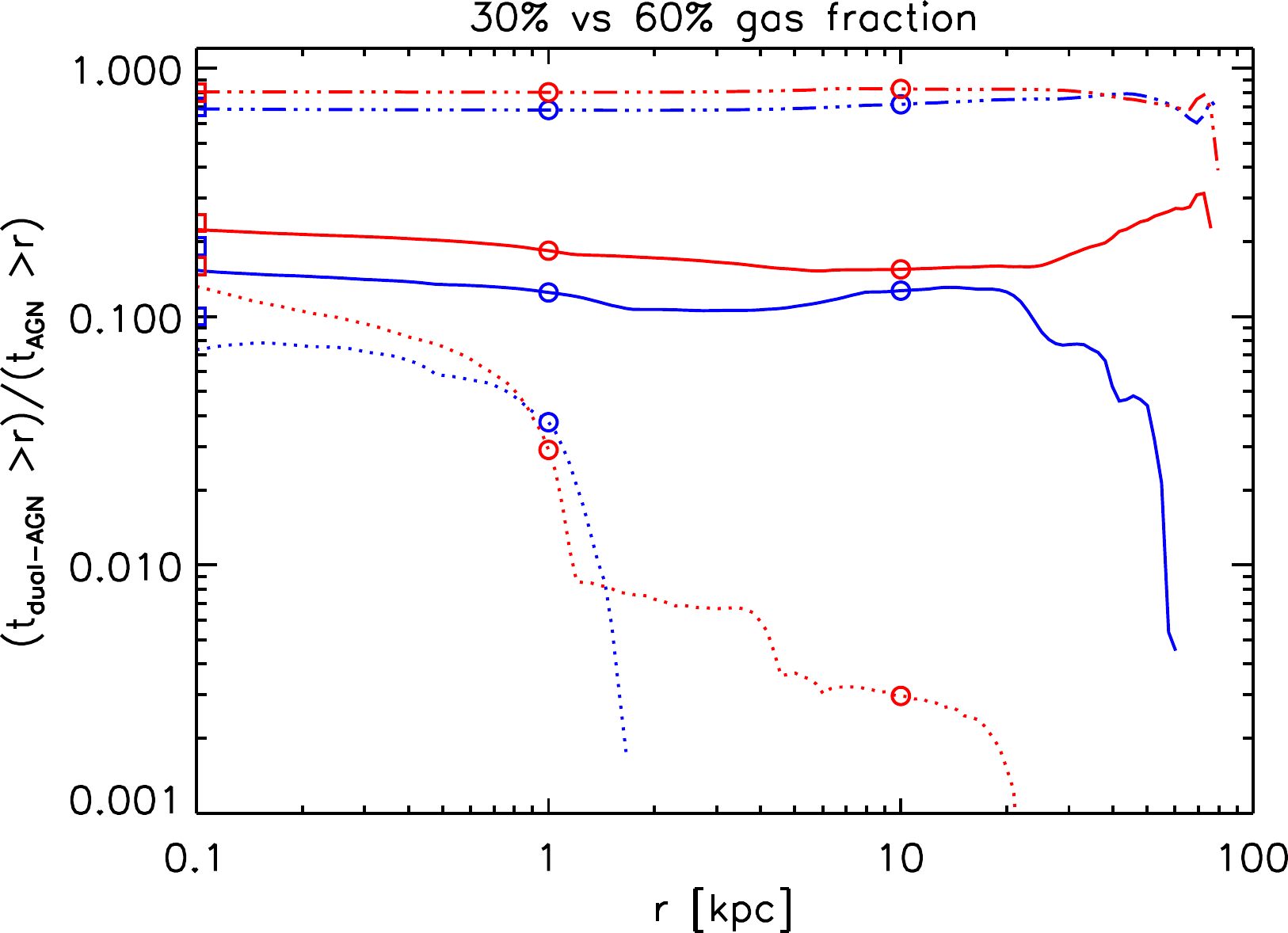}
\vspace{-5.0pt}
\caption[]{Same as Fig.~\ref{dualagnpaper:fig:dual_agn_Bolom_Mdot_deltat_proj_norm2_no3b_coppropro_onlyr_3thr}, but for the 1:2 coplanar, prograde--prograde merger with 30 (blue; total encounter time 1.158~Gyr) and 60 (red; 1.022~Gyr) per cent gas fraction, and standard BH mass and feedback efficiency. At $r = 10$~kpc, the normalized dual-activity time is highest for the 60-per-cent gas-fraction merger, regardless of the luminosity threshold shown.\\\\\\\\\\\\\\\\\\}
\label{dualagnpaper:fig:dual_agn_Bolom_Mdot_deltat_proj_norm2_no3b_lowvshighgasfrac_onlyr_3thr}
\end{figure}

\begin{figure}
\centering
\vspace{4.1pt}
\includegraphics[width=0.99\columnwidth,angle=0]{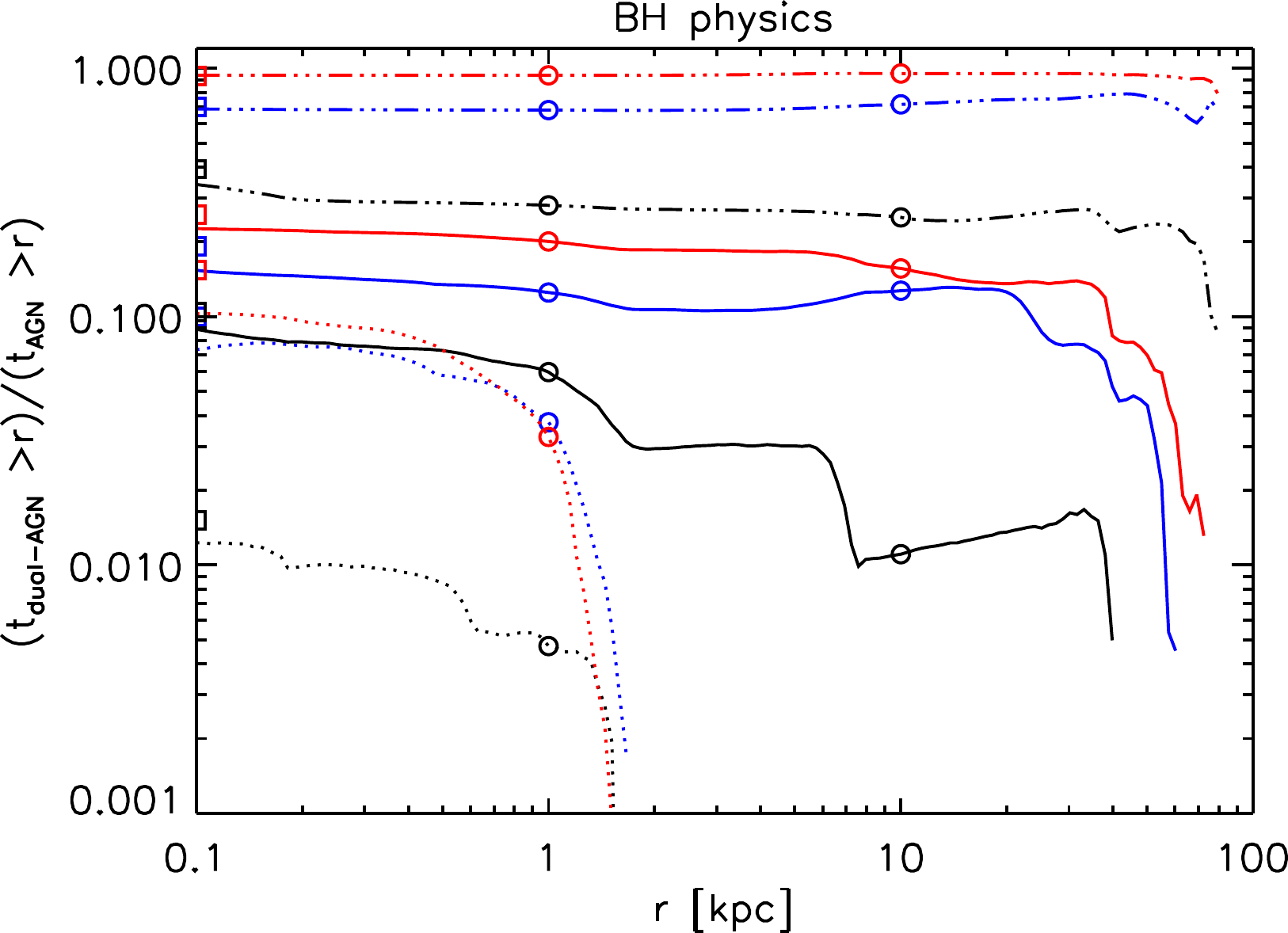}
\vspace{-5.0pt}
\caption[]{Same as Fig.~\ref{dualagnpaper:fig:dual_agn_Bolom_Mdot_deltat_proj_norm2_no3b_coppropro_onlyr_3thr}, but for the 1:2 coplanar, prograde--prograde mergers with 30 per cent gas fraction and different BH parameters (standard BH mass and feedback efficiency: blue, total encounter time 1.158~Gyr; standard BH mass and high BH feedback efficiency: black, 1.281~Gyr; and large BH mass and standard BH feedback efficiency: red, 1.090~Gyr). At $r = 0.1$~kpc, the normalized dual-activity time is highest for the merger with large BH mass and standard BH feedback efficiency, followed by the merger with standard BH mass and feedback efficiency and by that with standard BH mass and high BH feedback efficiency, regardless of the luminosity threshold shown.}
\label{dualagnpaper:fig:dual_agn_Bolom_Mdot_deltat_proj_norm2_no3b_bhphysics_onlyr_3thr}
\end{figure}

\clearpage

\begin{table*} \centering
\vspace{-3.5pt}
\caption[Simulation results]{Results for the 1:2 coplanar, prograde--prograde merger with 30 per cent gas fraction and standard BH mass and feedback efficiency (the default merger). The thresholds 41.3, 41.8, 42.3, 42.8, and 43.3 refer to log$_{10}(L_{\rm 2-10\;keV}$ [erg~s$^{-1}$]); 0.005, 0.05, and 0.5 refer to $f_{\rm Edd}$; and 42.0, 42.5, 43.0, 43.5, and 44.0 refer to log$_{10}(L_{\rm bol}$ [erg~s$^{-1}$]). All times are given in Gyr and are either total (bottom section) or on a per-stage basis (top three sections). The number in parenthesis in the first column of each section is the duration of the stage (or of the total merger). Similar tables, for all other mergers, can be found in the online-only material.
\label{dualagnpaper:tab:m2_hr_gf0_3_BHeff0_001_phi000000}}
\vspace{5pt}
\begin{tabular*}{0.94\textwidth}{c|m{40pt}m{40pt}m{40pt}m{40pt}|m{40pt}m{40pt}m{40pt}m{61pt}}
Run 02    & Threshold & BH$_1$ & BH$_2$ & Either BH & \multicolumn{4}{c}{Both BHs above the threshold} \B \\
         & L$_{\rm thr}$/$f_{\rm Edd}$ & above thr. & above thr. & above thr. & any $d$ or $v$ & $d > 1$ kpc & $d > 10$ kpc & $v > 150$ km s$^{-1}$ \B \\
\hline
Stochastic  & 41.30 &  0.591 &  0.442 &  0.719 &  0.315 &  0.312 &  0.278 &  0.060 \T \B \\
(0.849)  & 41.80 &  0.216 &  0.128 &  0.300 &  0.043 &  0.043 &  0.038 &  0.006 \B \\
         & 42.30 &  0.039 &  0.021 &  0.058 &  0.002 &  0.002 &  0.002 &  0.000 \B \\
         & 42.80 &  0.003 &  0.002 &  0.005 &  0.000 &  0.000 &  0.000 &  0.000 \B \\
         & 43.30 &  0.000 &  0.000 &  0.000 &  0.000 &  0.000 &  0.000 &  0.000 \B \\
         & 0.005 &  0.564 &  0.598 &  0.761 &  0.401 &  0.399 &  0.357 &  0.084 \B \\
         & 0.050 &  0.055 &  0.070 &  0.118 &  0.007 &  0.007 &  0.006 &  0.001 \B \\
         & 0.500 &  0.001 &  0.002 &  0.002 &  0.000 &  0.000 &  0.000 &  0.000 \B \\
         & 42.00 &  0.763 &  0.651 &  0.827 &  0.587 &  0.584 &  0.518 &  0.129 \B \\
         & 42.50 &  0.490 &  0.343 &  0.625 &  0.209 &  0.206 &  0.184 &  0.038 \B \\
         & 43.00 &  0.177 &  0.101 &  0.248 &  0.031 &  0.031 &  0.027 &  0.004 \B \\
         & 43.50 &  0.042 &  0.021 &  0.061 &  0.002 &  0.002 &  0.002 &  0.000 \B \\
         & 44.00 &  0.006 &  0.003 &  0.009 &  0.000 &  0.000 &  0.000 &  0.000 \B \\
\hline
Merger  & 41.30 &  0.199 &  0.103 &  0.200 &  0.102 &  0.033 &  0.000 &  0.014 \T \B \\
(0.206)  & 41.80 &  0.182 &  0.063 &  0.184 &  0.061 &  0.014 &  0.000 &  0.008 \B \\
         & 42.30 &  0.145 &  0.032 &  0.147 &  0.030 &  0.005 &  0.000 &  0.004 \B \\
         & 42.80 &  0.092 &  0.010 &  0.094 &  0.008 &  0.001 &  0.000 &  0.002 \B \\
         & 43.30 &  0.048 &  0.001 &  0.049 &  0.000 &  0.000 &  0.000 &  0.000 \B \\
         & 0.005 &  0.189 &  0.113 &  0.196 &  0.106 &  0.038 &  0.000 &  0.015 \B \\
         & 0.050 &  0.131 &  0.044 &  0.138 &  0.038 &  0.008 &  0.000 &  0.005 \B \\
         & 0.500 &  0.038 &  0.007 &  0.042 &  0.003 &  0.001 &  0.000 &  0.001 \B \\
         & 42.00 &  0.204 &  0.129 &  0.204 &  0.129 &  0.050 &  0.000 &  0.017 \B \\
         & 42.50 &  0.196 &  0.093 &  0.197 &  0.092 &  0.027 &  0.000 &  0.012 \B \\
         & 43.00 &  0.178 &  0.059 &  0.180 &  0.056 &  0.012 &  0.000 &  0.007 \B \\
         & 43.50 &  0.146 &  0.033 &  0.148 &  0.031 &  0.005 &  0.000 &  0.004 \B \\
         & 44.00 &  0.104 &  0.014 &  0.106 &  0.012 &  0.002 &  0.000 &  0.002 \B \\
\hline
Remnant & 41.30 &  0.097 &  0.062 &  0.101 &  0.057 &  0.000 &  0.000 &  0.000 \T \B \\
(0.103)  & 41.80 &  0.062 &  0.028 &  0.079 &  0.012 &  0.000 &  0.000 &  0.000 \B \\
         & 42.30 &  0.013 &  0.007 &  0.020 &  0.000 &  0.000 &  0.000 &  0.000 \B \\
         & 42.80 &  0.001 &  0.001 &  0.001 &  0.000 &  0.000 &  0.000 &  0.000 \B \\
         & 43.30 &  0.000 &  0.000 &  0.000 &  0.000 &  0.000 &  0.000 &  0.000 \B \\
         & 0.005 &  0.045 &  0.062 &  0.086 &  0.021 &  0.000 &  0.000 &  0.000 \B \\
         & 0.050 &  0.001 &  0.011 &  0.011 &  0.000 &  0.000 &  0.000 &  0.000 \B \\
         & 0.500 &  0.000 &  0.000 &  0.000 &  0.000 &  0.000 &  0.000 &  0.000 \B \\
         & 42.00 &  0.101 &  0.077 &  0.102 &  0.076 &  0.000 &  0.000 &  0.000 \B \\
         & 42.50 &  0.090 &  0.053 &  0.099 &  0.045 &  0.000 &  0.000 &  0.000 \B \\
         & 43.00 &  0.057 &  0.025 &  0.073 &  0.009 &  0.000 &  0.000 &  0.000 \B \\
         & 43.50 &  0.014 &  0.007 &  0.020 &  0.000 &  0.000 &  0.000 &  0.000 \B \\
         & 44.00 &  0.001 &  0.001 &  0.002 &  0.000 &  0.000 &  0.000 &  0.000 \B \\
\hline
Total    & 41.30 &  0.887 &  0.607 &  1.020 &  0.474 &  0.345 &  0.278 &  0.074 \T \B \\
(1.158)  & 41.80 &  0.460 &  0.219 &  0.563 &  0.116 &  0.056 &  0.038 &  0.014 \B \\
         & 42.30 &  0.197 &  0.059 &  0.224 &  0.033 &  0.007 &  0.002 &  0.005 \B \\
         & 42.80 &  0.096 &  0.013 &  0.100 &  0.008 &  0.001 &  0.000 &  0.002 \B \\
         & 43.30 &  0.048 &  0.001 &  0.049 &  0.000 &  0.000 &  0.000 &  0.000 \B \\
         & 0.005 &  0.798 &  0.773 &  1.043 &  0.529 &  0.437 &  0.357 &  0.099 \B \\
         & 0.050 &  0.187 &  0.125 &  0.267 &  0.045 &  0.015 &  0.006 &  0.006 \B \\
         & 0.500 &  0.039 &  0.008 &  0.045 &  0.003 &  0.001 &  0.000 &  0.001 \B \\
         & 42.00 &  1.068 &  0.857 &  1.134 &  0.792 &  0.634 &  0.518 &  0.147 \B \\
         & 42.50 &  0.777 &  0.490 &  0.921 &  0.345 &  0.234 &  0.184 &  0.050 \B \\
         & 43.00 &  0.412 &  0.185 &  0.501 &  0.096 &  0.042 &  0.027 &  0.012 \B \\
         & 43.50 &  0.201 &  0.061 &  0.229 &  0.034 &  0.007 &  0.002 &  0.005 \B \\
         & 44.00 &  0.111 &  0.018 &  0.117 &  0.012 &  0.002 &  0.000 &  0.002 \B \\
\hline
\end{tabular*}
\vspace{5pt}
\end{table*}

\clearpage

\onecolumn

\section{Online-only supplementary material}

\begin{figure*}
\centering
\vspace{2.0pt}
\includegraphics[width=0.815\columnwidth,angle=90]{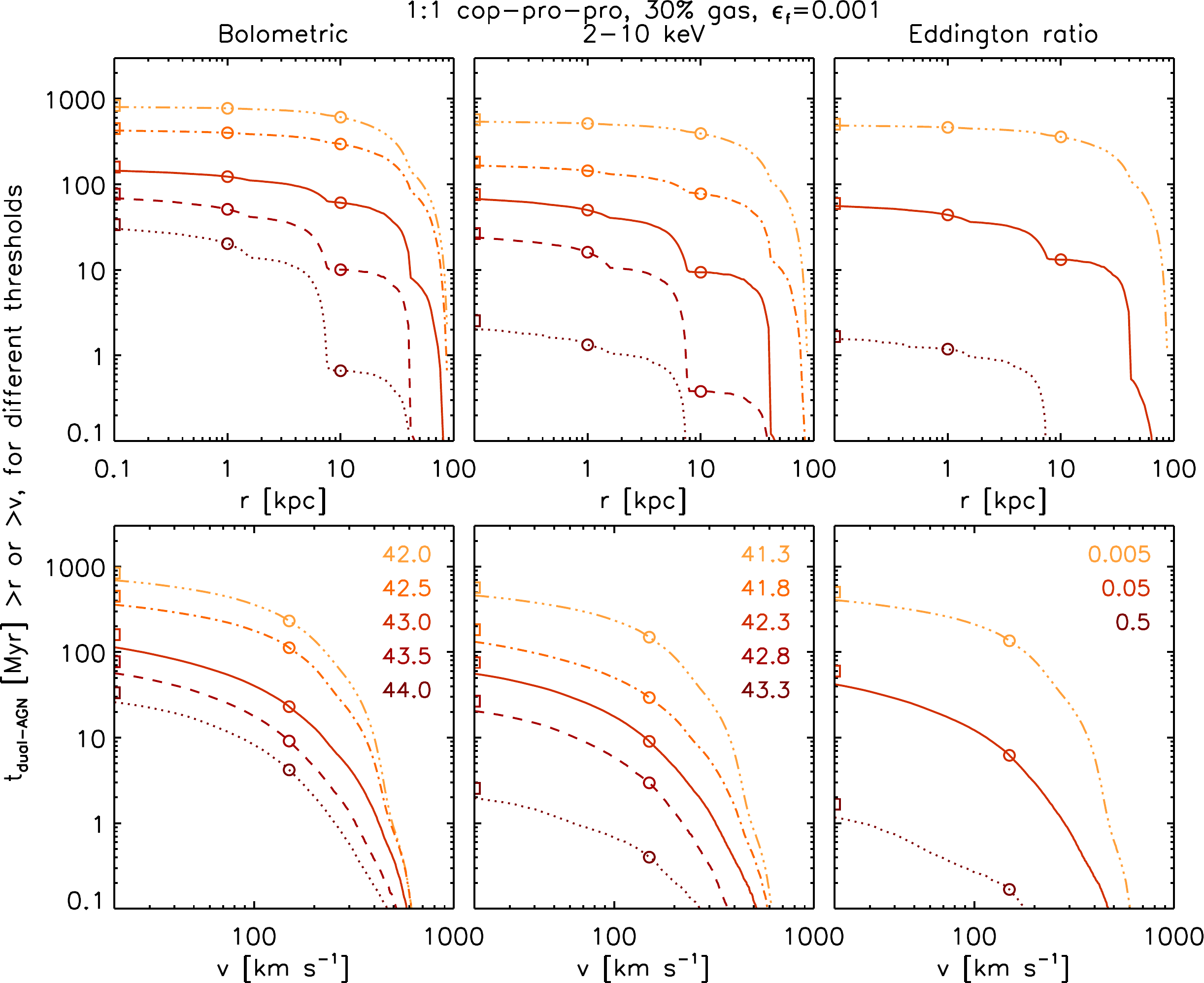}
\vspace{5.0pt}
\caption[]{Same as Fig.~\ref{dualagnpaper:fig:m2_hr_gf0_3_BHeff0_001_phi000000_dual_agn_Mdot_deltat_proj_no3b}, but for the 1:1 coplanar, prograde--prograde merger.}
\label{dualagnpaper:fig:m1_hr_gf0_3_BHeff0_001_phi000000_dual_agn_Mdot_deltat_proj_no3b}
\end{figure*}

\begin{figure*}
\centering
\vspace{2.0pt}
\includegraphics[width=0.815\columnwidth,angle=90]{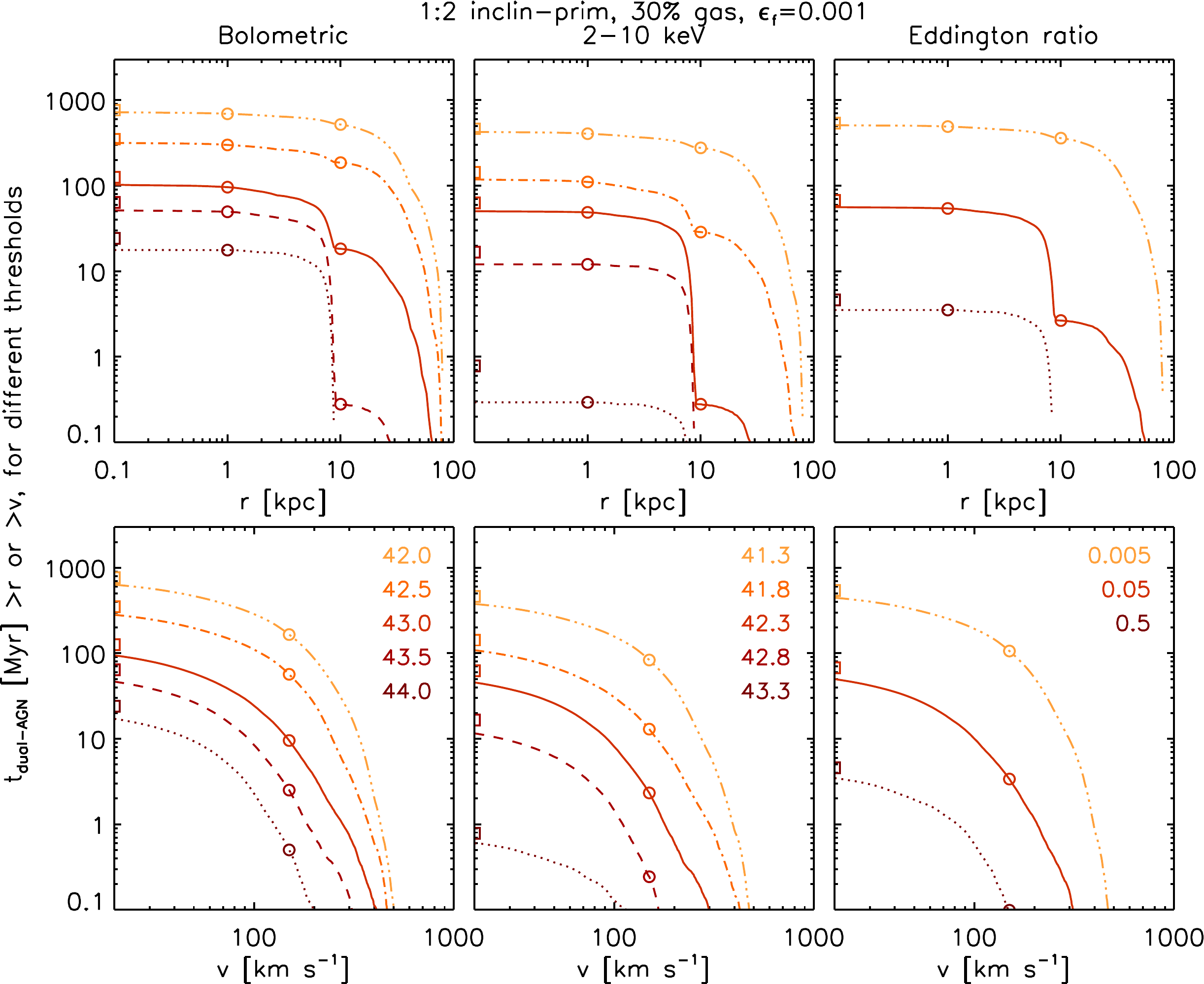}
\vspace{5.0pt}
\caption[]{Same as Fig.~\ref{dualagnpaper:fig:m2_hr_gf0_3_BHeff0_001_phi000000_dual_agn_Mdot_deltat_proj_no3b}, but for the 1:2 inclined-primary merger.}
\label{dualagnpaper:fig:m2_hr_gf0_3_BHeff0_001_phi045000_dual_agn_Mdot_deltat_proj_no3b}
\end{figure*}

\begin{figure*}
\centering
\vspace{2.0pt}
\includegraphics[width=0.815\columnwidth,angle=90]{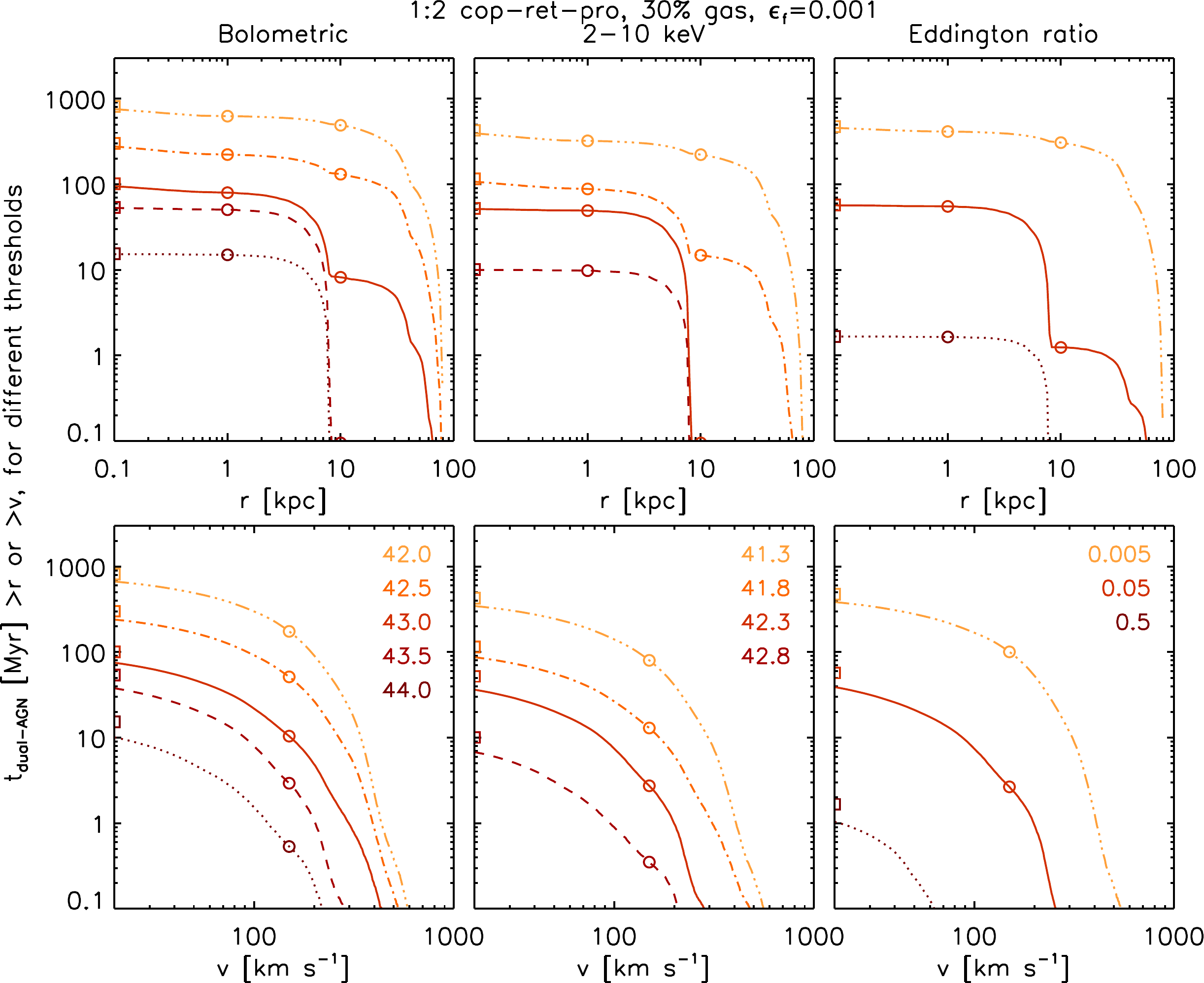}
\vspace{5.0pt}
\caption[]{Same as Fig.~\ref{dualagnpaper:fig:m2_hr_gf0_3_BHeff0_001_phi000000_dual_agn_Mdot_deltat_proj_no3b}, but for the 1:2 coplanar, retrograde--prograde merger.}
\label{dualagnpaper:fig:m2_hr_gf0_3_BHeff0_001_phi180000_dual_agn_Mdot_deltat_proj_no3b}
\end{figure*}

\begin{figure*}
\centering
\vspace{2.0pt}
\includegraphics[width=0.815\columnwidth,angle=90]{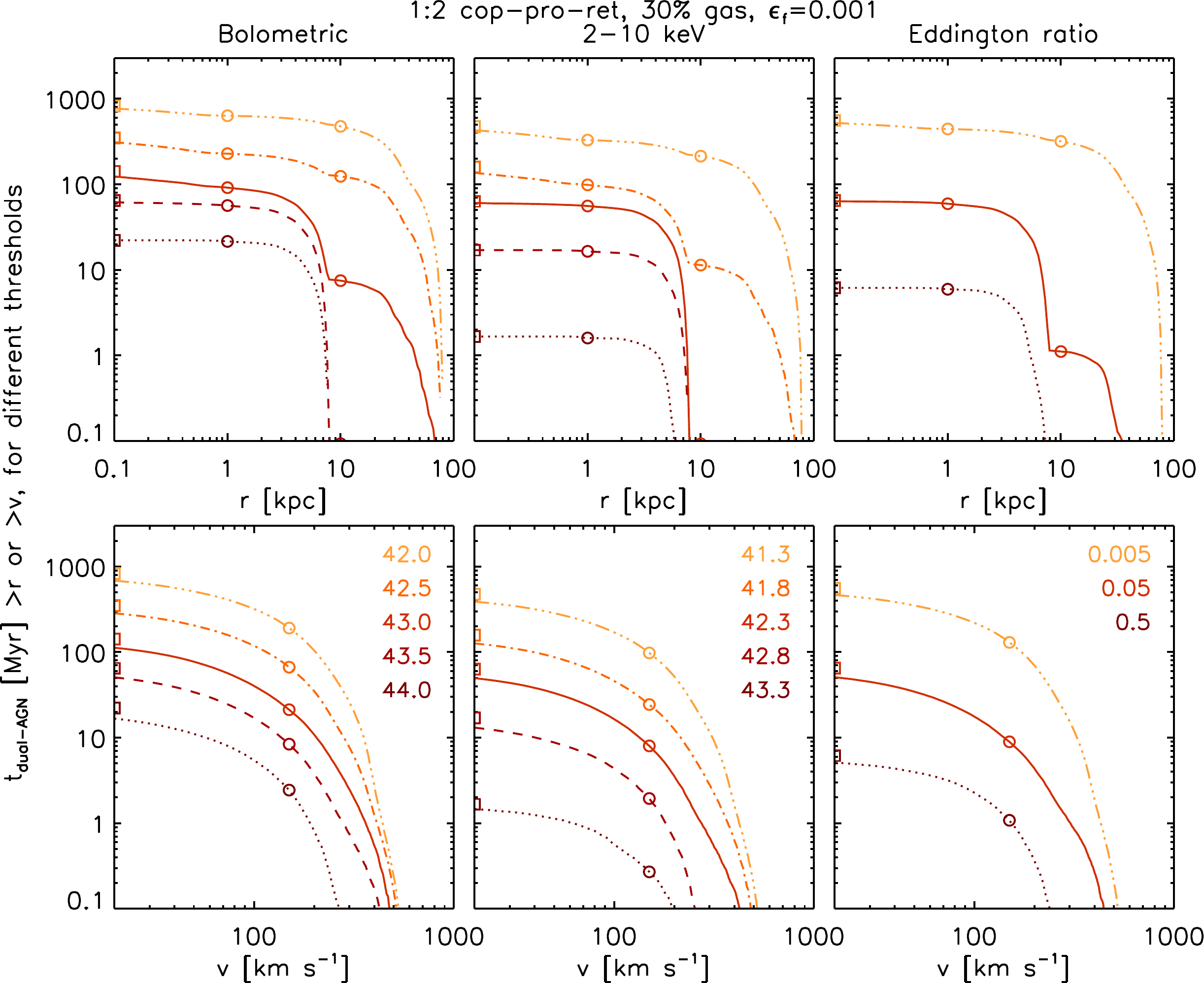}
\vspace{5.0pt}
\caption[]{Same as Fig.~\ref{dualagnpaper:fig:m2_hr_gf0_3_BHeff0_001_phi000000_dual_agn_Mdot_deltat_proj_no3b}, but for the 1:2 coplanar, prograde--retrograde merger.}
\label{dualagnpaper:fig:m2_hr_gf0_3_BHeff0_001_phi180000_dual_agn_Mdot_deltat_proj_no3b}
\end{figure*}

\begin{figure*}
\centering
\vspace{2.0pt}
\includegraphics[width=0.815\columnwidth,angle=90]{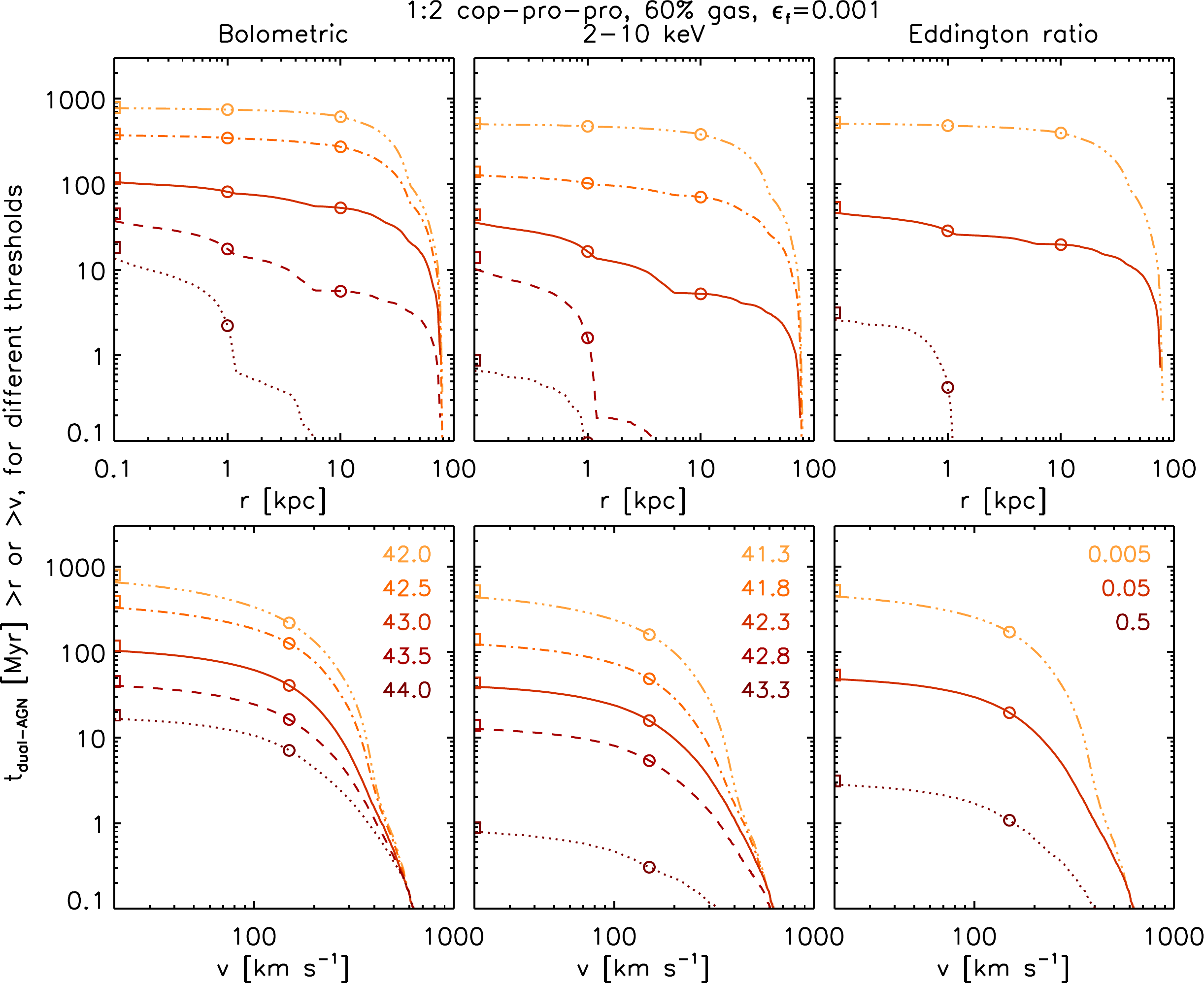}
\vspace{5.0pt}
\caption[]{Same as Fig.~\ref{dualagnpaper:fig:m2_hr_gf0_3_BHeff0_001_phi000000_dual_agn_Mdot_deltat_proj_no3b}, but for the 1:2 coplanar, prograde--prograde merger with 60 per cent gas fraction, standard BH mass, and standard BH feedback efficiency.}
\label{dualagnpaper:fig:m2_hr_gf0_6_BHeff0_001_phi000000_dual_agn_Mdot_deltat_proj_no3b}
\end{figure*}

\begin{figure*}
\centering
\vspace{2.0pt}
\includegraphics[width=0.815\columnwidth,angle=90]{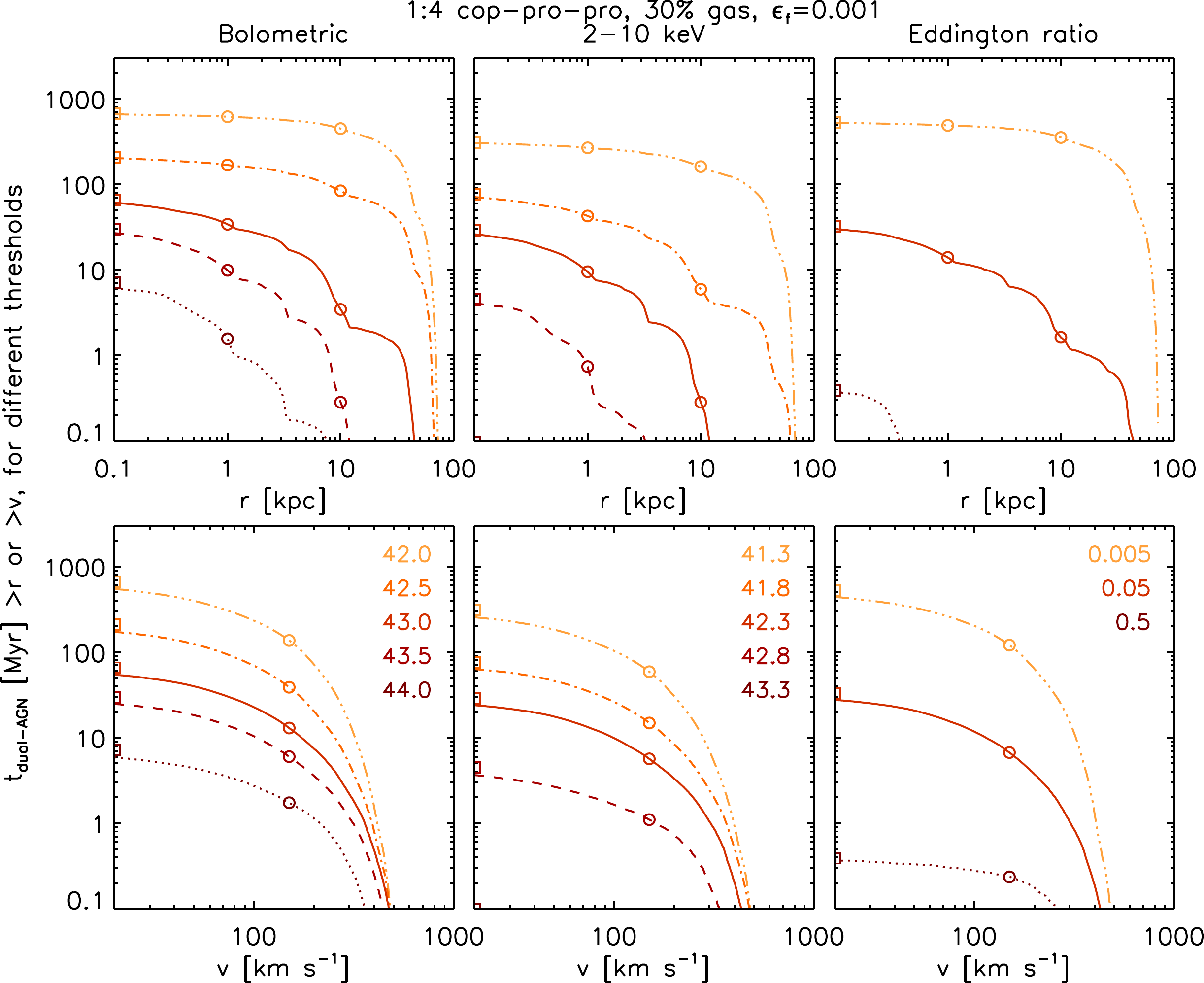}
\vspace{5.0pt}
\caption[]{Same as Fig.~\ref{dualagnpaper:fig:m2_hr_gf0_3_BHeff0_001_phi000000_dual_agn_Mdot_deltat_proj_no3b}, but for the 1:4 coplanar, prograde--prograde merger.}
\label{dualagnpaper:fig:m4_hr_gf0_3_BHeff0_001_phi000000_dual_agn_Mdot_deltat_proj_no3b}
\end{figure*}

\begin{figure*}
\centering
\vspace{2.0pt}
\includegraphics[width=0.815\columnwidth,angle=90]{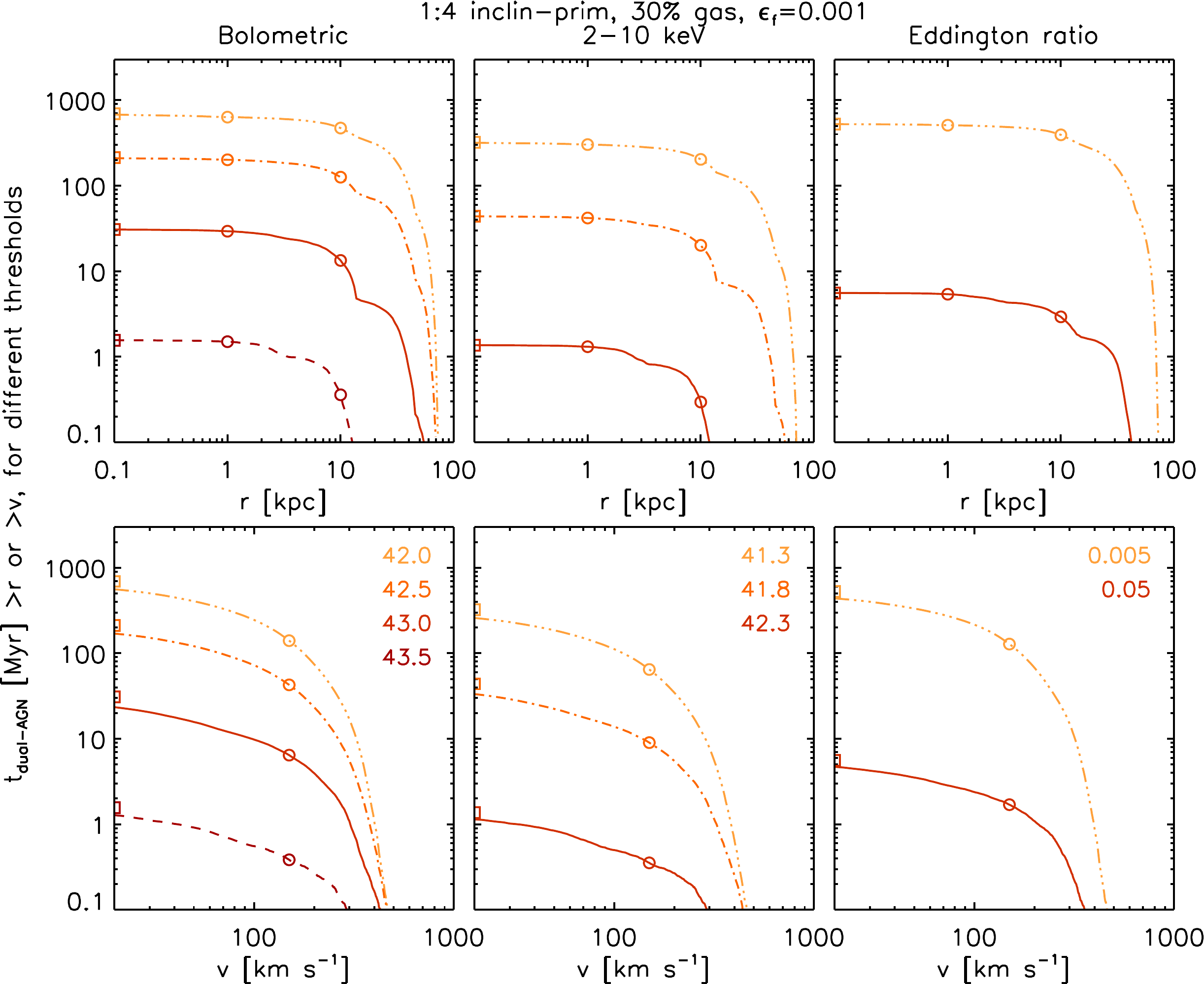}
\vspace{5.0pt}
\caption[]{Same as Fig.~\ref{dualagnpaper:fig:m2_hr_gf0_3_BHeff0_001_phi000000_dual_agn_Mdot_deltat_proj_no3b}, but for the 1:4 inclined-primary merger.}
\label{dualagnpaper:fig:m4_hr_gf0_3_BHeff0_001_phi045000_dual_agn_Mdot_deltat_proj_no3b}
\end{figure*}

\begin{figure*}
\centering
\vspace{2.0pt}
\includegraphics[width=0.815\columnwidth,angle=90]{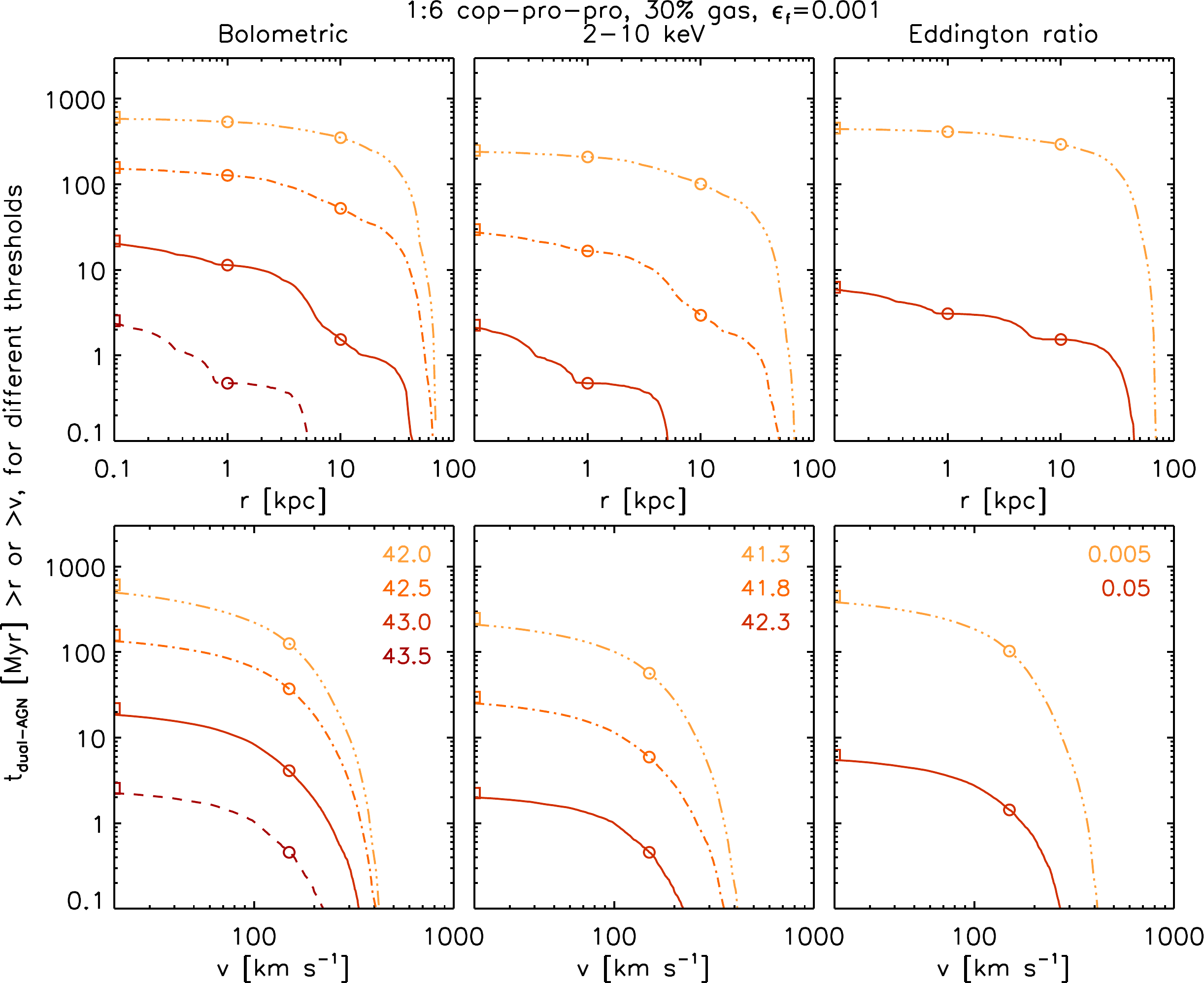}
\vspace{5.0pt}
\caption[]{Same as Fig.~\ref{dualagnpaper:fig:m2_hr_gf0_3_BHeff0_001_phi000000_dual_agn_Mdot_deltat_proj_no3b}, but for the 1:6 coplanar, prograde--prograde merger.}
\label{dualagnpaper:fig:m6_hr_gf0_3_BHeff0_001_phi000000_dual_agn_Mdot_deltat_proj_no3b}
\end{figure*}

\begin{figure*}
\centering
\vspace{2.0pt}
\includegraphics[width=0.815\columnwidth,angle=90]{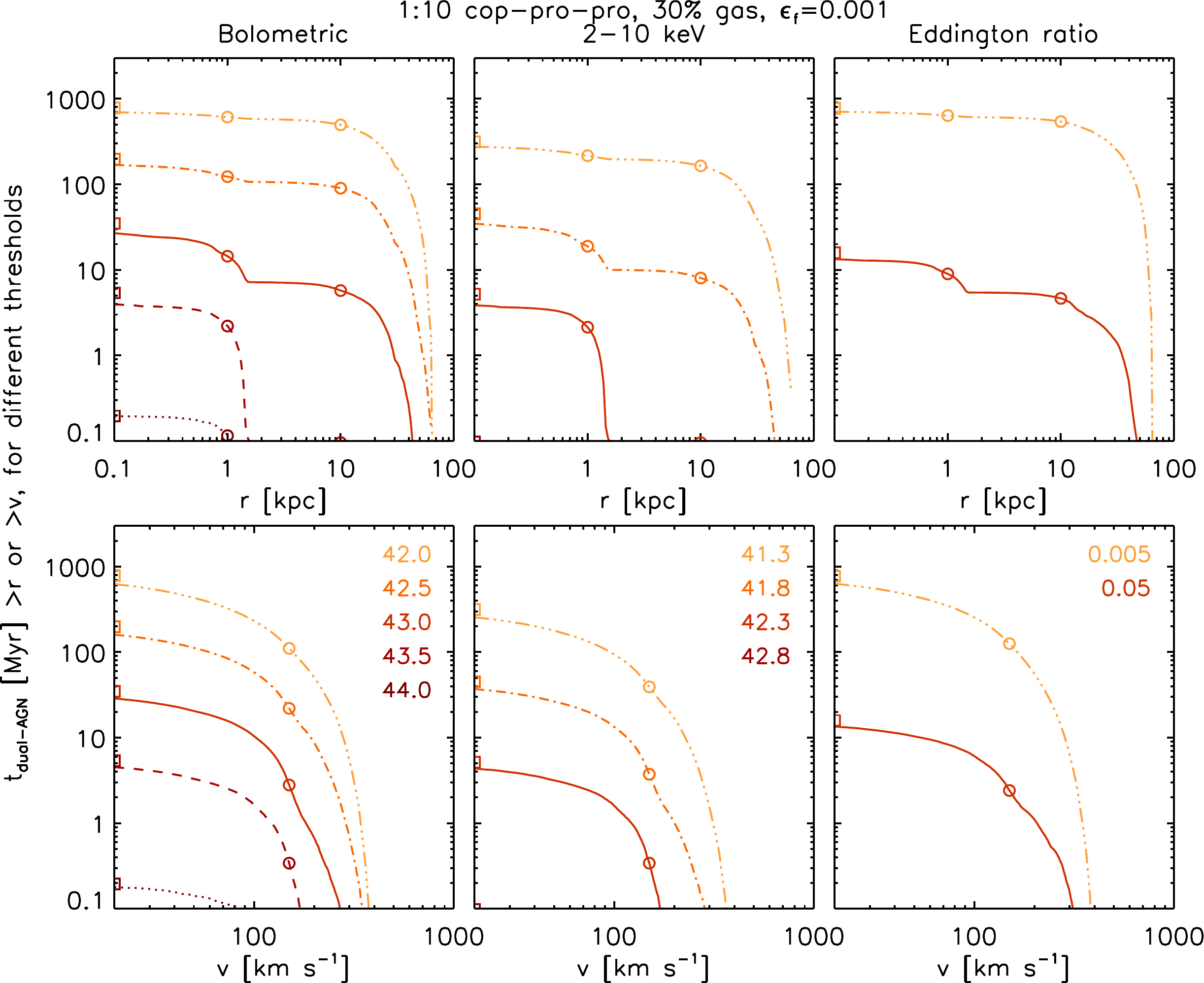}
\vspace{5.0pt}
\caption[]{Same as Fig.~\ref{dualagnpaper:fig:m2_hr_gf0_3_BHeff0_001_phi000000_dual_agn_Mdot_deltat_proj_no3b}, but for the 1:10 coplanar, prograde--prograde merger.}
\label{dualagnpaper:fig:m10_hr_gf0_3_BHeff0_001_phi000000_dual_agn_Mdot_deltat_proj_no3b}
\end{figure*}

\begin{figure*}
\centering
\vspace{2.0pt}
\includegraphics[width=0.815\columnwidth,angle=90]{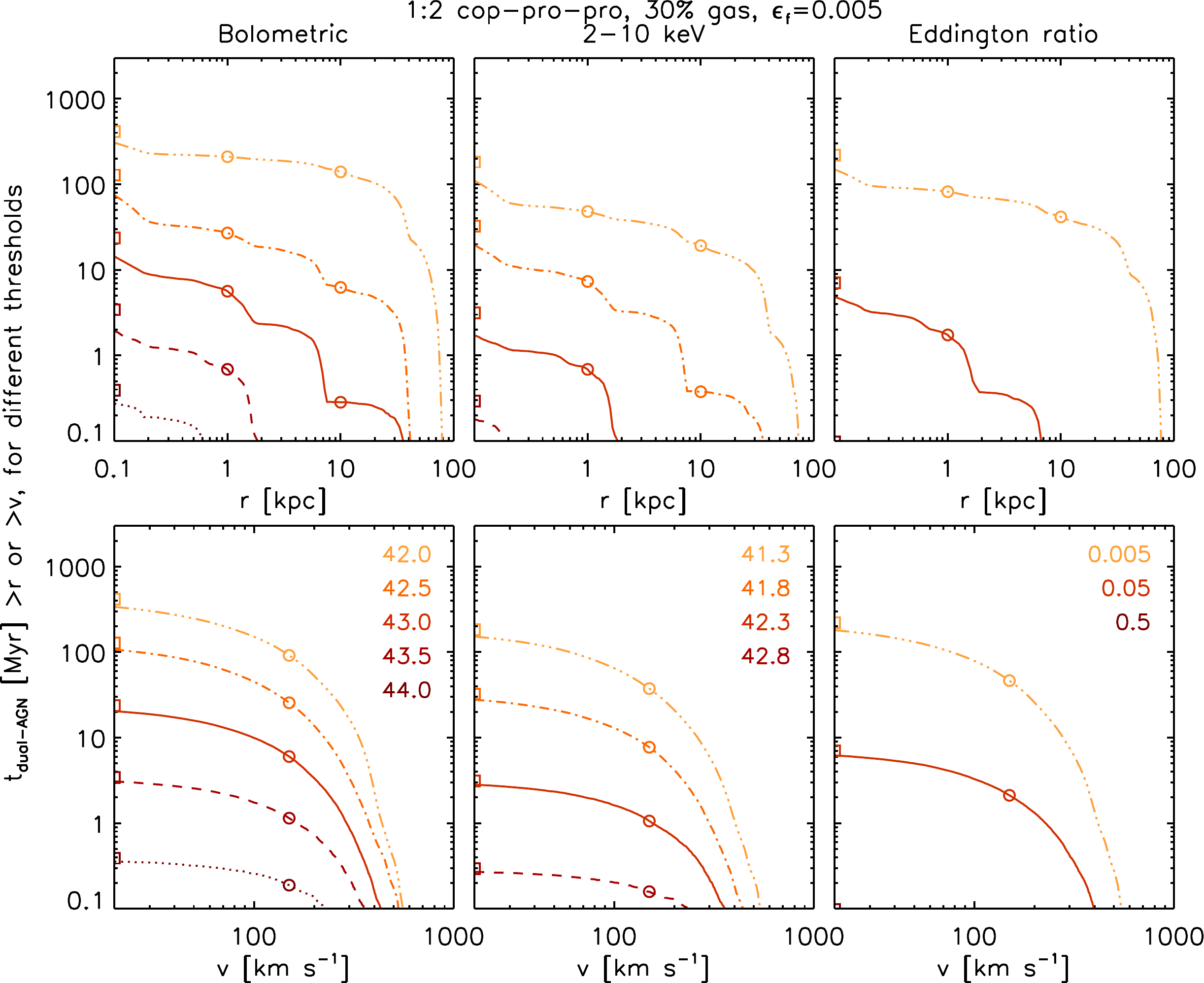}
\vspace{5.0pt}
\caption[]{Same as Fig.~\ref{dualagnpaper:fig:m2_hr_gf0_3_BHeff0_001_phi000000_dual_agn_Mdot_deltat_proj_no3b}, but for the 1:2 coplanar, prograde--prograde merger with 30 per cent gas fraction, standard BH mass, and high BH feedback efficiency.}
\label{dualagnpaper:fig:m2_hr_gf0_3_BHeff0_005_phi000000_dual_agn_Mdot_deltat_proj_no3b}
\end{figure*}

\begin{figure*}
\centering
\vspace{2.0pt}
\includegraphics[width=0.815\columnwidth,angle=90]{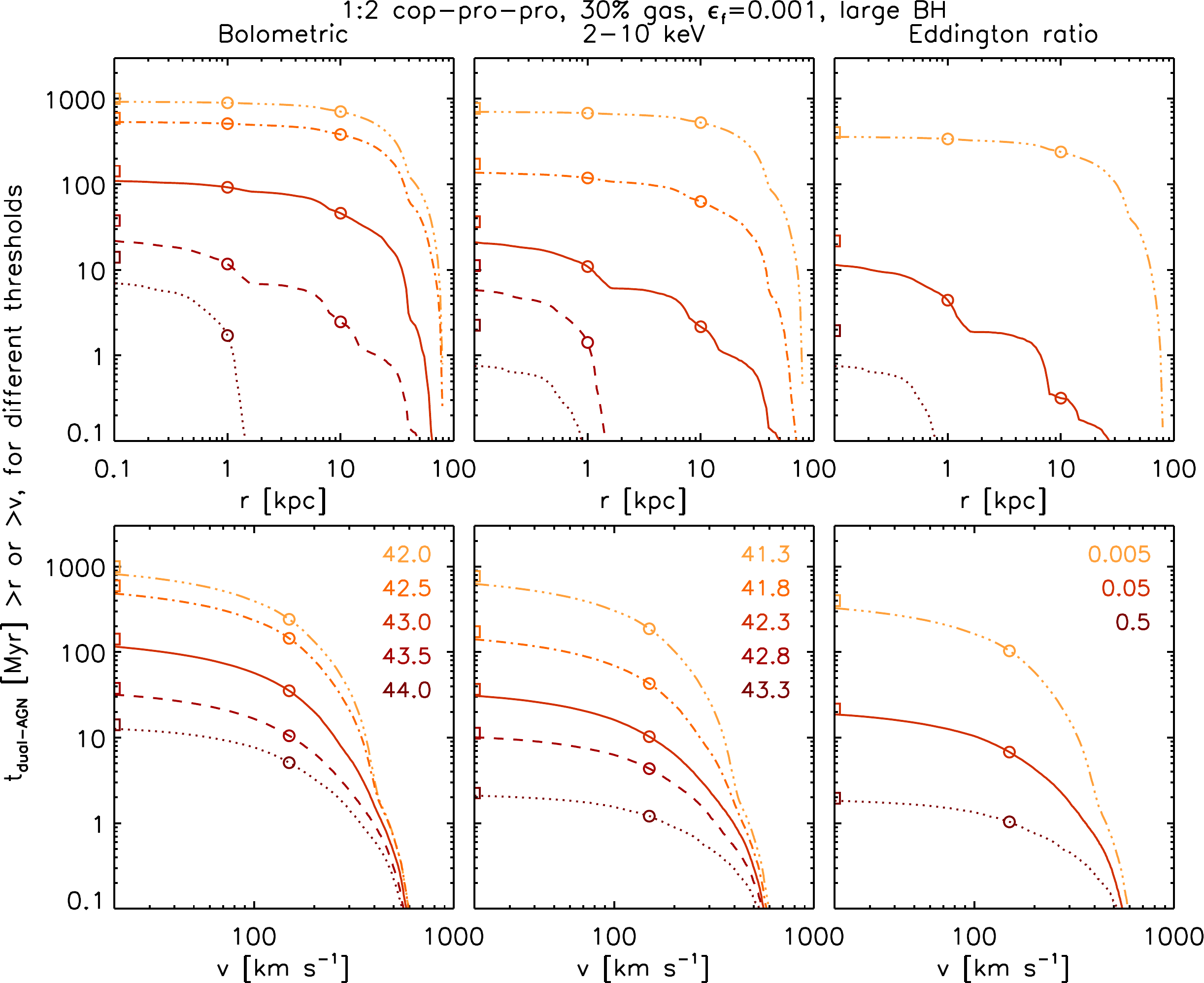}
\vspace{5.0pt}
\caption[]{Same as Fig.~\ref{dualagnpaper:fig:m2_hr_gf0_3_BHeff0_001_phi000000_dual_agn_Mdot_deltat_proj_no3b}, but for the 1:2 coplanar, prograde--prograde merger with 30 per cent gas fraction, standard BH feedback efficiency, and large BH mass.}
\label{dualagnpaper:fig:m2_hr_gf0_3_BHeff0_001_phi000000_largerBH_dual_agn_Mdot_deltat_proj_no3b}
\end{figure*}

\clearpage

\begin{table*} \centering
\vspace{-3.5pt}
\caption[Simulation results]{Same as Table~\ref{dualagnpaper:tab:m2_hr_gf0_3_BHeff0_001_phi000000}, but for the 1:1 coplanar, prograde--prograde merger.
\label{dualagnpaper:tab:m1_hr_gf0_3_BHeff0_001_phi000000}}
\vspace{10.5pt}
\vspace{5pt}
\begin{tabular*}{0.94\textwidth}{c|m{40pt}m{40pt}m{40pt}m{40pt}|m{40pt}m{40pt}m{40pt}m{61pt}}
Run 01    & Threshold & BH$_1$ & BH$_2$ & Either BH & \multicolumn{4}{c}{Both BHs above the threshold} \B \\
         & L$_{\rm thr}$/$f_{\rm Edd}$ & above thr. & above thr. & above thr. & any $d$ or $v$ & $d > 1$ kpc & $d > 10$ kpc & $v > 150$ km s$^{-1}$ \B \\
\hline
Stochastic  & 41.30 &  0.652 &  0.538 &  0.747 &  0.444 &  0.442 &  0.392 &  0.131 \T \B \\
(0.805)  & 41.80 &  0.339 &  0.177 &  0.430 &  0.086 &  0.085 &  0.078 &  0.016 \B \\
         & 42.30 &  0.144 &  0.030 &  0.164 &  0.010 &  0.010 &  0.009 &  0.000 \B \\
         & 42.80 &  0.049 &  0.004 &  0.053 &  0.000 &  0.000 &  0.000 &  0.000 \B \\
         & 43.30 &  0.008 &  0.000 &  0.009 &  0.000 &  0.000 &  0.000 &  0.000 \B \\
         & 0.005 &  0.625 &  0.506 &  0.728 &  0.403 &  0.401 &  0.358 &  0.121 \B \\
\hline
\end{tabular*}
\vspace{5pt}
\end{table*}

\clearpage

\begin{table*} \centering
\vspace{-3.5pt}
\caption[Simulation results]{Same as Table~\ref{dualagnpaper:tab:m2_hr_gf0_3_BHeff0_001_phi000000}, but for the 1:2 inclined-primary merger.
\label{dualagnpaper:tab:m2_hr_gf0_3_BHeff0_001_phi045000}}
\vspace{10.5pt}
\vspace{5pt}
\begin{tabular*}{0.94\textwidth}{c|m{40pt}m{40pt}m{40pt}m{40pt}|m{40pt}m{40pt}m{40pt}m{61pt}}
Run 03    & Threshold & BH$_1$ & BH$_2$ & Either BH & \multicolumn{4}{c}{Both BHs above the threshold} \B \\
         & L$_{\rm thr}$/$f_{\rm Edd}$ & above thr. & above thr. & above thr. & any $d$ or $v$ & $d > 1$ kpc & $d > 10$ kpc & $v > 150$ km s$^{-1}$ \B \\
\hline
Stochastic  & 41.30 &  0.622 &  0.411 &  0.728 &  0.304 &  0.303 &  0.277 &  0.070 \T \B \\
(0.859)  & 41.80 &  0.202 &  0.111 &  0.281 &  0.032 &  0.031 &  0.029 &  0.007 \B \\
         & 42.30 &  0.023 &  0.018 &  0.041 &  0.000 &  0.000 &  0.000 &  0.000 \B \\
         & 42.80 &  0.002 &  0.002 &  0.004 &  0.000 &  0.000 &  0.000 &  0.000 \B \\
         & 43.30 &  0.000 &  0.000 &  0.000 &  0.000 &  0.000 &  0.000 &  0.000 \B \\
         & 0.005 &  0.586 &  0.571 &  0.763 &  0.394 &  0.393 &  0.360 &  0.094 \B \\
\hline
\end{tabular*}
\vspace{5pt}
\end{table*}

\clearpage

\begin{table*} \centering
\vspace{-3.5pt}
\caption[Simulation results]{Same as Table~\ref{dualagnpaper:tab:m2_hr_gf0_3_BHeff0_001_phi000000}, but for the 1:2 coplanar, retrograde--prograde merger.
\label{dualagnpaper:tab:m2_hr_gf0_3_BHeff0_001_phi180000}}
\vspace{10.5pt}
\vspace{5pt}
\begin{tabular*}{0.94\textwidth}{c|m{40pt}m{40pt}m{40pt}m{40pt}|m{40pt}m{40pt}m{40pt}m{61pt}}
Run 04    & Threshold & BH$_1$ & BH$_2$ & Either BH & \multicolumn{4}{c}{Both BHs above the threshold} \B \\
         & L$_{\rm thr}$/$f_{\rm Edd}$ & above thr. & above thr. & above thr. & any $d$ or $v$ & $d > 1$ kpc & $d > 10$ kpc & $v > 150$ km s$^{-1}$ \B \\
\hline
Stochastic  & 41.30 &  0.487 &  0.458 &  0.704 &  0.241 &  0.241 &  0.222 &  0.058 \T \B \\
(0.876)  & 41.80 &  0.127 &  0.122 &  0.233 &  0.016 &  0.016 &  0.015 &  0.003 \B \\
         & 42.30 &  0.011 &  0.018 &  0.029 &  0.000 &  0.000 &  0.000 &  0.000 \B \\
         & 42.80 &  0.000 &  0.002 &  0.002 &  0.000 &  0.000 &  0.000 &  0.000 \B \\
         & 43.30 &  0.000 &  0.000 &  0.000 &  0.000 &  0.000 &  0.000 &  0.000 \B \\
         & 0.005 &  0.466 &  0.628 &  0.762 &  0.332 &  0.331 &  0.307 &  0.085 \B \\
\hline
\end{tabular*}
\vspace{5pt}
\end{table*}

\clearpage

\begin{table*} \centering
\vspace{-3.5pt}
\caption[Simulation results]{Same as Table~\ref{dualagnpaper:tab:m2_hr_gf0_3_BHeff0_001_phi000000}, but for the 1:2 coplanar, prograde--retrograde merger.
\label{dualagnpaper:tab:m2_hr_gf0_3_BHeff0_001_phi000180}}
\vspace{10.5pt}
\vspace{5pt}
\begin{tabular*}{0.94\textwidth}{c|m{40pt}m{40pt}m{40pt}m{40pt}|m{40pt}m{40pt}m{40pt}m{61pt}}
Run 05    & Threshold & BH$_1$ & BH$_2$ & Either BH & \multicolumn{4}{c}{Both BHs above the threshold} \B \\
         & L$_{\rm thr}$/$f_{\rm Edd}$ & above thr. & above thr. & above thr. & any $d$ or $v$ & $d > 1$ kpc & $d > 10$ kpc & $v > 150$ km s$^{-1}$ \B \\
\hline
Stochastic  & 41.30 &  0.555 &  0.348 &  0.667 &  0.236 &  0.234 &  0.213 &  0.064 \T \B \\
(0.839)  & 41.80 &  0.181 &  0.059 &  0.227 &  0.013 &  0.013 &  0.011 &  0.004 \B \\
         & 42.30 &  0.039 &  0.004 &  0.043 &  0.000 &  0.000 &  0.000 &  0.000 \B \\
         & 42.80 &  0.006 &  0.000 &  0.006 &  0.000 &  0.000 &  0.000 &  0.000 \B \\
         & 43.30 &  0.000 &  0.000 &  0.000 &  0.000 &  0.000 &  0.000 &  0.000 \B \\
         & 0.005 &  0.532 &  0.538 &  0.718 &  0.352 &  0.350 &  0.318 &  0.101 \B \\
\hline
\end{tabular*}
\vspace{5pt}
\end{table*}

\clearpage

\begin{table*} \centering
\vspace{-3.5pt}
\caption[Simulation results]{Same as Table~\ref{dualagnpaper:tab:m2_hr_gf0_3_BHeff0_001_phi000000}, but for the 1:2 coplanar, prograde--prograde merger with 60 per cent gas fraction, standard BH mass, and standard BH feedback efficiency.
\label{dualagnpaper:tab:m2_hr_gf0_6_BHeff0_001_phi000000}}
\vspace{5pt}
\begin{tabular*}{0.94\textwidth}{c|m{40pt}m{40pt}m{40pt}m{40pt}|m{40pt}m{40pt}m{40pt}m{61pt}}
Run 06    & Threshold & BH$_1$ & BH$_2$ & Either BH & \multicolumn{4}{c}{Both BHs above the threshold} \B \\
         & L$_{\rm thr}$/$f_{\rm Edd}$ & above thr. & above thr. & above thr. & any $d$ or $v$ & $d > 1$ kpc & $d > 10$ kpc & $v > 150$ km s$^{-1}$ \B \\
\hline
Stochastic  & 41.30 &  0.698 &  0.541 &  0.795 &  0.444 &  0.441 &  0.383 &  0.136 \T \B \\
(0.856)  & 41.80 &  0.324 &  0.180 &  0.427 &  0.078 &  0.077 &  0.071 &  0.028 \B \\
         & 42.30 &  0.099 &  0.029 &  0.122 &  0.005 &  0.005 &  0.005 &  0.002 \B \\
         & 42.80 &  0.018 &  0.001 &  0.019 &  0.000 &  0.000 &  0.000 &  0.000 \B \\
         & 43.30 &  0.001 &  0.000 &  0.001 &  0.000 &  0.000 &  0.000 &  0.000 \B \\
         & 0.005 &  0.575 &  0.656 &  0.778 &  0.453 &  0.450 &  0.398 &  0.148 \B \\
\hline
\end{tabular*}
\vspace{5pt}
\end{table*}

\clearpage

\begin{table*} \centering
\vspace{-3.5pt}
\caption[Simulation results]{Same as Table~\ref{dualagnpaper:tab:m2_hr_gf0_3_BHeff0_001_phi000000}, but for the 1:4 coplanar, prograde--prograde merger.
\label{dualagnpaper:tab:m4_hr_gf0_3_BHeff0_001_phi000000}}
\vspace{10.5pt}
\vspace{5pt}
\begin{tabular*}{0.94\textwidth}{c|m{40pt}m{40pt}m{40pt}m{40pt}|m{40pt}m{40pt}m{40pt}m{61pt}}
Run 07    & Threshold & BH$_1$ & BH$_2$ & Either BH & \multicolumn{4}{c}{Both BHs above the threshold} \B \\
         & L$_{\rm thr}$/$f_{\rm Edd}$ & above thr. & above thr. & above thr. & any $d$ or $v$ & $d > 1$ kpc & $d > 10$ kpc & $v > 150$ km s$^{-1}$ \B \\
\hline
Stochastic  & 41.30 &  0.595 &  0.291 &  0.718 &  0.168 &  0.167 &  0.152 &  0.033 \T \B \\
(0.973)  & 41.80 &  0.152 &  0.059 &  0.206 &  0.005 &  0.005 &  0.004 &  0.001 \B \\
         & 42.30 &  0.017 &  0.010 &  0.027 &  0.000 &  0.000 &  0.000 &  0.000 \B \\
         & 42.80 &  0.001 &  0.001 &  0.002 &  0.000 &  0.000 &  0.000 &  0.000 \B \\
         & 43.30 &  0.000 &  0.000 &  0.000 &  0.000 &  0.000 &  0.000 &  0.000 \B \\
         & 0.005 &  0.555 &  0.670 &  0.840 &  0.386 &  0.385 &  0.342 &  0.096 \B \\
\hline
\end{tabular*}
\vspace{5pt}
\end{table*}

\clearpage

\begin{table*} \centering
\vspace{-3.5pt}
\caption[Simulation results]{Same as Table~\ref{dualagnpaper:tab:m2_hr_gf0_3_BHeff0_001_phi000000}, but for the 1:4 inclined-primary merger.
\label{dualagnpaper:tab:m4_hr_gf0_3_BHeff0_001_phi045000}}
\vspace{10.5pt}
\vspace{5pt}
\begin{tabular*}{0.94\textwidth}{c|m{40pt}m{40pt}m{40pt}m{40pt}|m{40pt}m{40pt}m{40pt}m{61pt}}
Run 08    & Threshold & BH$_1$ & BH$_2$ & Either BH & \multicolumn{4}{c}{Both BHs above the threshold} \B \\
         & L$_{\rm thr}$/$f_{\rm Edd}$ & above thr. & above thr. & above thr. & any $d$ or $v$ & $d > 1$ kpc & $d > 10$ kpc & $v > 150$ km s$^{-1}$ \B \\
\hline
Stochastic  & 41.30 &  0.545 &  0.287 &  0.656 &  0.176 &  0.176 &  0.159 &  0.047 \T \B \\
(0.983)  & 41.80 &  0.142 &  0.052 &  0.184 &  0.010 &  0.010 &  0.009 &  0.003 \B \\
         & 42.30 &  0.014 &  0.004 &  0.018 &  0.000 &  0.000 &  0.000 &  0.000 \B \\
         & 42.80 &  0.001 &  0.000 &  0.001 &  0.000 &  0.000 &  0.000 &  0.000 \B \\
         & 43.30 &  0.000 &  0.000 &  0.000 &  0.000 &  0.000 &  0.000 &  0.000 \B \\
         & 0.005 &  0.512 &  0.704 &  0.822 &  0.394 &  0.393 &  0.353 &  0.111 \B \\
\hline
\end{tabular*}
\vspace{5pt}
\end{table*}

\clearpage

\begin{table*} \centering
\vspace{-3.5pt}
\caption[Simulation results]{Same as Table~\ref{dualagnpaper:tab:m2_hr_gf0_3_BHeff0_001_phi000000}, but for the 1:6 coplanar, prograde--prograde merger.
\label{dualagnpaper:tab:m6_hr_gf0_3_BHeff0_001_phi000000}}
\vspace{10.5pt}
\vspace{5pt}
\begin{tabular*}{0.94\textwidth}{c|m{40pt}m{40pt}m{40pt}m{40pt}|m{40pt}m{40pt}m{40pt}m{61pt}}
Run 09    & Threshold & BH$_1$ & BH$_2$ & Either BH & \multicolumn{4}{c}{Both BHs above the threshold} \B \\
         & L$_{\rm thr}$/$f_{\rm Edd}$ & above thr. & above thr. & above thr. & any $d$ or $v$ & $d > 1$ kpc & $d > 10$ kpc & $v > 150$ km s$^{-1}$ \B \\
\hline
Stochastic  & 41.30 &  0.439 &  0.179 &  0.533 &  0.084 &  0.084 &  0.079 &  0.024 \T \B \\
(1.091)  & 41.80 &  0.062 &  0.026 &  0.086 &  0.002 &  0.002 &  0.002 &  0.001 \B \\
         & 42.30 &  0.005 &  0.002 &  0.007 &  0.000 &  0.000 &  0.000 &  0.000 \B \\
         & 42.80 &  0.000 &  0.000 &  0.000 &  0.000 &  0.000 &  0.000 &  0.000 \B \\
         & 43.30 &  0.000 &  0.000 &  0.000 &  0.000 &  0.000 &  0.000 &  0.000 \B \\
         & 0.005 &  0.413 &  0.704 &  0.828 &  0.290 &  0.289 &  0.271 &  0.071 \B \\
\hline
\end{tabular*}
\vspace{5pt}
\end{table*}

\clearpage

\begin{table*} \centering
\vspace{-3.5pt}
\caption[Simulation results]{Same as Table~\ref{dualagnpaper:tab:m2_hr_gf0_3_BHeff0_001_phi000000}, but for the 1:10 coplanar, prograde--prograde merger.
\label{dualagnpaper:tab:m10_hr_gf0_3_BHeff0_001_phi000000}}
\vspace{10.5pt}
\vspace{5pt}
\begin{tabular*}{0.94\textwidth}{c|m{40pt}m{40pt}m{40pt}m{40pt}|m{40pt}m{40pt}m{40pt}m{61pt}}
Run 10    & Threshold & BH$_1$ & BH$_2$ & Either BH & \multicolumn{4}{c}{Both BHs above the threshold} \B \\
         & L$_{\rm thr}$/$f_{\rm Edd}$ & above thr. & above thr. & above thr. & any $d$ or $v$ & $d > 1$ kpc & $d > 10$ kpc & $v > 150$ km s$^{-1}$ \B \\
\hline
Stochastic  & 41.30 &  0.755 &  0.129 &  0.798 &  0.086 &  0.085 &  0.079 &  0.021 \T \B \\
(1.277)  & 41.80 &  0.164 &  0.015 &  0.176 &  0.003 &  0.003 &  0.003 &  0.001 \B \\
         & 42.30 &  0.021 &  0.001 &  0.022 &  0.000 &  0.000 &  0.000 &  0.000 \B \\
         & 42.80 &  0.002 &  0.000 &  0.002 &  0.000 &  0.000 &  0.000 &  0.000 \B \\
         & 43.30 &  0.000 &  0.000 &  0.000 &  0.000 &  0.000 &  0.000 &  0.000 \B \\
         & 0.005 &  0.676 &  0.888 &  1.082 &  0.483 &  0.482 &  0.450 &  0.098 \B \\
\hline
\end{tabular*}
\vspace{5pt}
\end{table*}

\clearpage

\begin{table*} \centering
\vspace{-3.5pt}
\caption[Simulation results]{Same as Table~\ref{dualagnpaper:tab:m2_hr_gf0_3_BHeff0_001_phi000000}, but for the 1:2 coplanar, prograde--prograde merger with 30 per cent gas fraction, standard BH mass, and high BH feedback efficiency.
\label{dualagnpaper:tab:m2_hr_gf0_3_BHeff0_005_phi000000}}
\vspace{5pt}
\begin{tabular*}{0.94\textwidth}{c|m{40pt}m{40pt}m{40pt}m{40pt}|m{40pt}m{40pt}m{40pt}m{61pt}}
Run C3    & Threshold & BH$_1$ & BH$_2$ & Either BH & \multicolumn{4}{c}{Both BHs above the threshold} \B \\
         & L$_{\rm thr}$/$f_{\rm Edd}$ & above thr. & above thr. & above thr. & any $d$ or $v$ & $d > 1$ kpc & $d > 10$ kpc & $v > 150$ km s$^{-1}$ \B \\
\hline
Stochastic  & 41.30 &  0.203 &  0.127 &  0.305 &  0.025 &  0.024 &  0.019 &  0.007 \T \B \\
(0.849)  & 41.80 &  0.022 &  0.019 &  0.040 &  0.001 &  0.000 &  0.000 &  0.000 \B \\
         & 42.30 &  0.002 &  0.001 &  0.004 &  0.000 &  0.000 &  0.000 &  0.000 \B \\
         & 42.80 &  0.000 &  0.000 &  0.000 &  0.000 &  0.000 &  0.000 &  0.000 \B \\
         & 43.30 &  0.000 &  0.000 &  0.000 &  0.000 &  0.000 &  0.000 &  0.000 \B \\
         & 0.005 &  0.198 &  0.275 &  0.421 &  0.052 &  0.050 &  0.042 &  0.015 \B \\
\hline
\end{tabular*}
\vspace{5pt}
\end{table*}

\clearpage

\begin{table*} \centering
\vspace{-3.5pt}
\caption[Simulation results]{Same as Table~\ref{dualagnpaper:tab:m2_hr_gf0_3_BHeff0_001_phi000000}, but for the 1:2 coplanar, prograde--prograde merger with 30 per cent gas fraction, standard BH feedback efficiency, and large BH mass.
\label{dualagnpaper:tab:m2_hr_gf0_3_BHeff0_001_phi000000_largerBH}}
\vspace{5pt}
\begin{tabular*}{0.94\textwidth}{c|m{40pt}m{40pt}m{40pt}m{40pt}|m{40pt}m{40pt}m{40pt}m{61pt}}
Run C4    & Threshold & BH$_1$ & BH$_2$ & Either BH & \multicolumn{4}{c}{Both BHs above the threshold} \B \\
         & L$_{\rm thr}$/$f_{\rm Edd}$ & above thr. & above thr. & above thr. & any $d$ or $v$ & $d > 1$ kpc & $d > 10$ kpc & $v > 150$ km s$^{-1}$ \B \\
\hline
Stochastic  & 41.30 &  0.754 &  0.690 &  0.834 &  0.610 &  0.606 &  0.526 &  0.160 \T \B \\
(0.850)  & 41.80 &  0.263 &  0.255 &  0.430 &  0.087 &  0.086 &  0.063 &  0.025 \B \\
         & 42.30 &  0.028 &  0.050 &  0.074 &  0.004 &  0.004 &  0.002 &  0.001 \B \\
         & 42.80 &  0.002 &  0.006 &  0.009 &  0.000 &  0.000 &  0.000 &  0.000 \B \\
         & 43.30 &  0.000 &  0.000 &  0.000 &  0.000 &  0.000 &  0.000 &  0.000 \B \\
         & 0.005 &  0.408 &  0.597 &  0.714 &  0.290 &  0.288 &  0.239 &  0.084 \B \\
\hline
\end{tabular*}
\vspace{5pt}
\end{table*}

\clearpage

\begin{figure*}
\centering
\vspace{-1.0pt}
\begin{overpic}[width=0.78\columnwidth,angle=0]{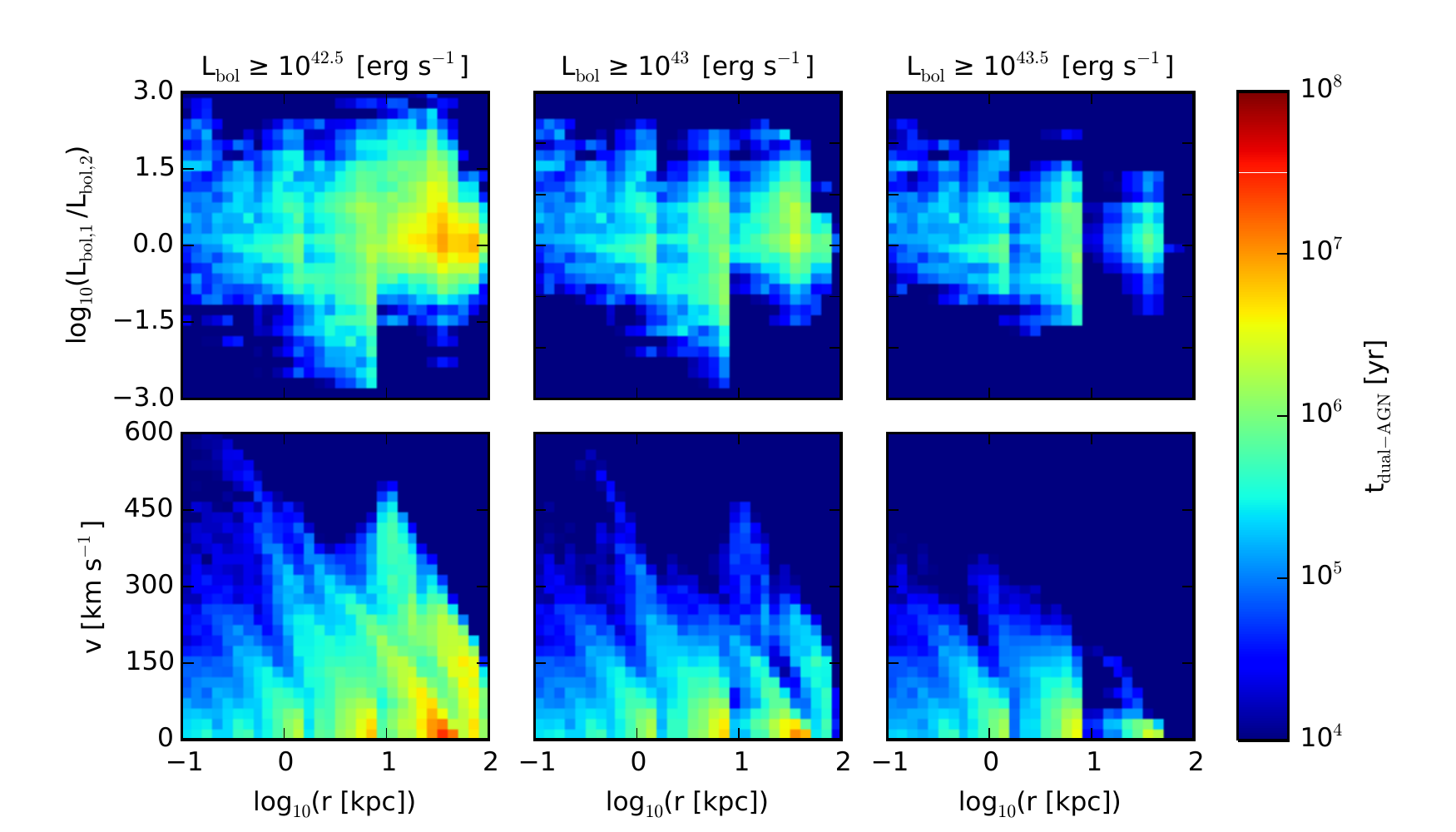}
\put (29.7,54.5) {\textcolor{black}{1:1 cop-pro-pro, 30\% gas, $\epsilon_{\rm f}=0.001$}}
\end{overpic}\\
\vspace{-4.0pt}
\begin{overpic}[width=0.78\columnwidth,angle=0]{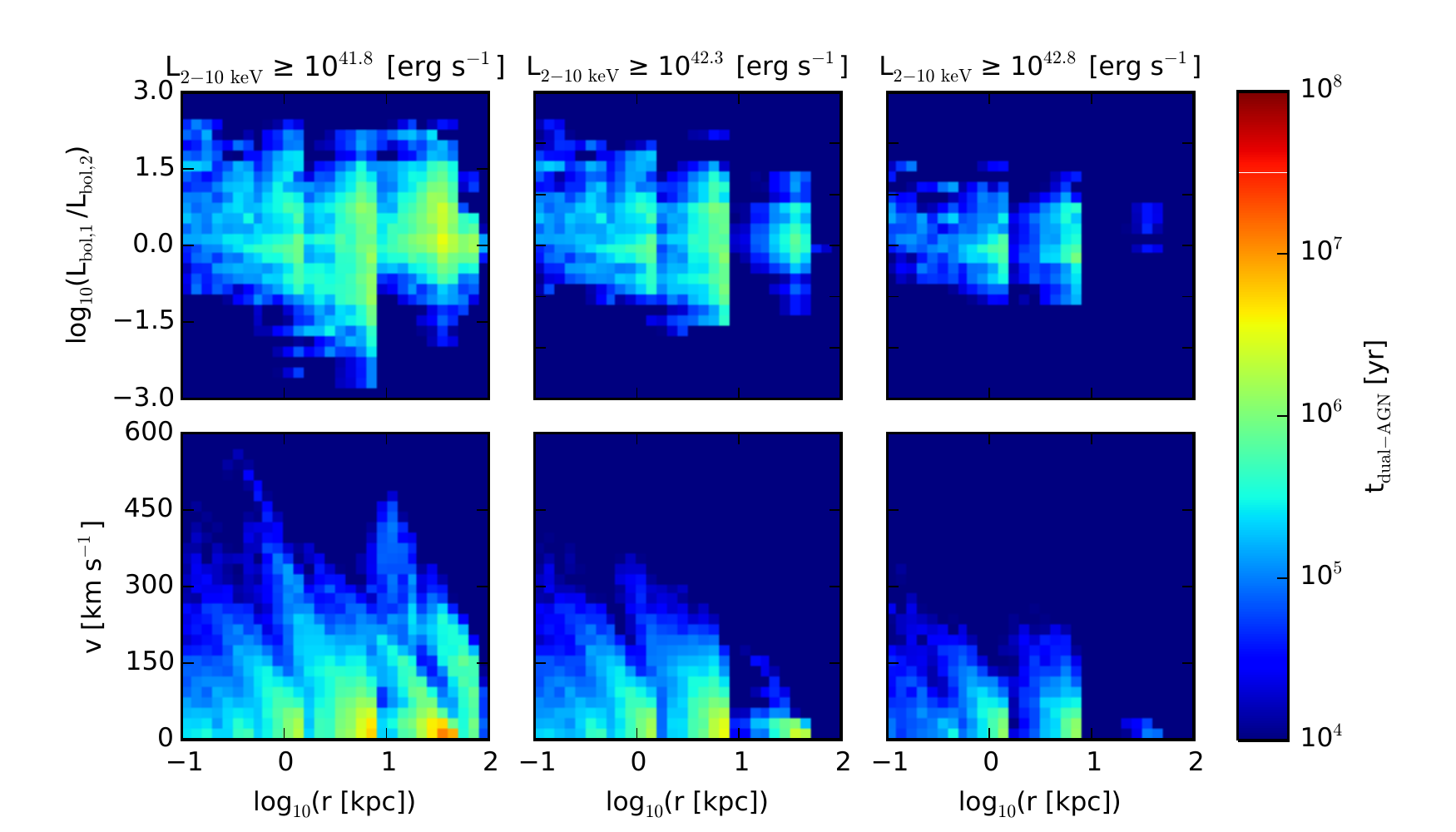}
\end{overpic}\\
\vspace{-4.0pt}
\begin{overpic}[width=0.78\columnwidth,angle=0]{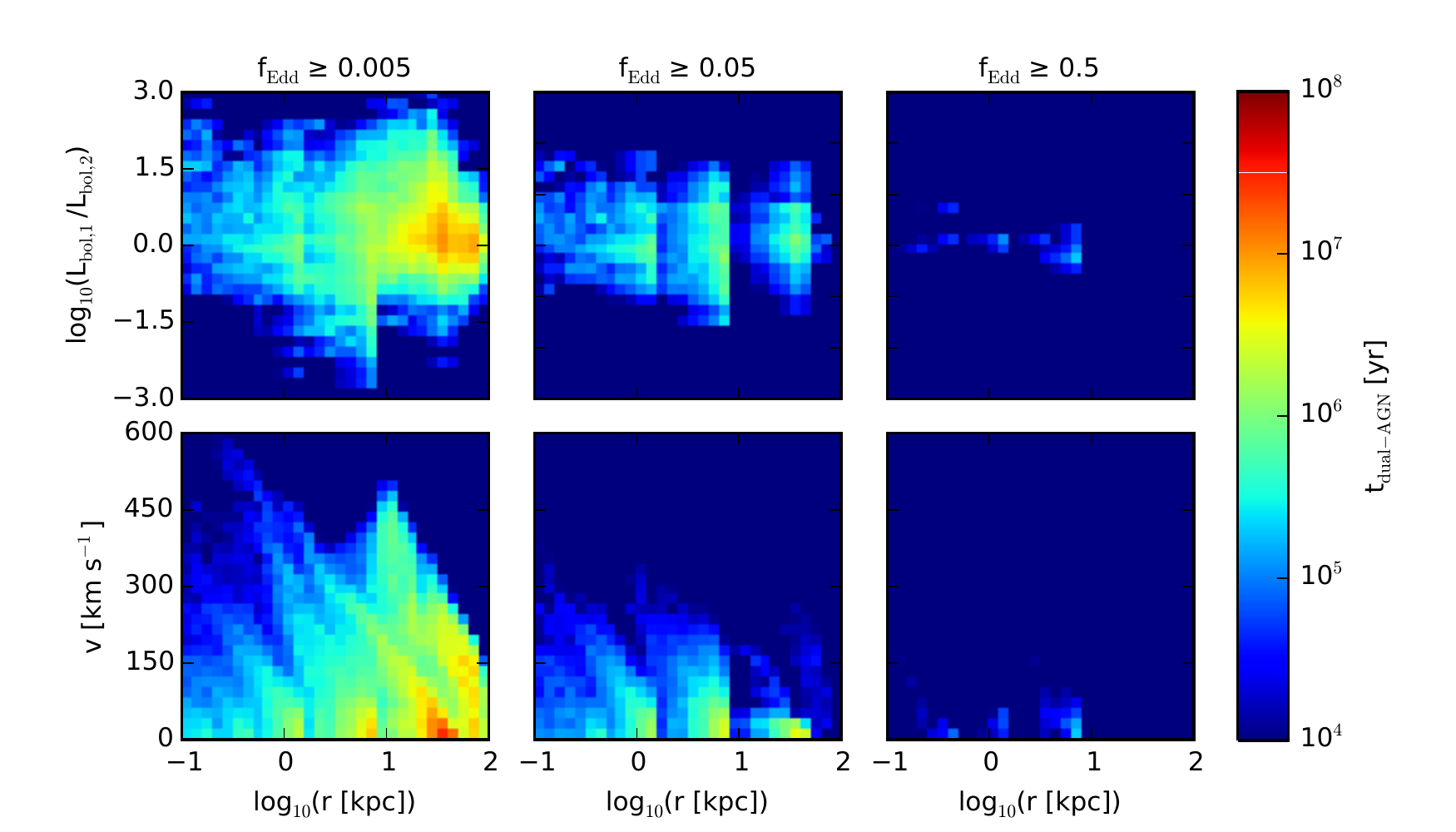}
\end{overpic}
\caption[]{Same as Fig.~\ref{dualagnpaper:fig:m2_hr_gf0_3_BHeff0_001_phi000000_dual_agn_map_proj_100los_stageall_loglog_no3b}, but for the 1:1 coplanar, prograde--prograde merger.}
\label{dualagnpaper:fig:m1_hr_gf0_3_BHeff0_001_phi000000_dual_agn_map_proj_100los_stageall_loglog_no3b}
\end{figure*}

\clearpage

\begin{figure*}
\centering
\vspace{-1.0pt}
\begin{overpic}[width=0.78\columnwidth,angle=0]{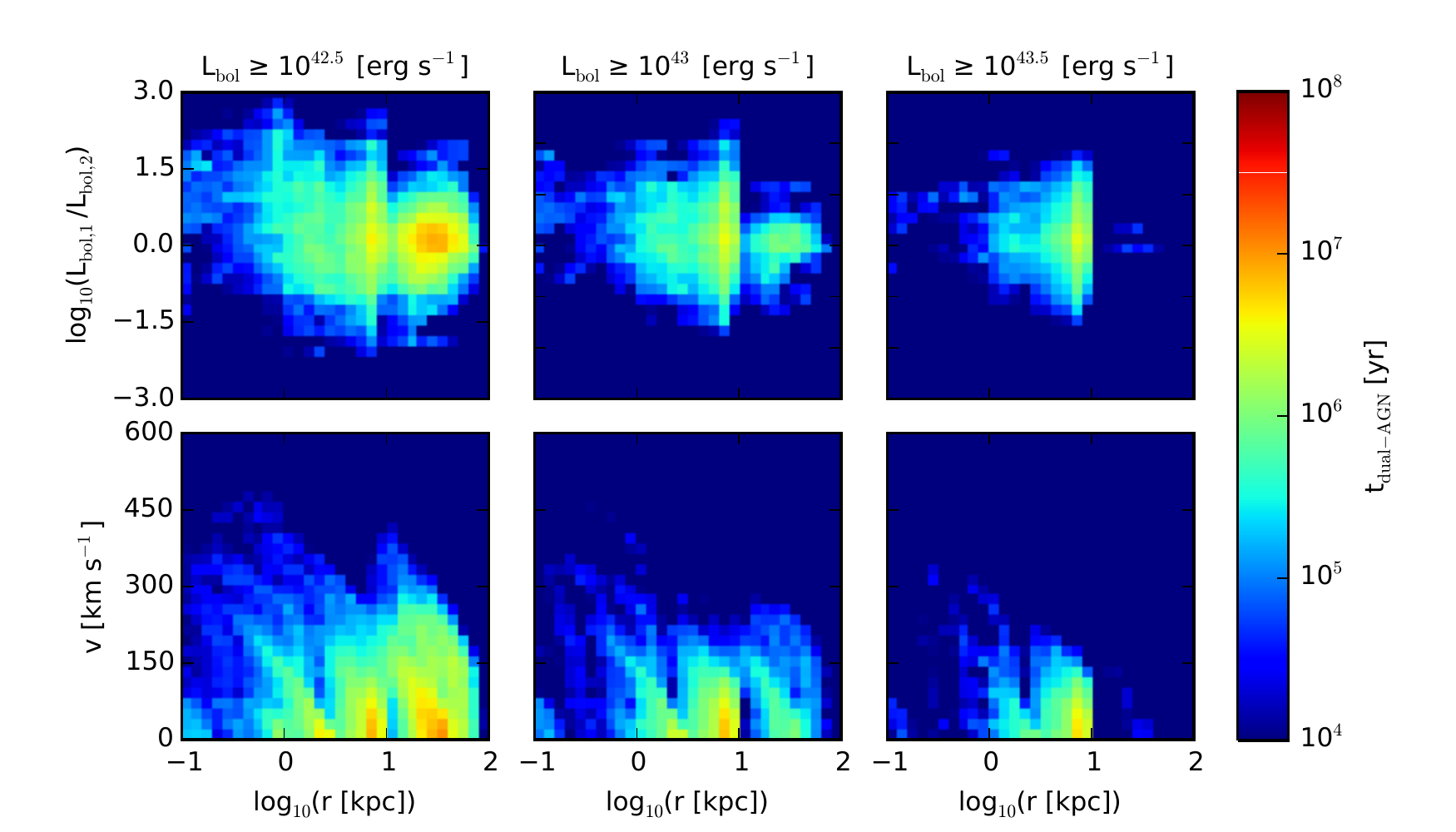}
\put (29.0,54.5) {\textcolor{black}{1:2 inclined-prim, 30\% gas, $\epsilon_{\rm f}=0.001$}}
\end{overpic}\\
\vspace{-4.0pt}
\begin{overpic}[width=0.78\columnwidth,angle=0]{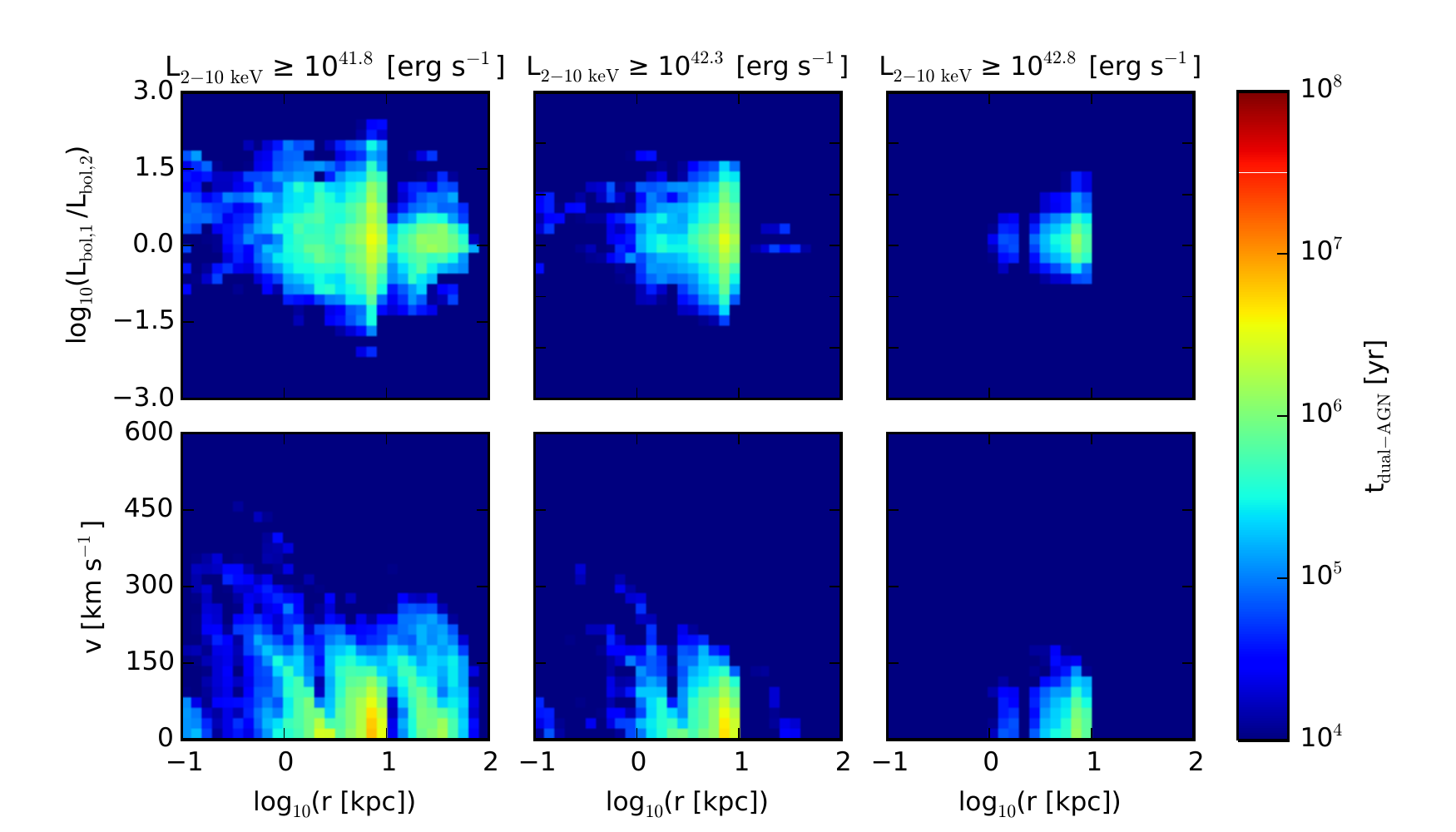}
\end{overpic}\\
\vspace{-4.0pt}
\begin{overpic}[width=0.78\columnwidth,angle=0]{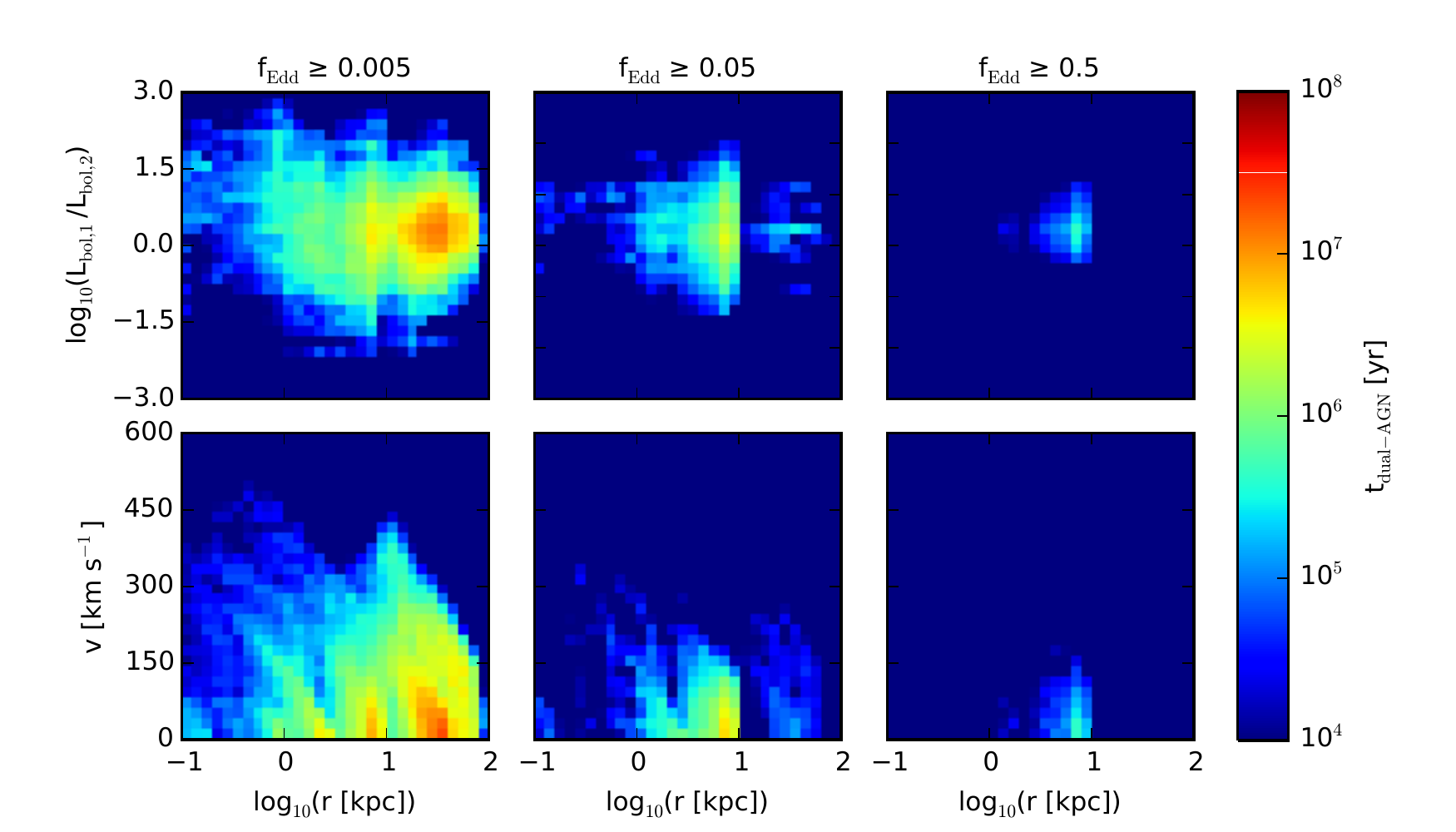}
\end{overpic}
\caption[]{Same as Fig.~\ref{dualagnpaper:fig:m2_hr_gf0_3_BHeff0_001_phi000000_dual_agn_map_proj_100los_stageall_loglog_no3b}, but for the 1:2 inclined-primary merger.}
\label{dualagnpaper:fig:m2_hr_gf0_3_BHeff0_001_phi045000_dual_agn_map_proj_100los_stageall_loglog_no3b}
\end{figure*}

\clearpage

\begin{figure*}
\centering
\vspace{-1.0pt}
\begin{overpic}[width=0.78\columnwidth,angle=0]{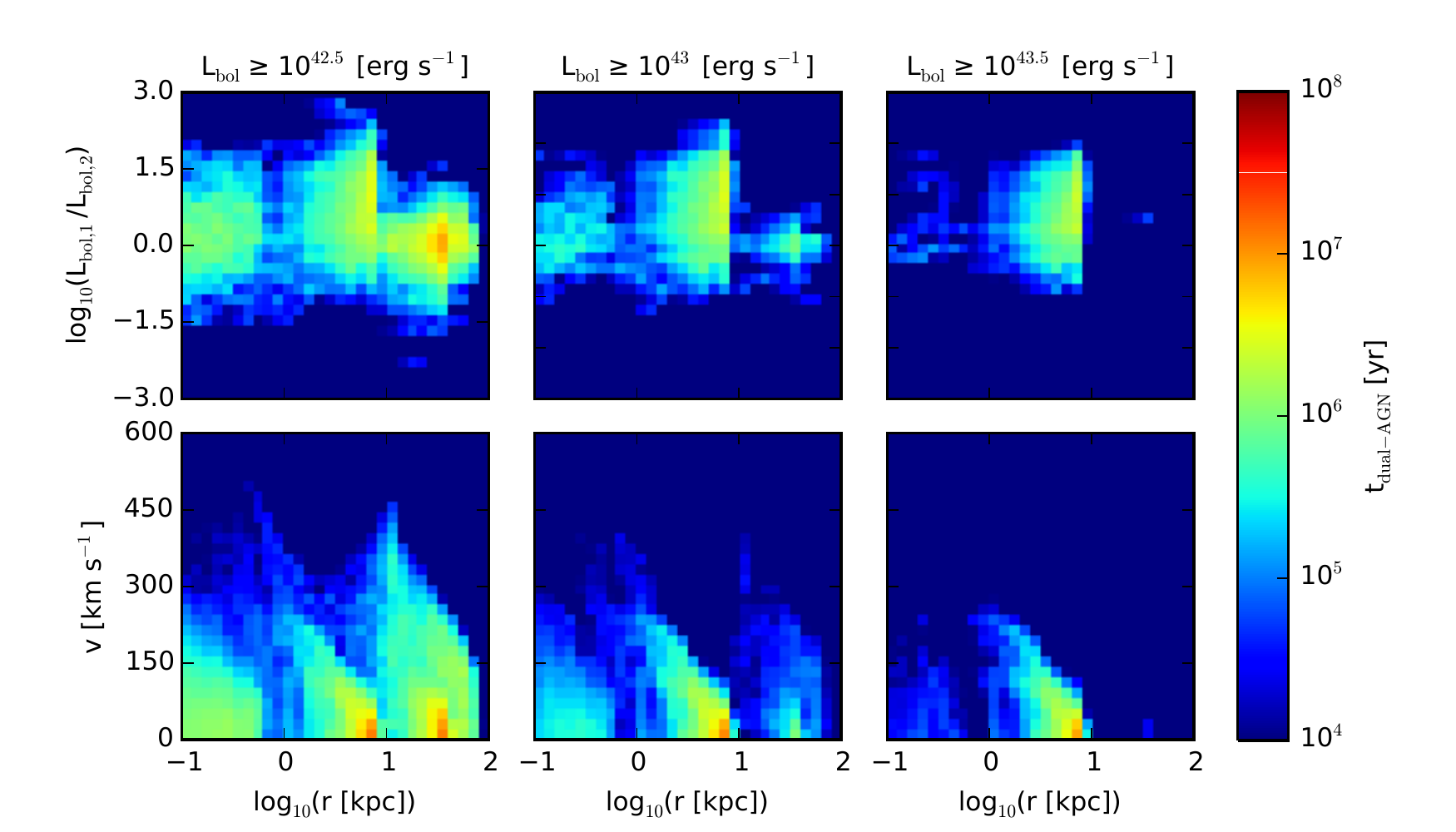}
\put (30.0,54.5) {\textcolor{black}{1:2 cop-ret-pro, 30\% gas, $\epsilon_{\rm f}=0.001$}}
\end{overpic}\\
\vspace{-4.0pt}
\begin{overpic}[width=0.78\columnwidth,angle=0]{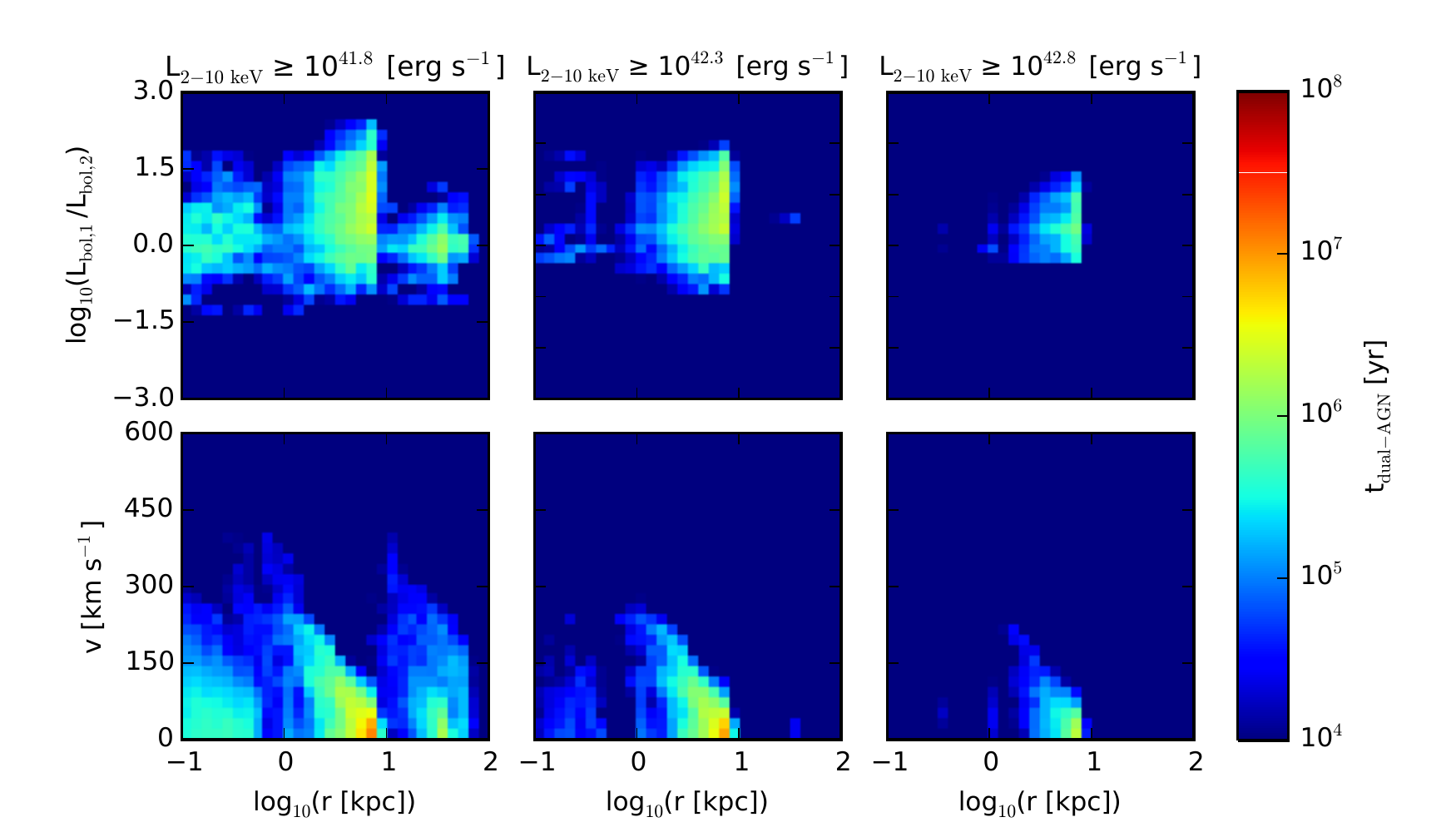}
\end{overpic}\\
\vspace{-4.0pt}
\begin{overpic}[width=0.78\columnwidth,angle=0]{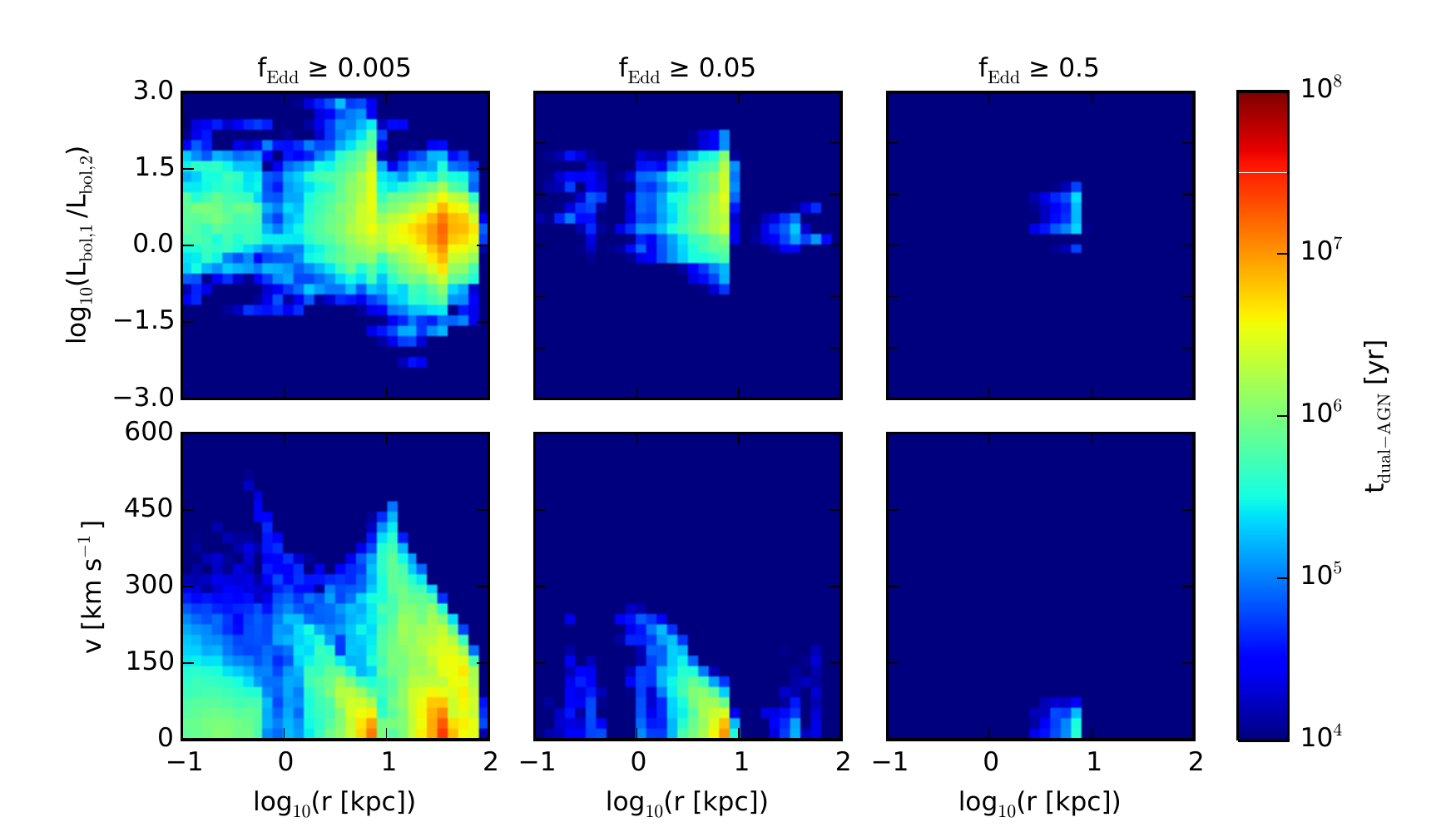}
\end{overpic}
\caption[]{Same as Fig.~\ref{dualagnpaper:fig:m2_hr_gf0_3_BHeff0_001_phi000000_dual_agn_map_proj_100los_stageall_loglog_no3b}, but for the 1:2 coplanar, retrograde--prograde merger.}
\label{dualagnpaper:fig:m2_hr_gf0_3_BHeff0_001_phi180000_dual_agn_map_proj_100los_stageall_loglog_no3b}
\end{figure*}

\clearpage

\begin{figure*}
\centering
\vspace{-1.0pt}
\begin{overpic}[width=0.78\columnwidth,angle=0]{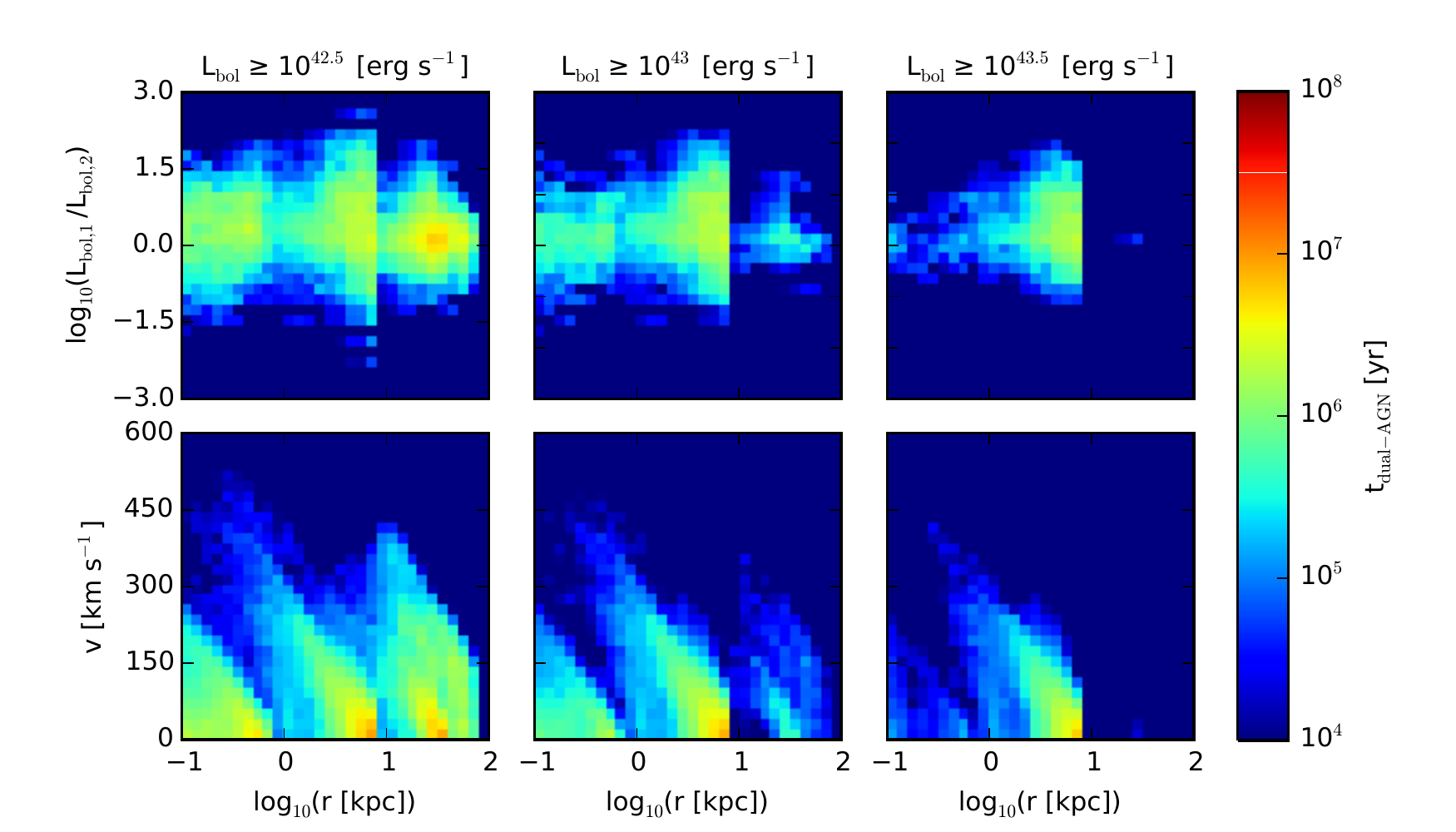}
\put (30.0,54.5) {\textcolor{black}{1:2 cop-pro-ret, 30\% gas, $\epsilon_{\rm f}=0.001$}}
\end{overpic}\\
\vspace{-4.0pt}
\begin{overpic}[width=0.78\columnwidth,angle=0]{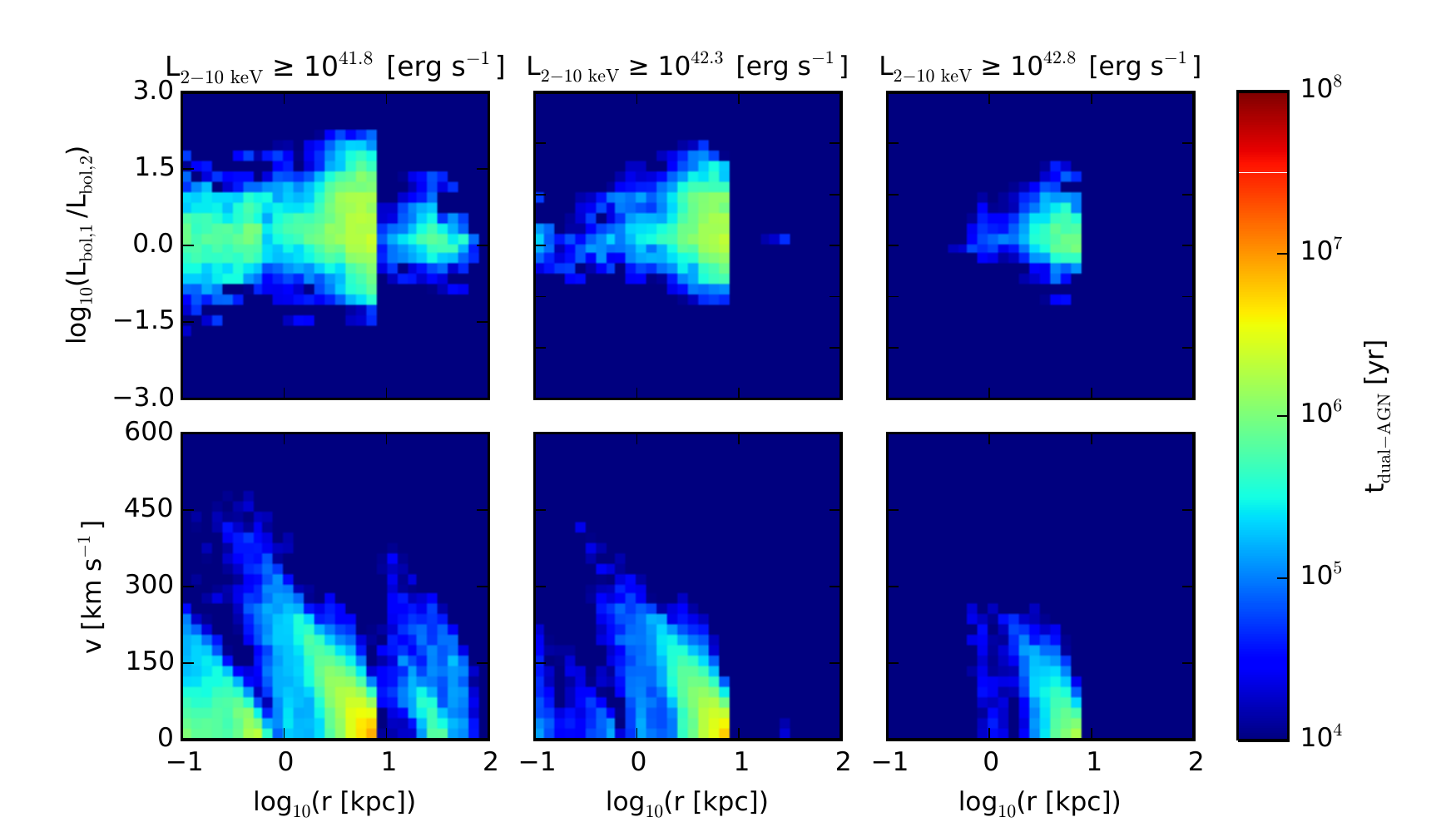}
\end{overpic}\\
\vspace{-4.0pt}
\begin{overpic}[width=0.78\columnwidth,angle=0]{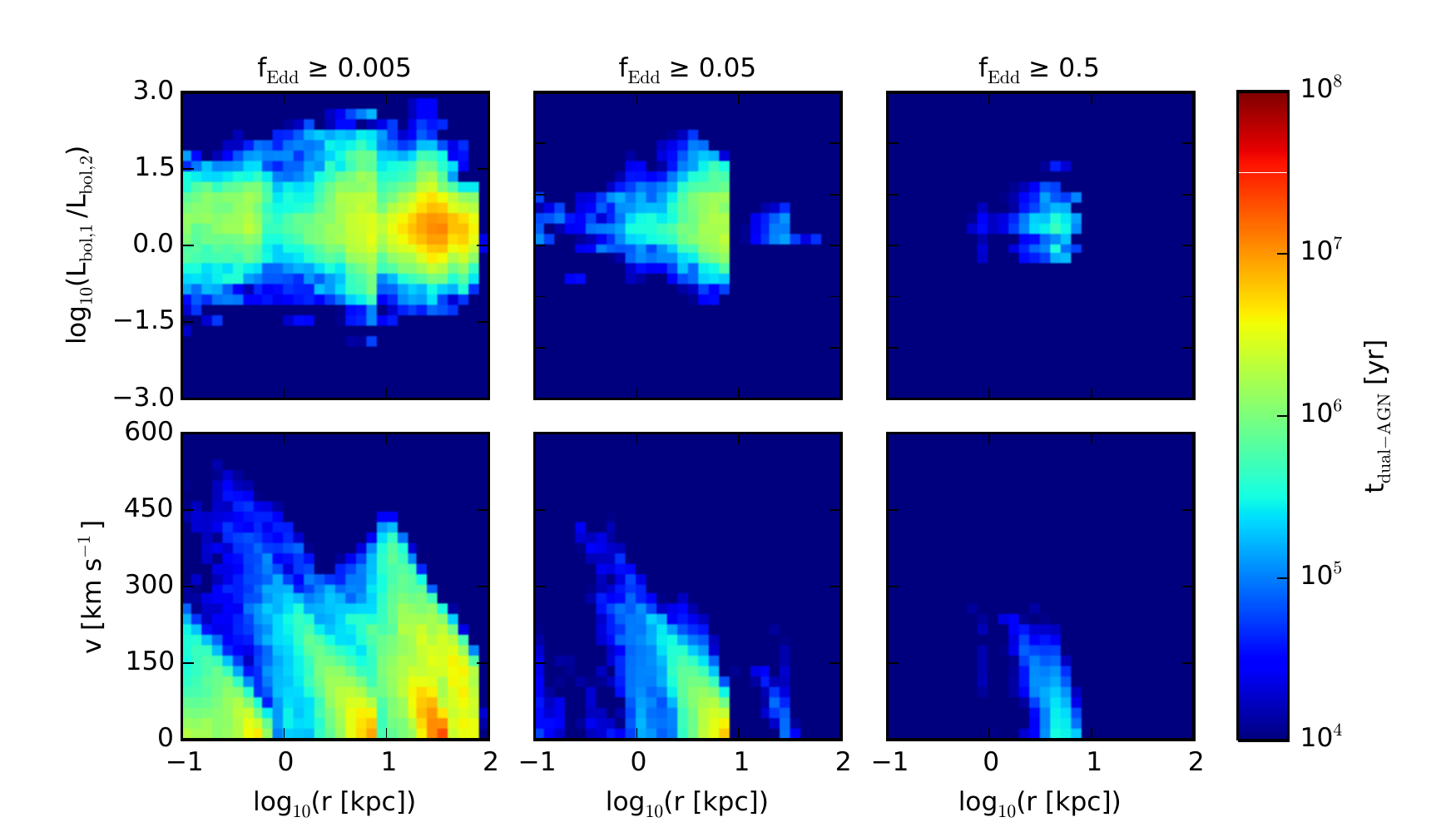}
\end{overpic}
\caption[]{Same as Fig.~\ref{dualagnpaper:fig:m2_hr_gf0_3_BHeff0_001_phi000000_dual_agn_map_proj_100los_stageall_loglog_no3b}, but for the 1:2 coplanar, prograde--retrograde merger.}
\label{dualagnpaper:fig:m2_hr_gf0_3_BHeff0_001_phi000180_dual_agn_map_proj_100los_stageall_loglog_no3b}
\end{figure*}

\clearpage

\begin{figure*}
\centering
\vspace{-1.0pt}
\begin{overpic}[width=0.78\columnwidth,angle=0]{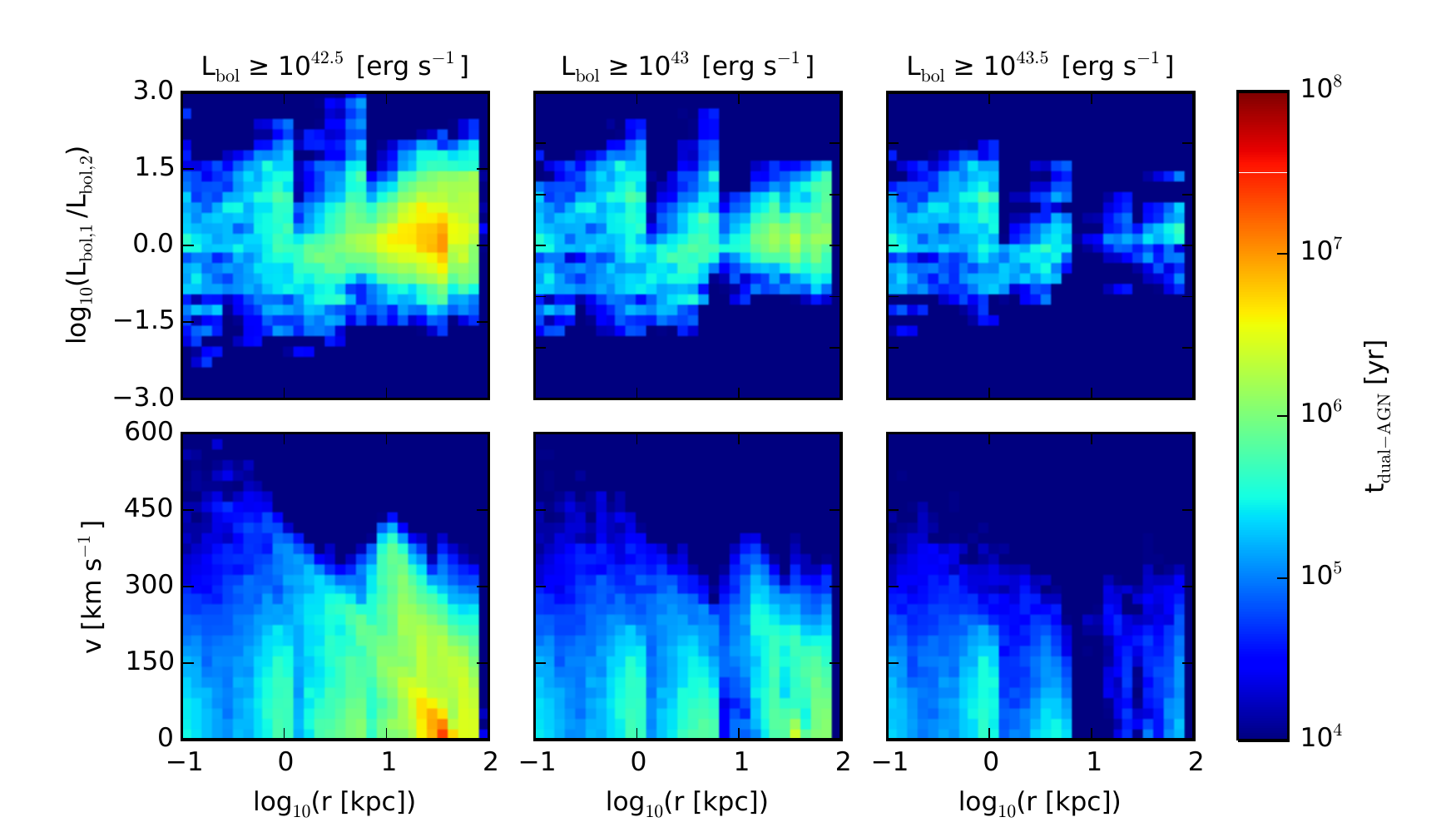}
\put (30.0,54.5) {\textcolor{black}{1:2 cop-pro-pro, 60\% gas, $\epsilon_{\rm f}=0.001$}}
\end{overpic}\\
\vspace{-4.0pt}
\begin{overpic}[width=0.78\columnwidth,angle=0]{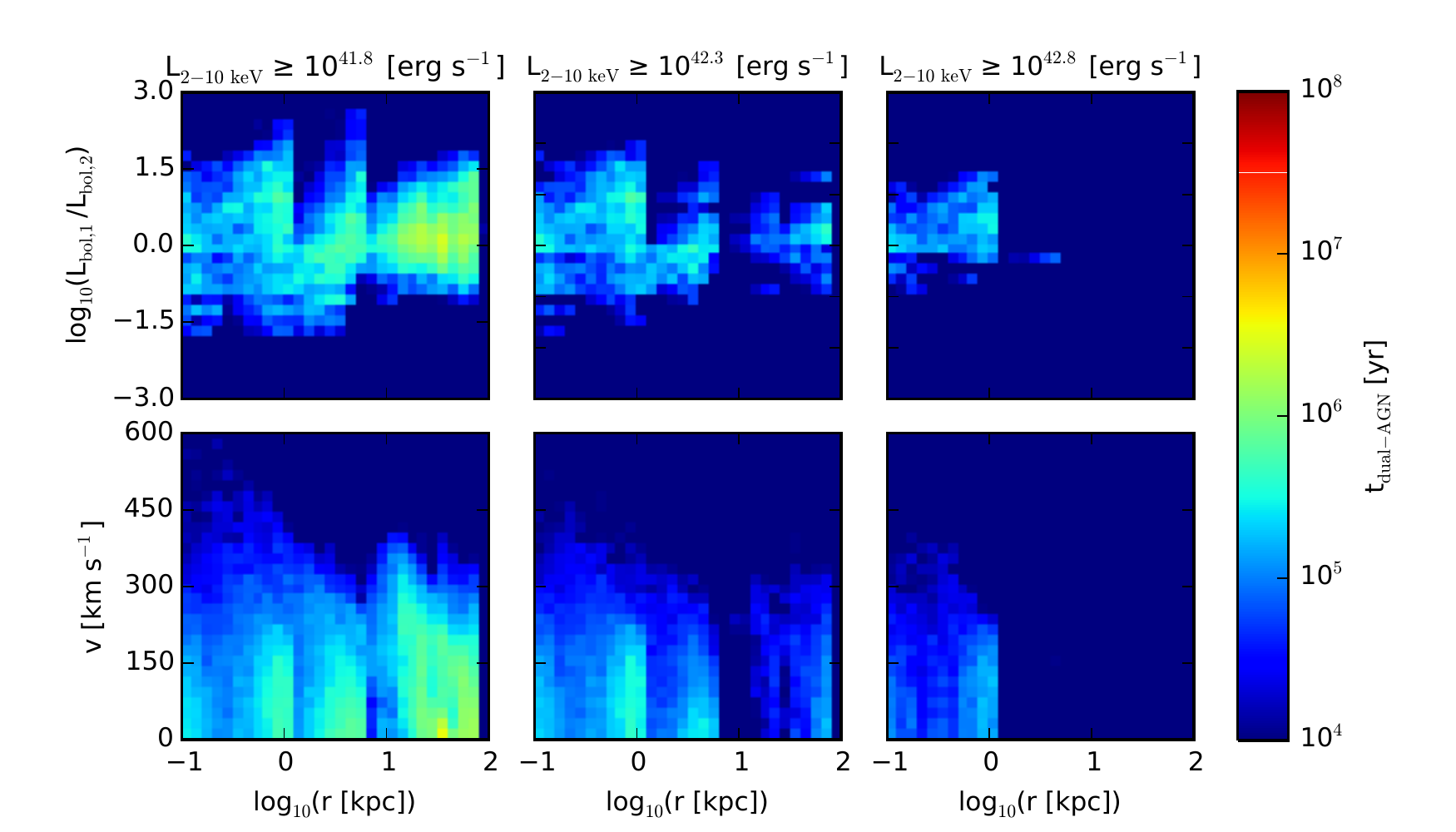}
\end{overpic}\\
\vspace{-4.0pt}
\begin{overpic}[width=0.78\columnwidth,angle=0]{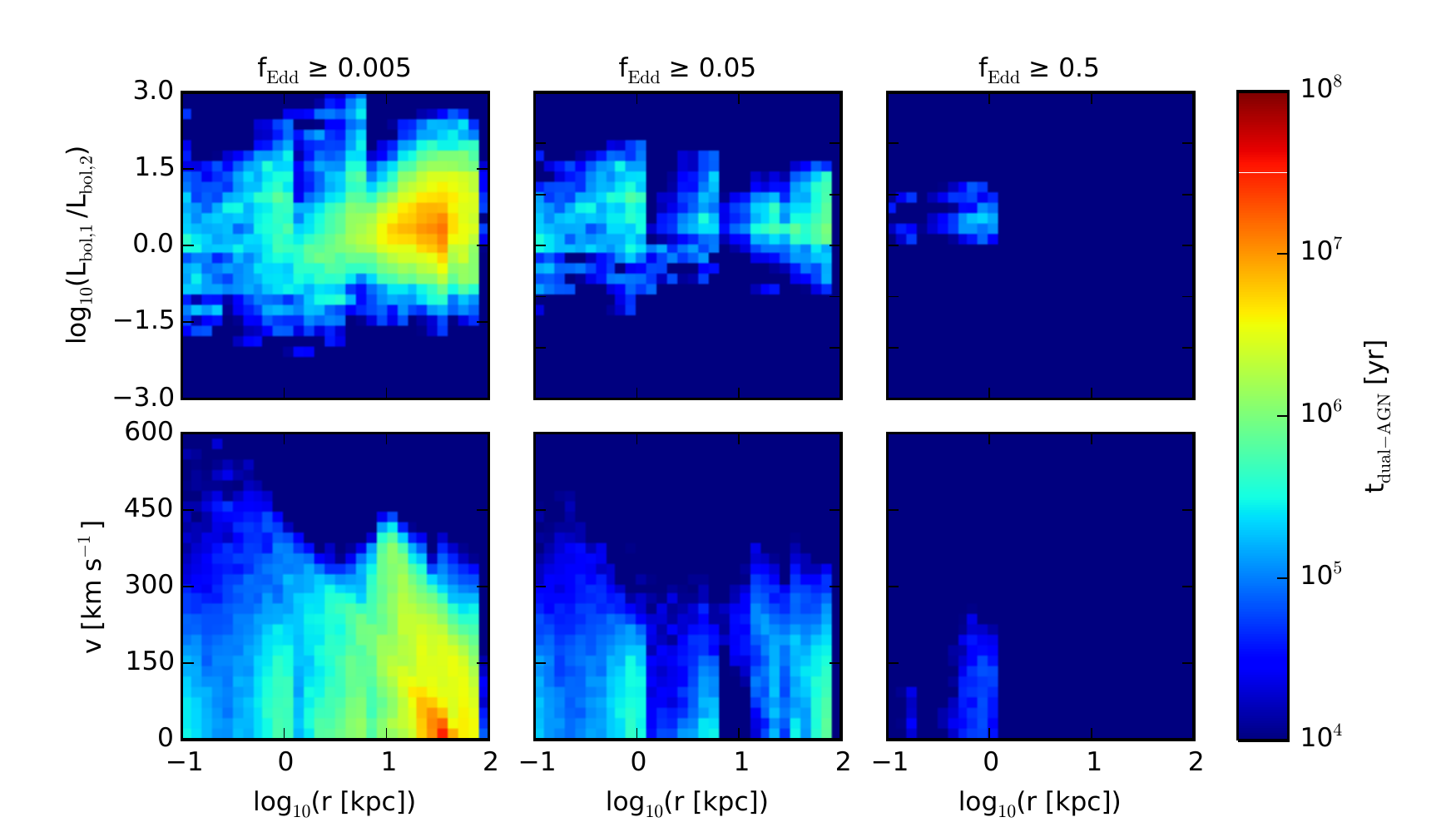}
\end{overpic}
\caption[]{Same as Fig.~\ref{dualagnpaper:fig:m2_hr_gf0_3_BHeff0_001_phi000000_dual_agn_map_proj_100los_stageall_loglog_no3b}, but for the 1:2 coplanar, prograde--prograde merger with 60 per cent gas fraction, standard BH mass, and standard BH feedback efficiency.}
\label{dualagnpaper:fig:m2_hr_gf0_6_BHeff0_001_phi000000_dual_agn_map_proj_100los_stageall_loglog_no3b}
\end{figure*}

\clearpage

\begin{figure*}
\centering
\vspace{-1.0pt}
\begin{overpic}[width=0.78\columnwidth,angle=0]{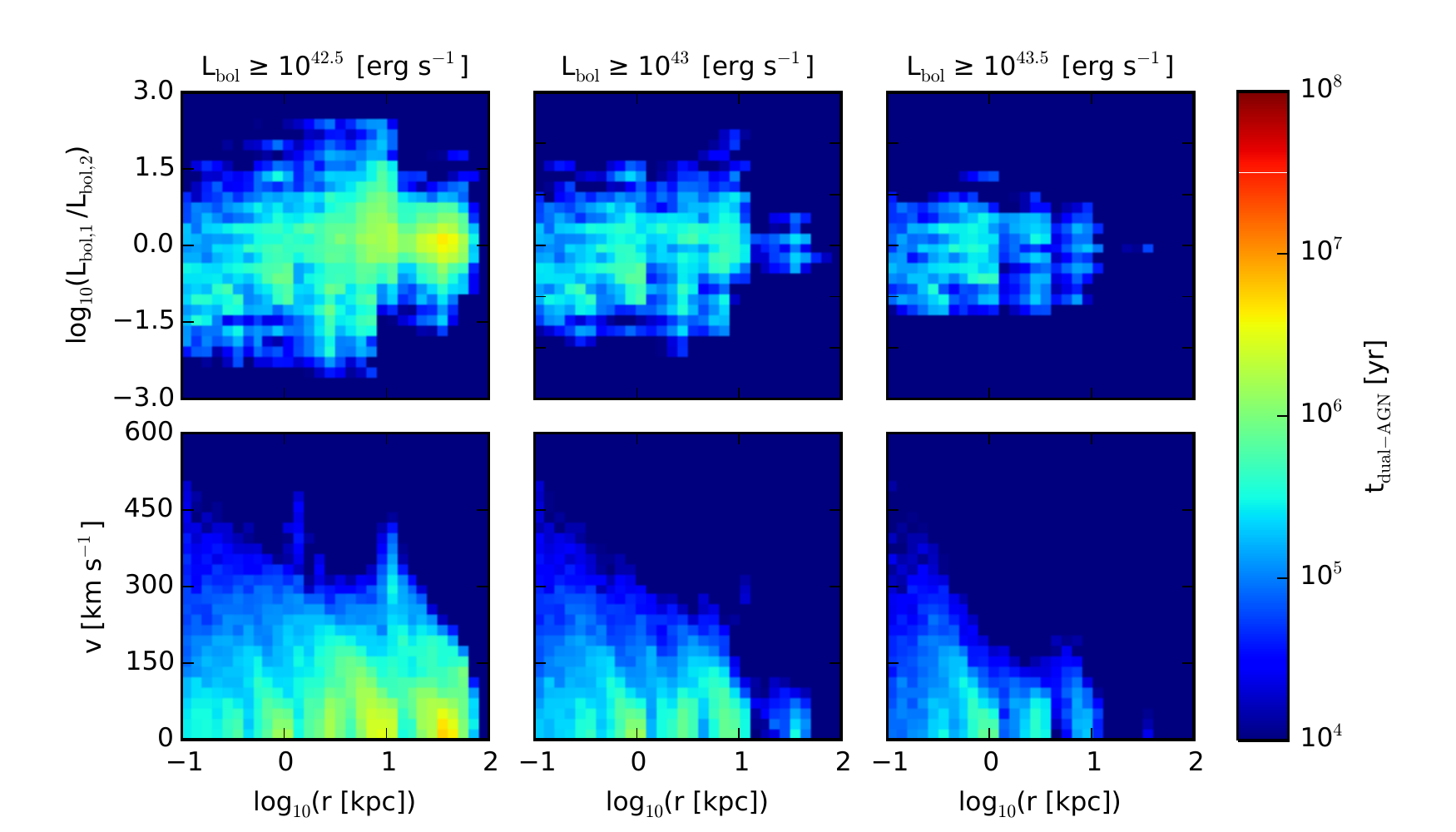}
\put (30.0,54.5) {\textcolor{black}{1:4 cop-pro-pro, 30\% gas, $\epsilon_{\rm f}=0.001$}}
\end{overpic}\\
\vspace{-4.0pt}
\begin{overpic}[width=0.78\columnwidth,angle=0]{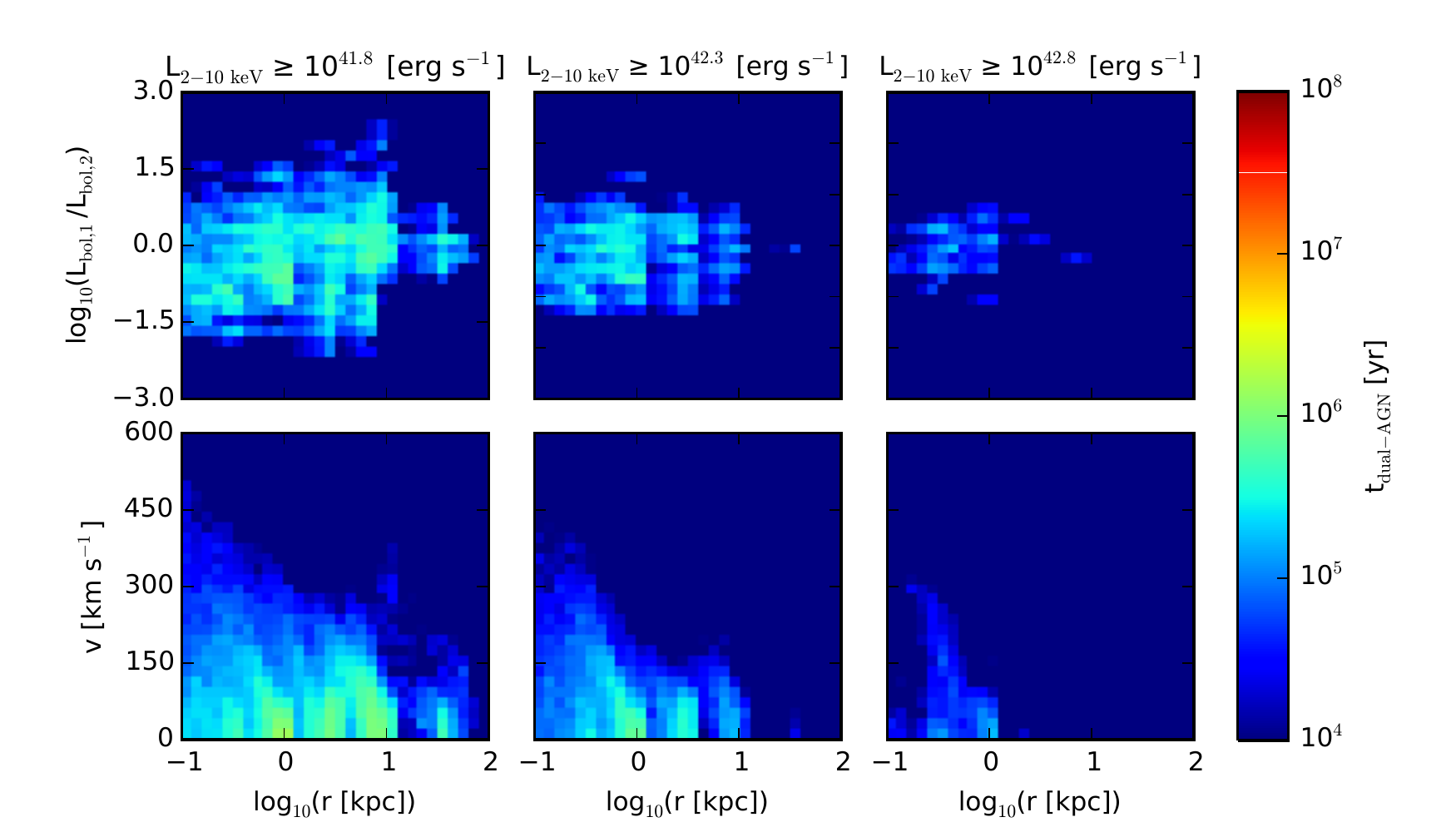}
\end{overpic}\\
\vspace{-4.0pt}
\begin{overpic}[width=0.78\columnwidth,angle=0]{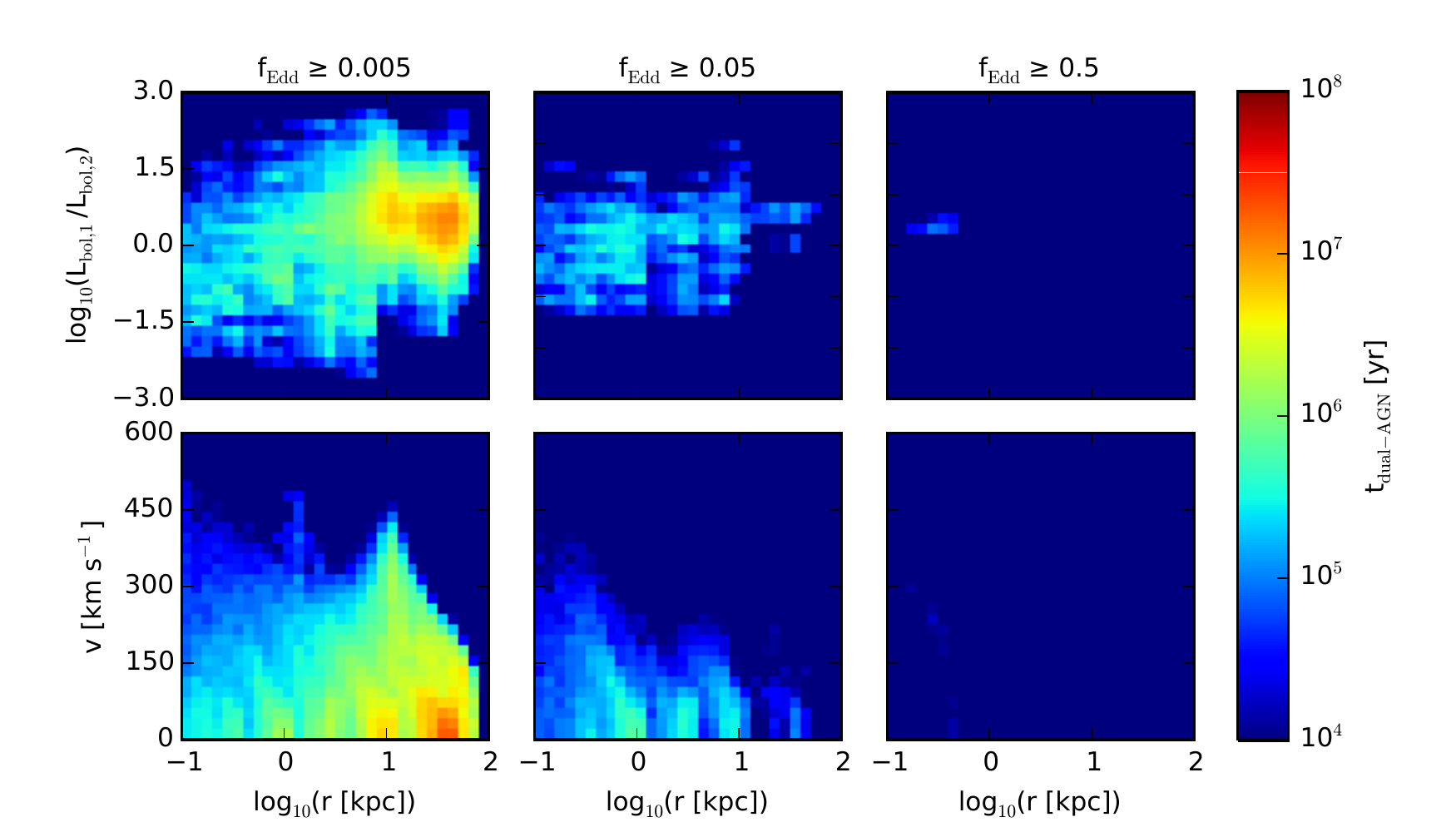}
\end{overpic}
\caption[]{Same as Fig.~\ref{dualagnpaper:fig:m2_hr_gf0_3_BHeff0_001_phi000000_dual_agn_map_proj_100los_stageall_loglog_no3b}, but for the 1:4 coplanar, prograde--prograde merger.}
\label{dualagnpaper:fig:m4_hr_gf0_3_BHeff0_001_phi000000_dual_agn_map_proj_100los_stageall_loglog_no3b}
\end{figure*}

\clearpage

\begin{figure*}
\centering
\vspace{-1.0pt}
\begin{overpic}[width=0.78\columnwidth,angle=0]{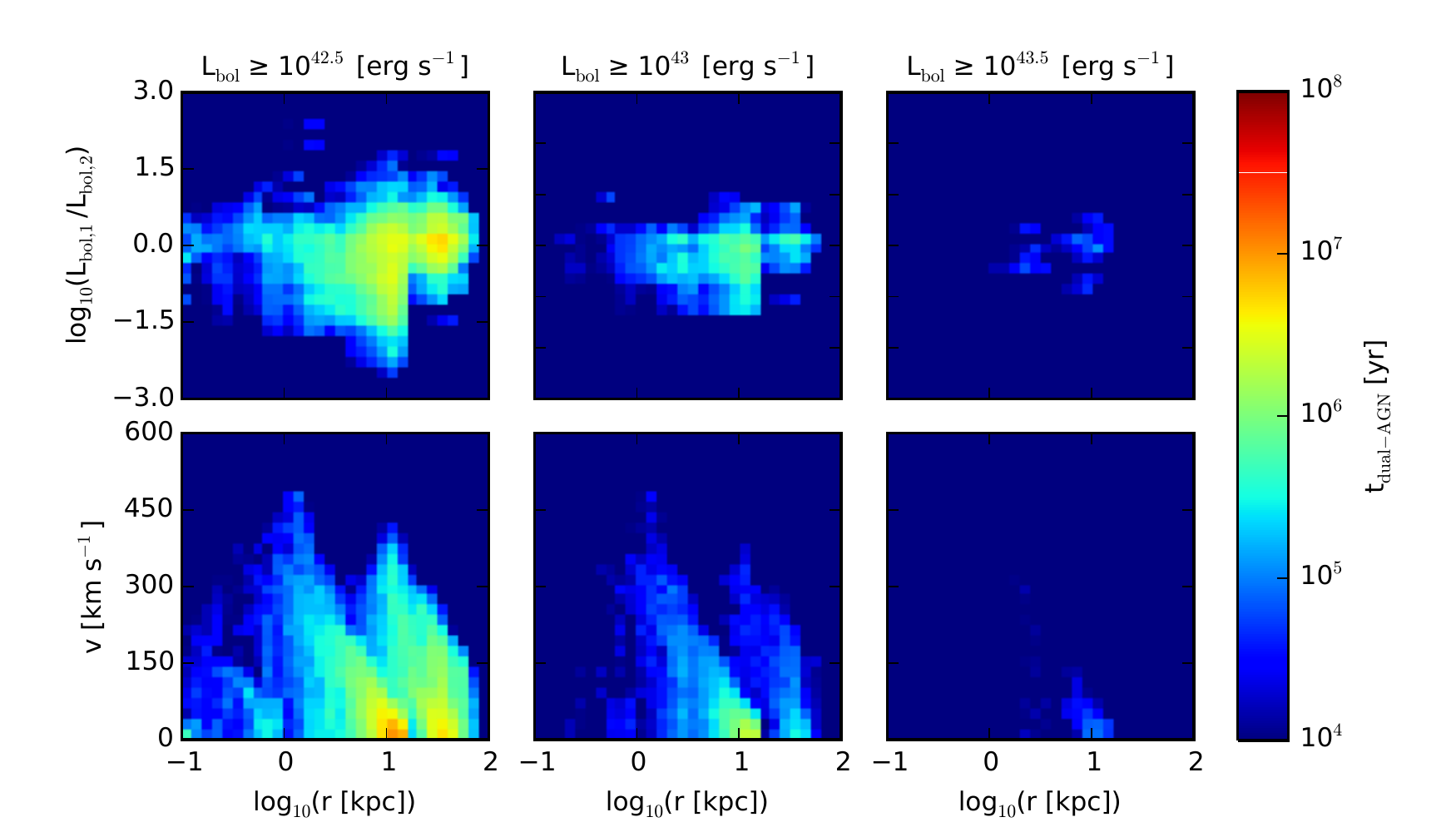}
\put (29.0,54.5) {\textcolor{black}{1:4 inclined-prim, 30\% gas, $\epsilon_{\rm f}=0.001$}}
\end{overpic}\\
\vspace{-2.0pt}
\begin{overpic}[width=0.55\columnwidth,angle=0]{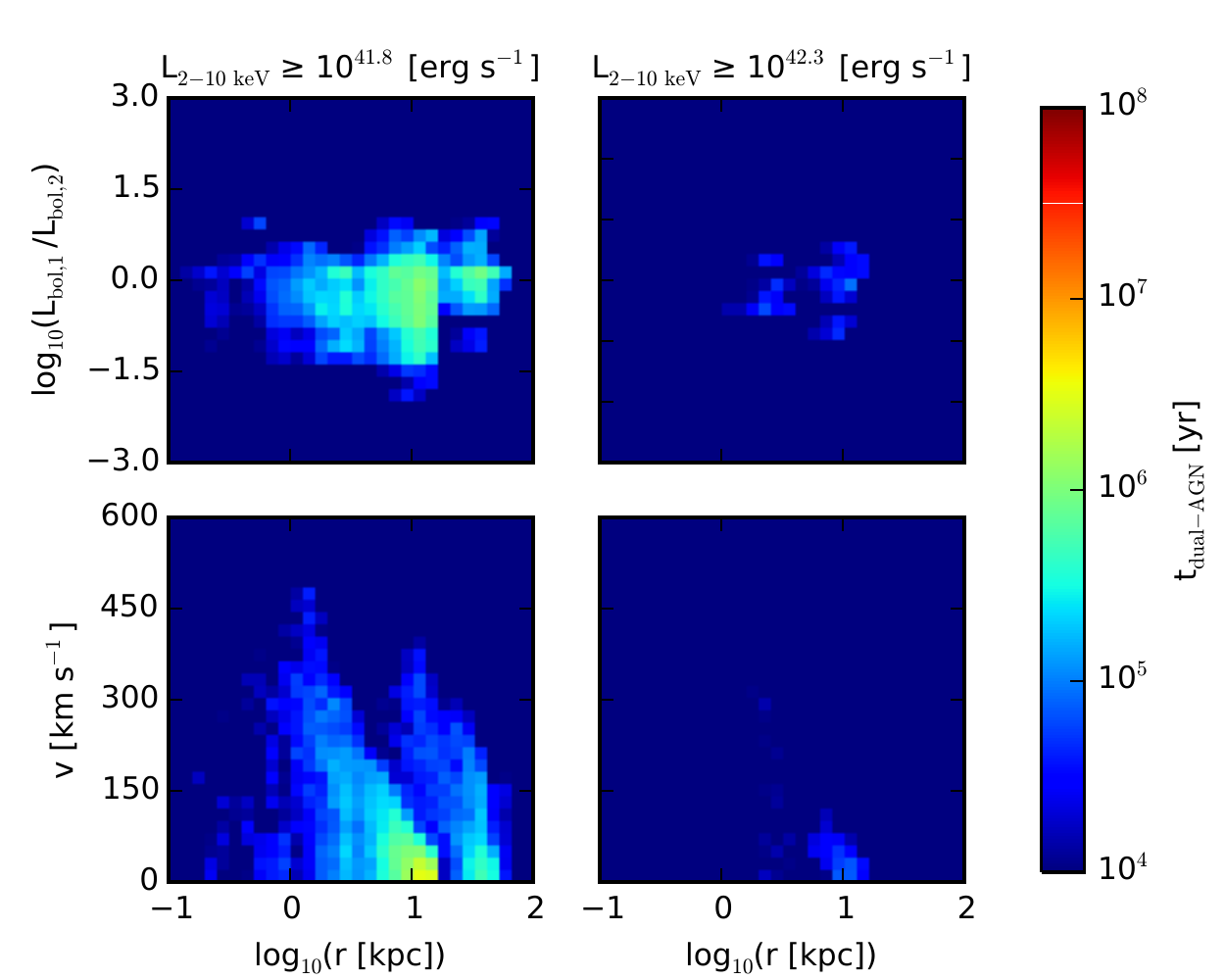}
\end{overpic}\\
\vspace{1.0pt}
\begin{overpic}[width=0.55\columnwidth,angle=0]{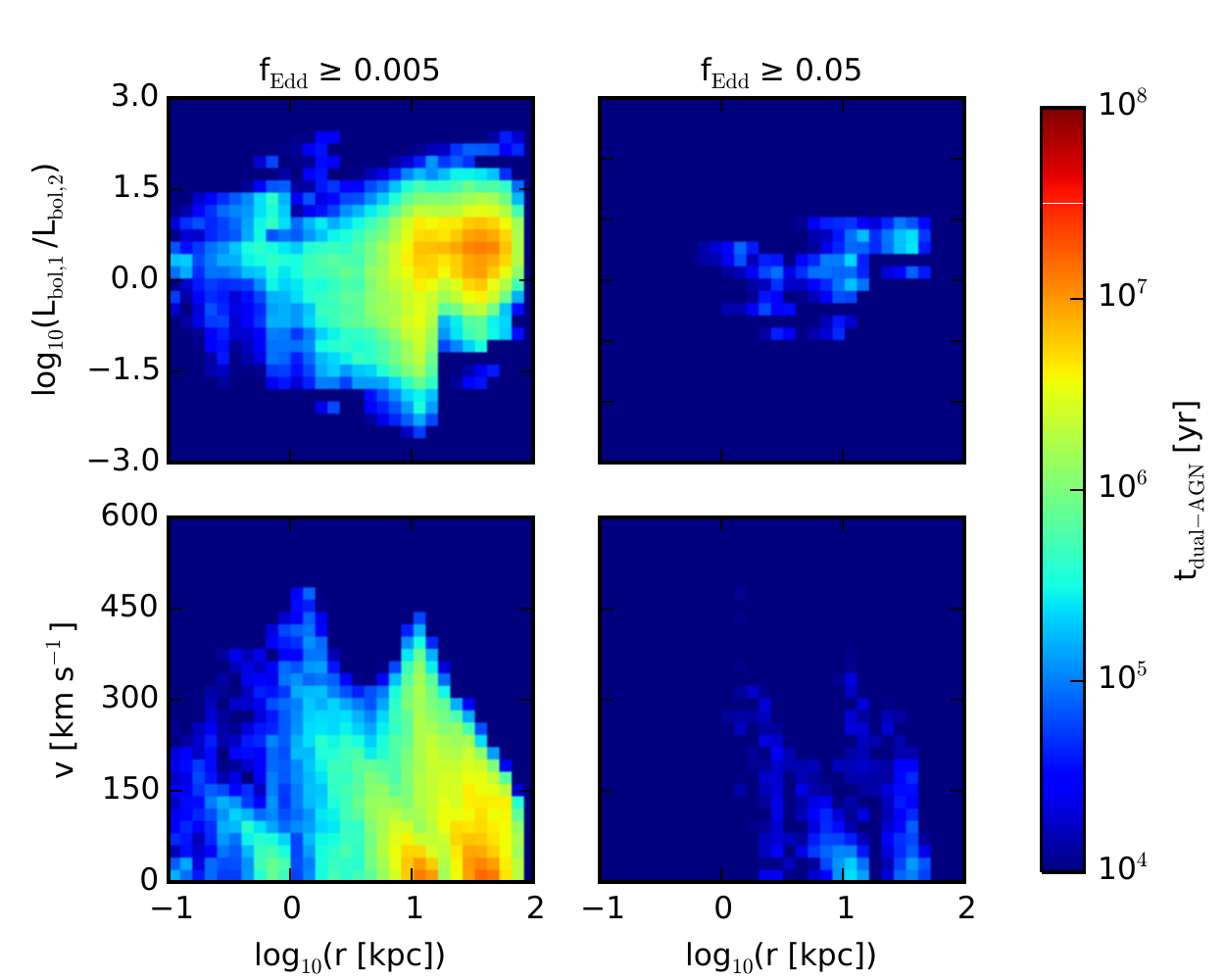}
\end{overpic}
\caption[]{Same as Fig.~\ref{dualagnpaper:fig:m2_hr_gf0_3_BHeff0_001_phi000000_dual_agn_map_proj_100los_stageall_loglog_no3b}, but for the 1:4 inclined-primary merger.}
\label{dualagnpaper:fig:m4_hr_gf0_3_BHeff0_001_phi045000_dual_agn_map_proj_100los_stageall_loglog_no3b}
\end{figure*}

\clearpage

\begin{figure*}
\centering
\vspace{-1.0pt}
\begin{overpic}[width=0.78\columnwidth,angle=0]{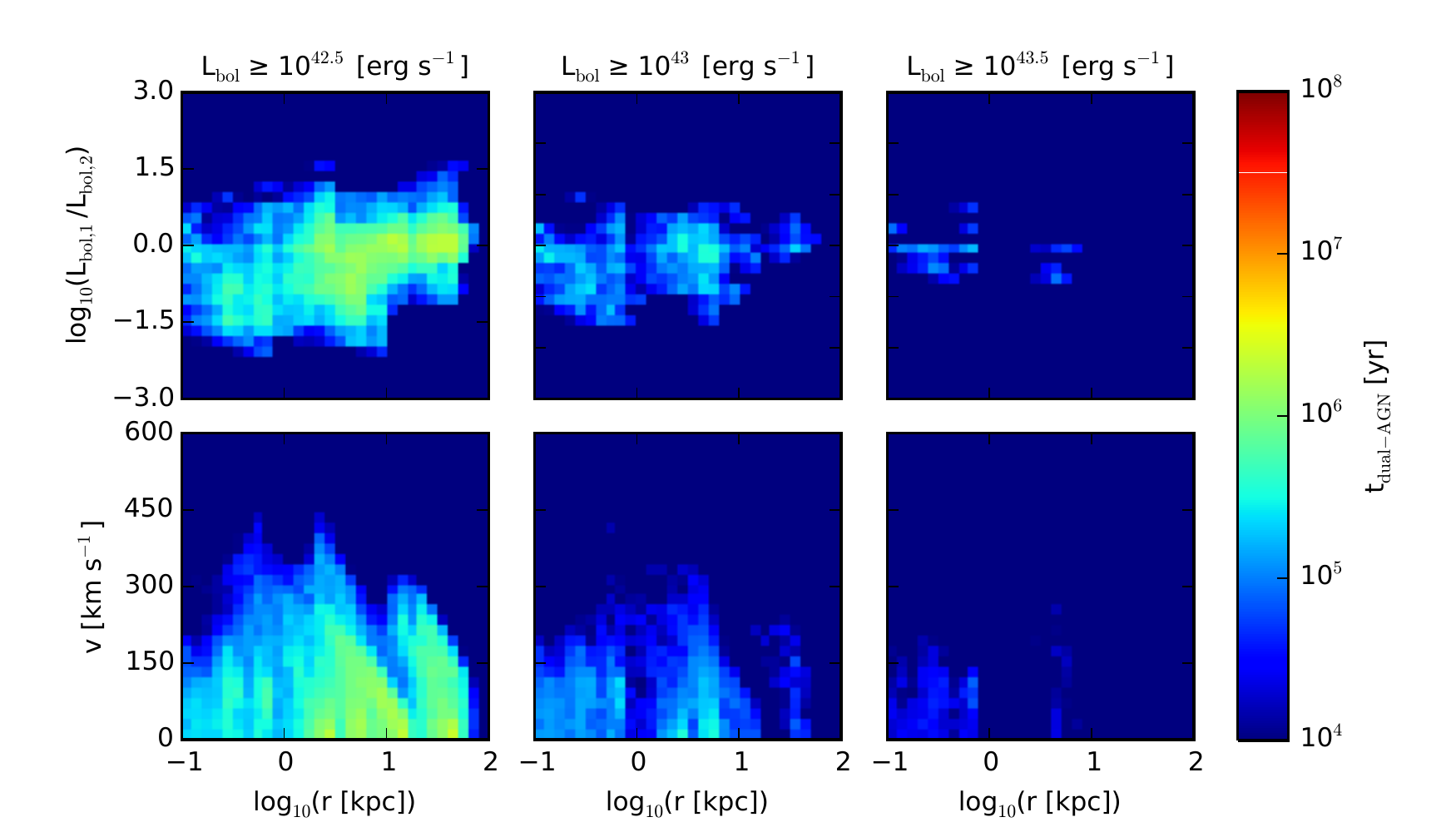}
\put (30.0,54.5) {\textcolor{black}{1:6 cop-pro-pro, 30\% gas, $\epsilon_{\rm f}=0.001$}}
\end{overpic}\\
\vspace{-2.0pt}
\begin{overpic}[width=0.55\columnwidth,angle=0]{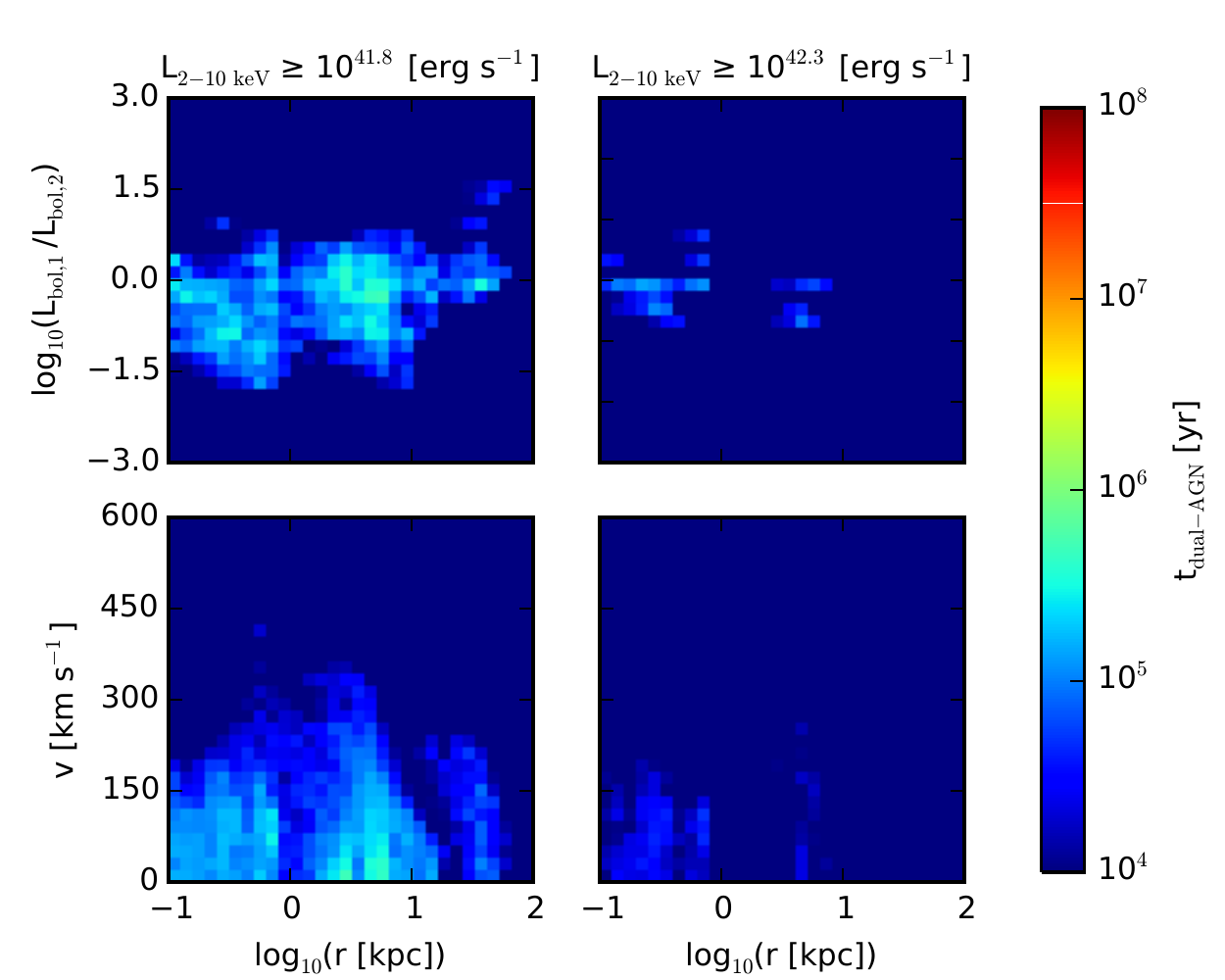}
\end{overpic}\\
\vspace{1.0pt}
\begin{overpic}[width=0.55\columnwidth,angle=0]{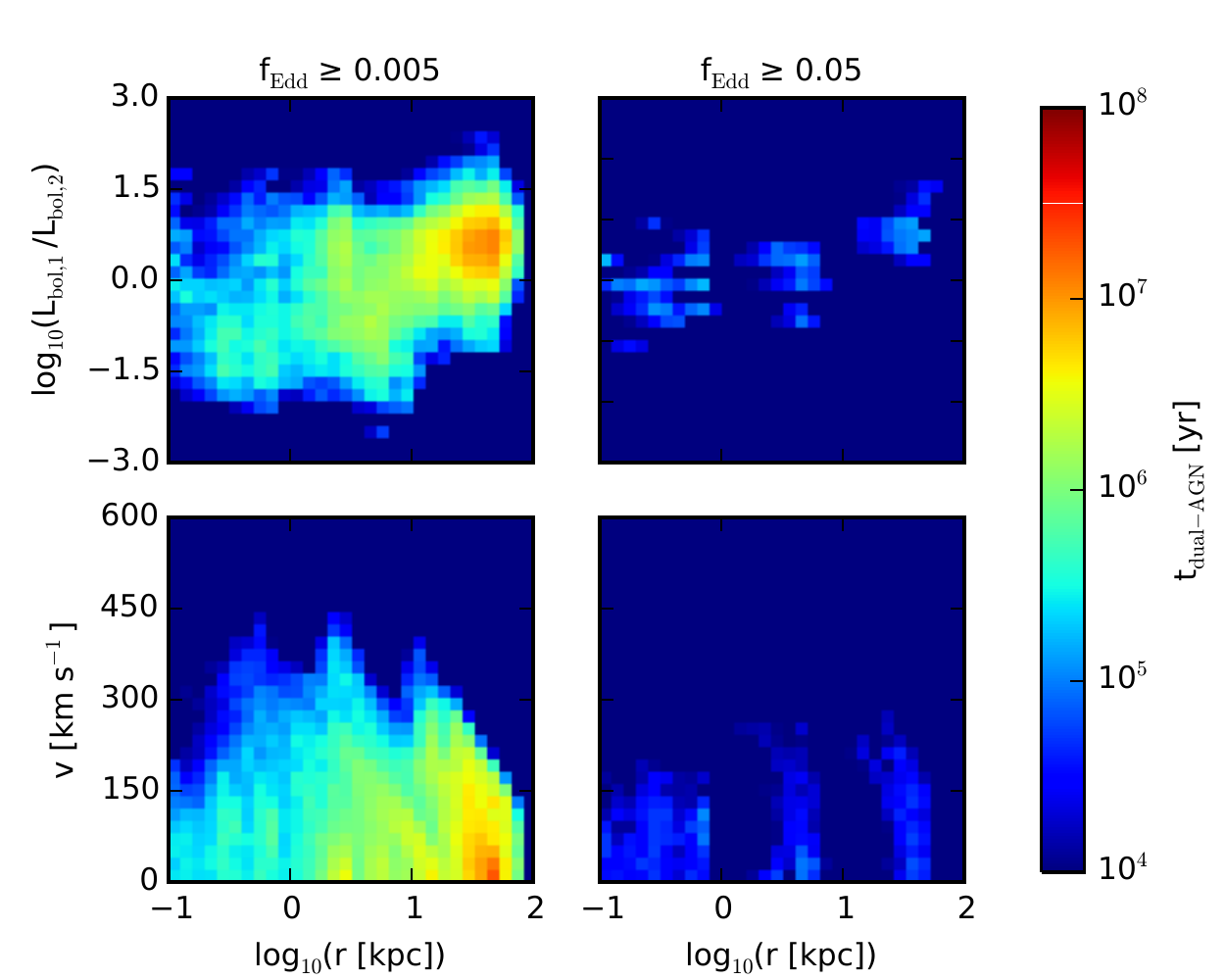}
\end{overpic}
\caption[]{Same as Fig.~\ref{dualagnpaper:fig:m2_hr_gf0_3_BHeff0_001_phi000000_dual_agn_map_proj_100los_stageall_loglog_no3b}, but for the 1:6 coplanar, prograde--prograde merger.}
\label{dualagnpaper:fig:m6_hr_gf0_3_BHeff0_001_phi000000_dual_agn_map_proj_100los_stageall_loglog_no3b}
\end{figure*}

\clearpage

\begin{figure*}
\centering
\vspace{-1.0pt}
\begin{overpic}[width=0.78\columnwidth,angle=0]{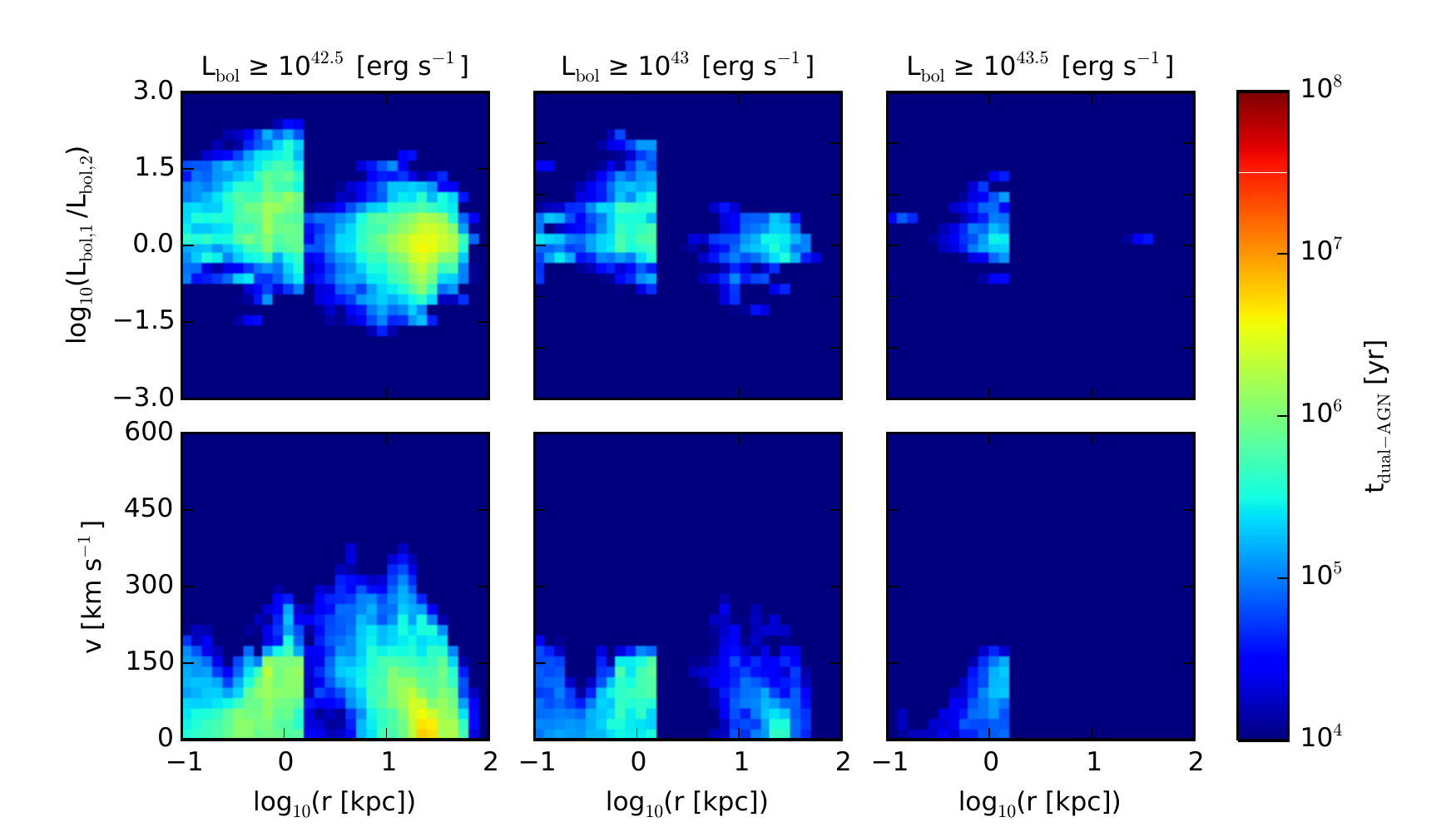}
\put (29.0,54.5) {\textcolor{black}{1:10 cop-pro-pro, 30\% gas, $\epsilon_{\rm f}=0.001$}}
\end{overpic}\\
\vspace{-2.0pt}
\begin{overpic}[width=0.55\columnwidth,angle=0]{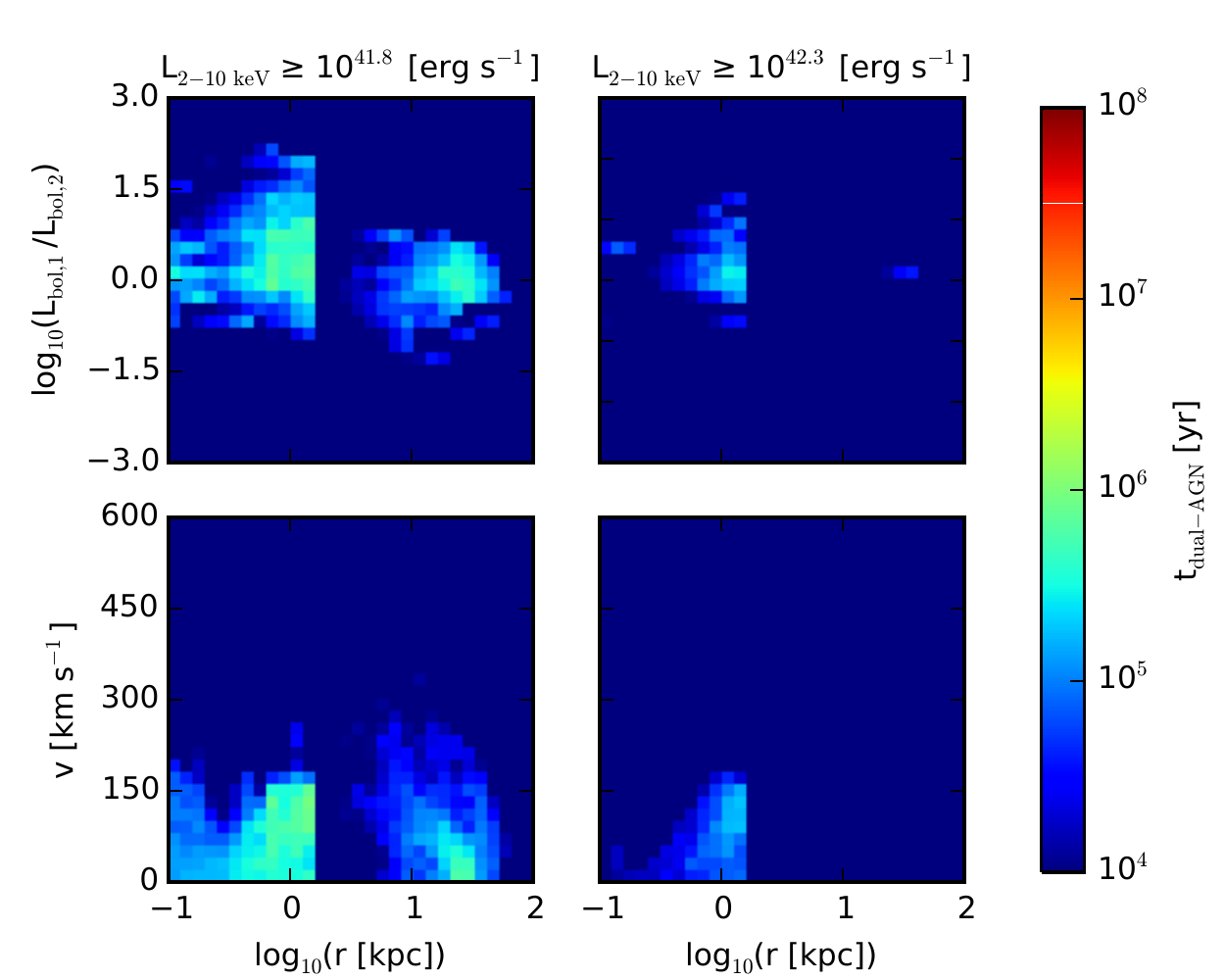}
\end{overpic}\\
\vspace{1.0pt}
\begin{overpic}[width=0.55\columnwidth,angle=0]{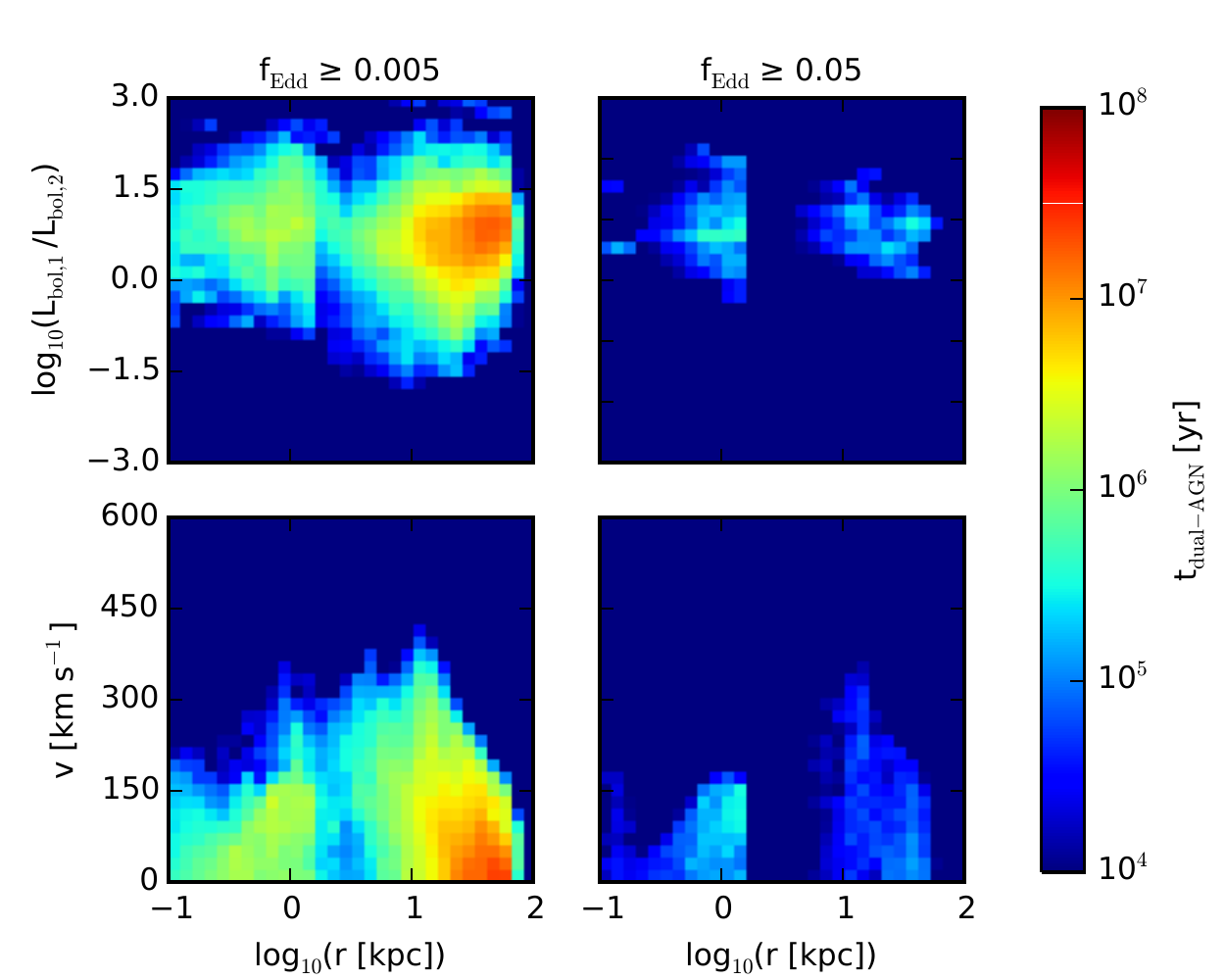}
\end{overpic}
\caption[]{Same as Fig.~\ref{dualagnpaper:fig:m2_hr_gf0_3_BHeff0_001_phi000000_dual_agn_map_proj_100los_stageall_loglog_no3b}, but for the 1:10 coplanar, prograde--prograde merger.}
\label{dualagnpaper:fig:m10_hr_gf0_3_BHeff0_001_phi000000_dual_agn_map_proj_100los_stageall_loglog_no3b}
\end{figure*}

\clearpage

\begin{figure*}
\centering
\vspace{-1.0pt}
\begin{overpic}[width=0.78\columnwidth,angle=0]{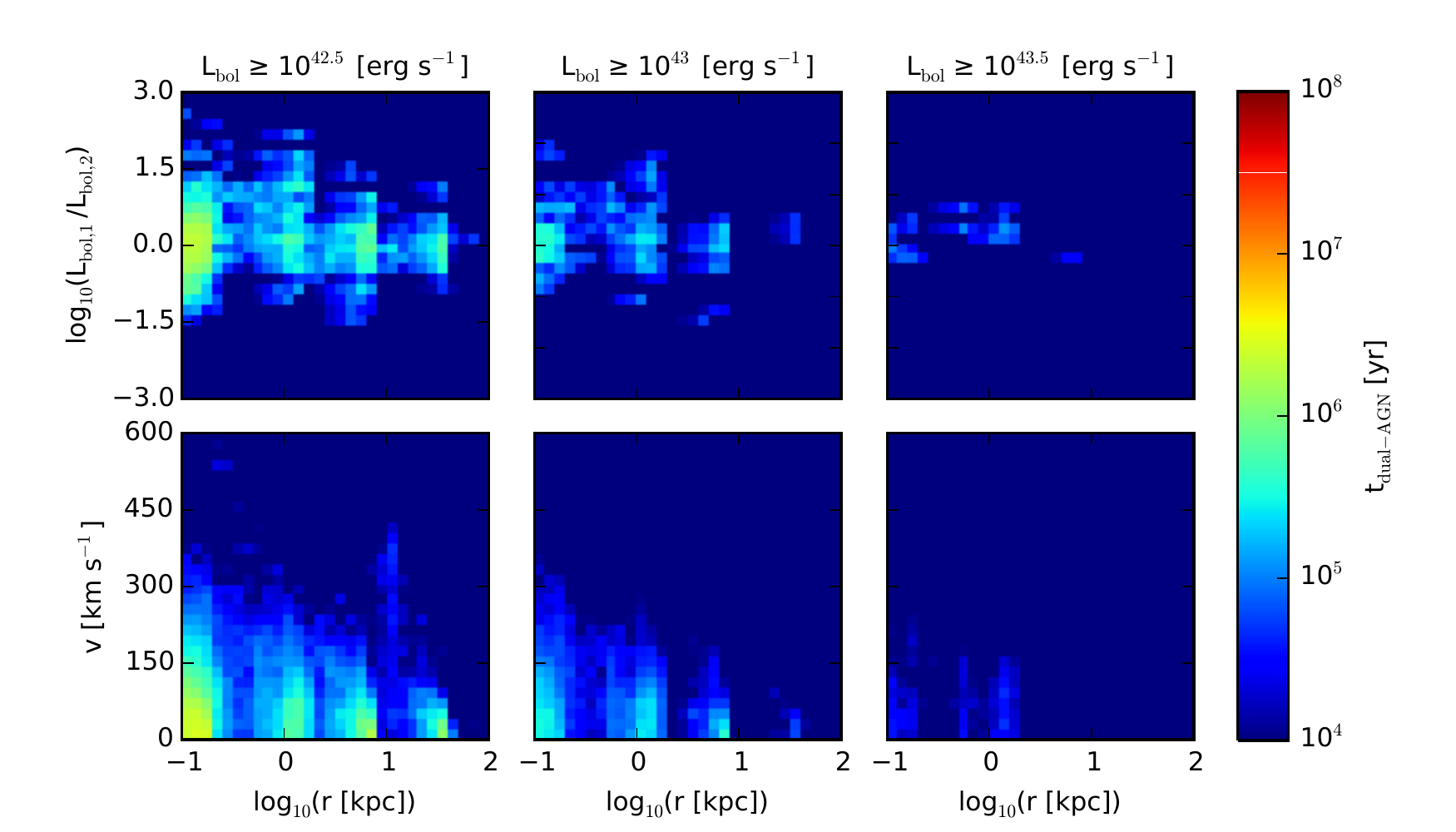}
\put (30.0,54.5) {\textcolor{black}{1:2 cop-pro-pro, 30\% gas, $\epsilon_{\rm f}=0.005$}}
\end{overpic}\\
\vspace{-2.0pt}
\begin{overpic}[width=0.55\columnwidth,angle=0]{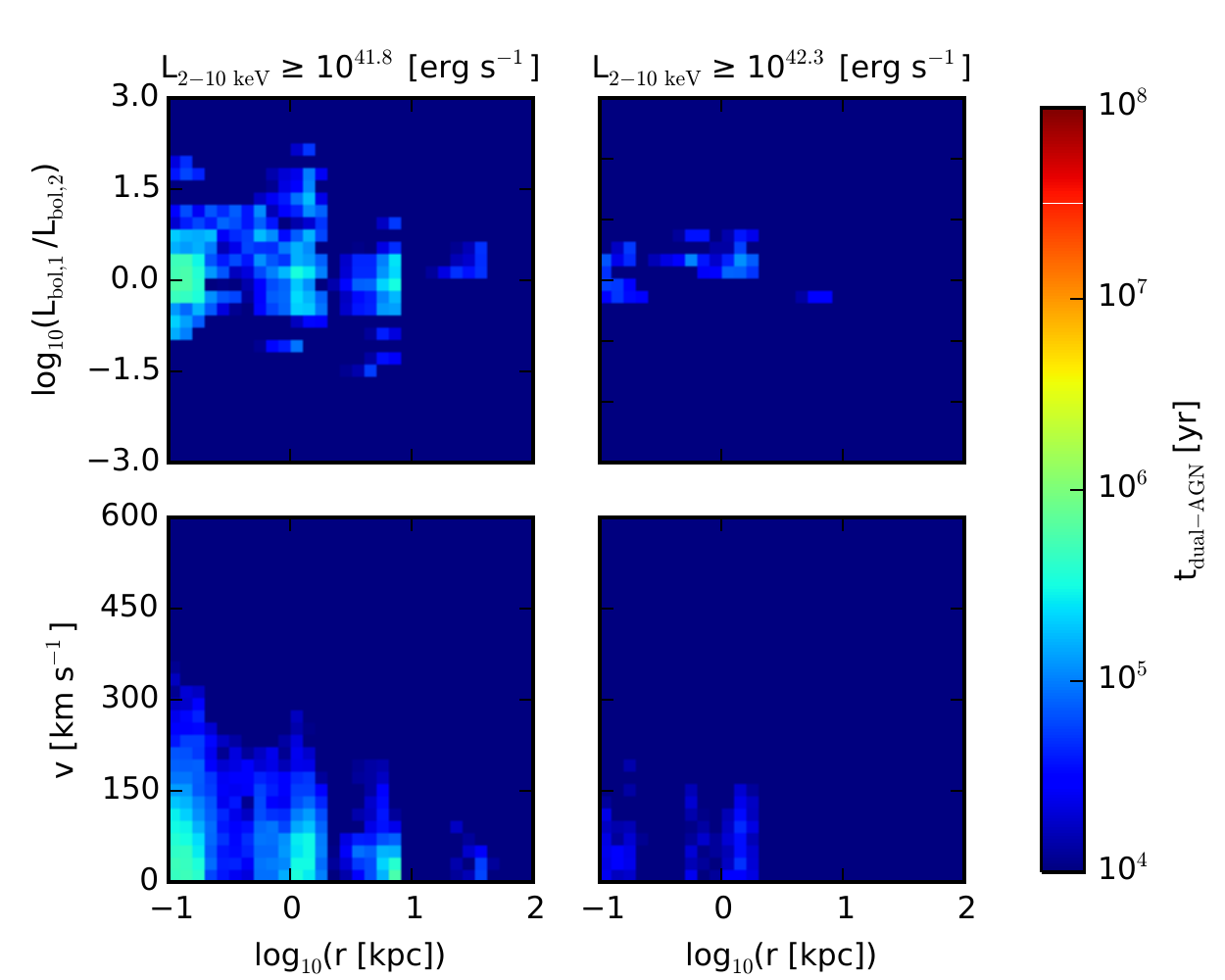}
\end{overpic}\\
\vspace{1.0pt}
\begin{overpic}[width=0.55\columnwidth,angle=0]{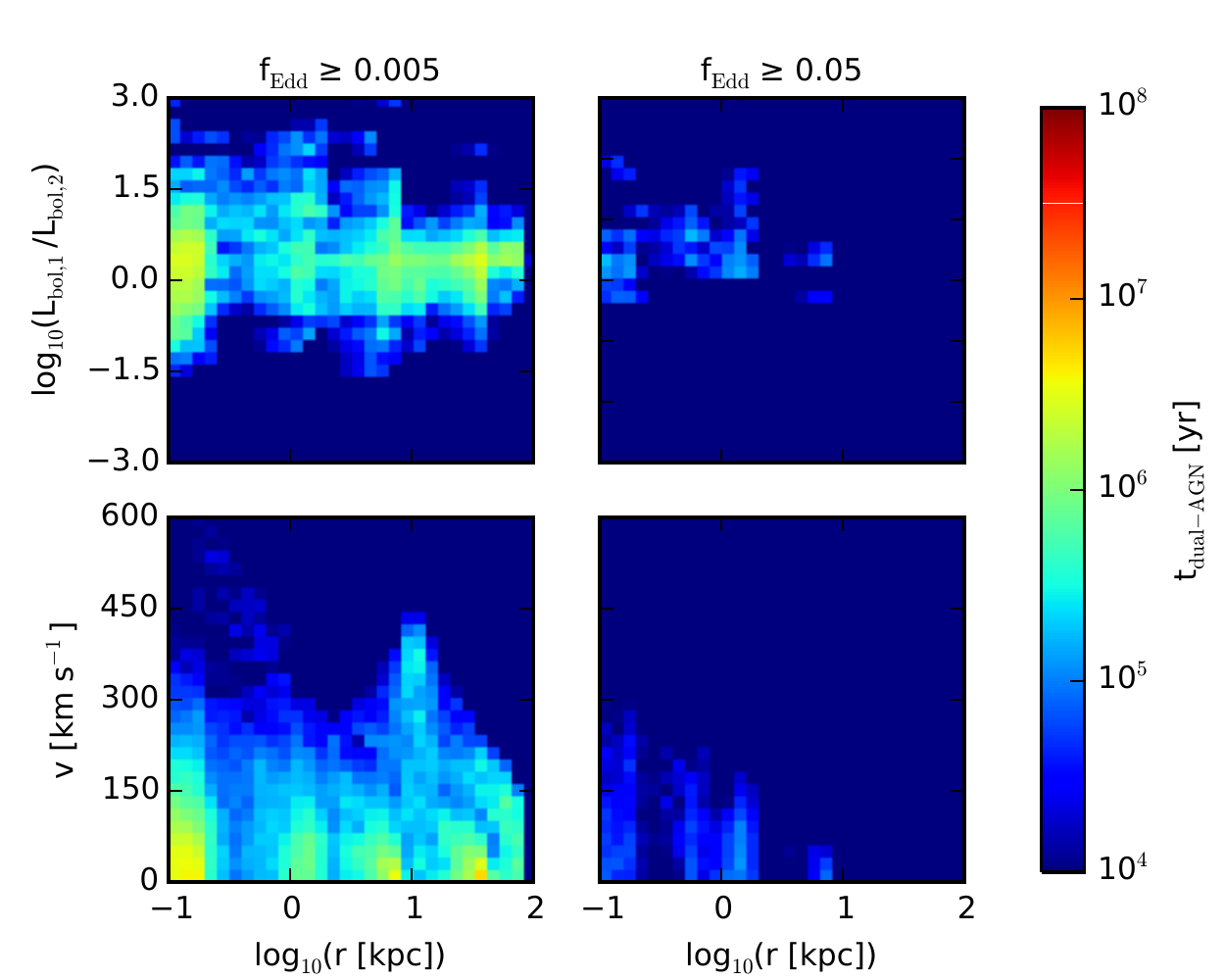}
\end{overpic}
\caption[]{Same as Fig.~\ref{dualagnpaper:fig:m2_hr_gf0_3_BHeff0_001_phi000000_dual_agn_map_proj_100los_stageall_loglog_no3b}, but for the 1:2 coplanar, prograde--prograde merger with 30 per cent gas fraction, standard BH mass, and high BH feedback efficiency.}
\label{dualagnpaper:fig:m2_hr_gf0_3_BHeff0_005_phi000000_dual_agn_map_proj_100los_stageall_loglog_no3b}
\end{figure*}

\clearpage

\begin{figure*}
\centering
\vspace{-1.0pt}
\begin{overpic}[width=0.78\columnwidth,angle=0]{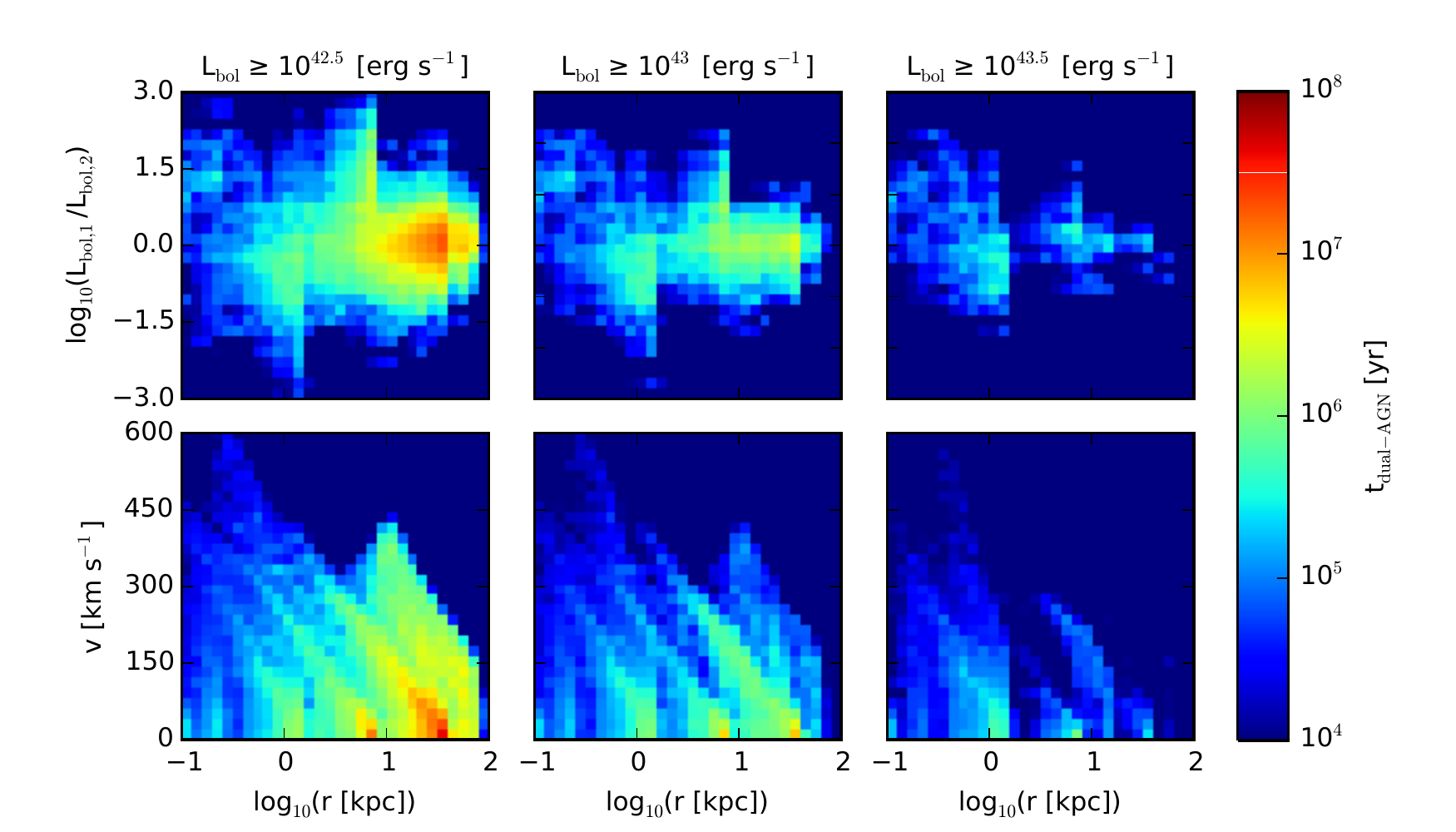}
\put (21.5,54.5) {\textcolor{black}{1:2 cop-pro-pro, 30\% gas, $\epsilon_{\rm f}=0.001$, larger-mass BHs}}
\end{overpic}\\
\vspace{-4.0pt}
\begin{overpic}[width=0.78\columnwidth,angle=0]{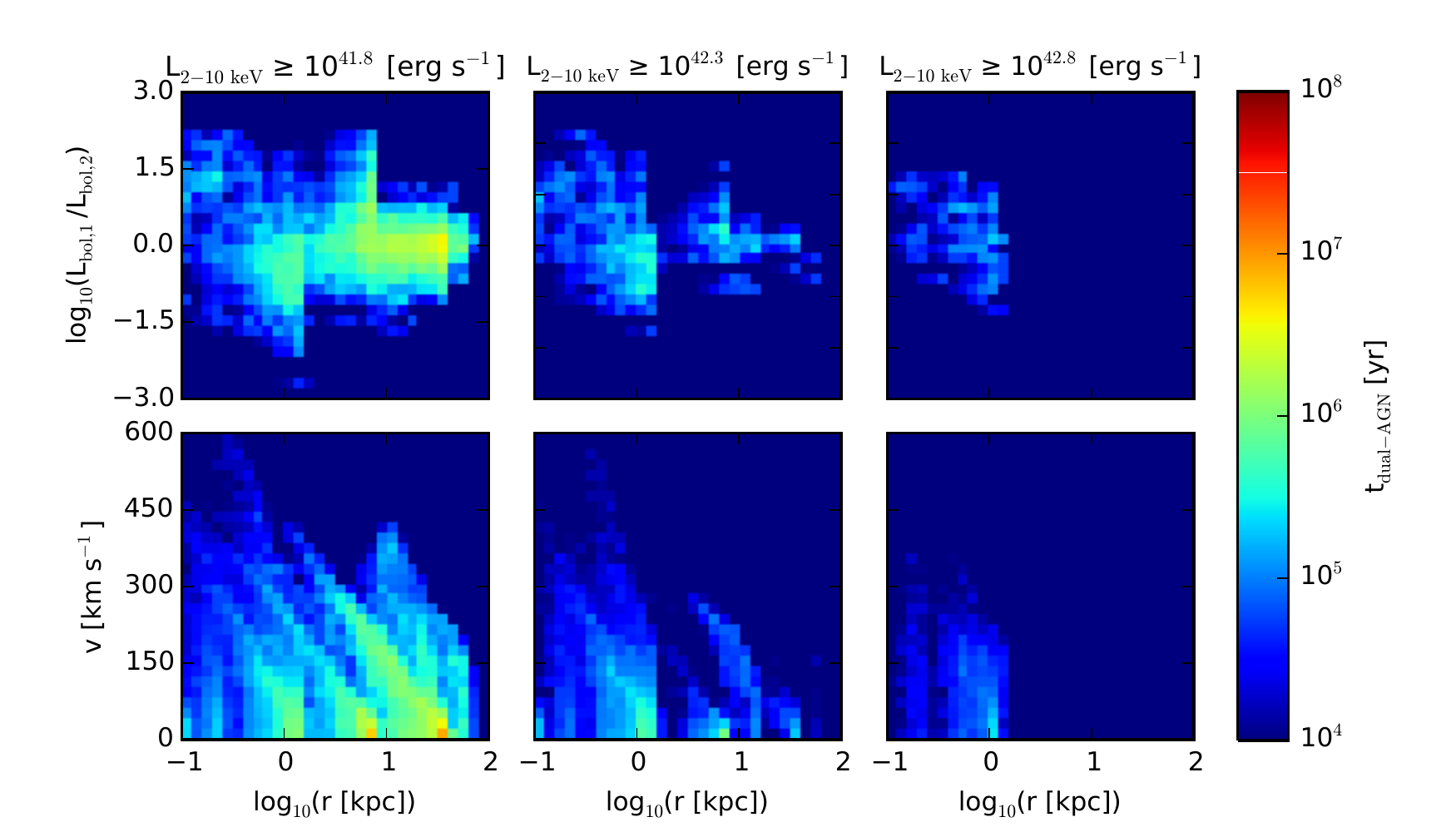}
\end{overpic}\\
\vspace{-4.0pt}
\begin{overpic}[width=0.78\columnwidth,angle=0]{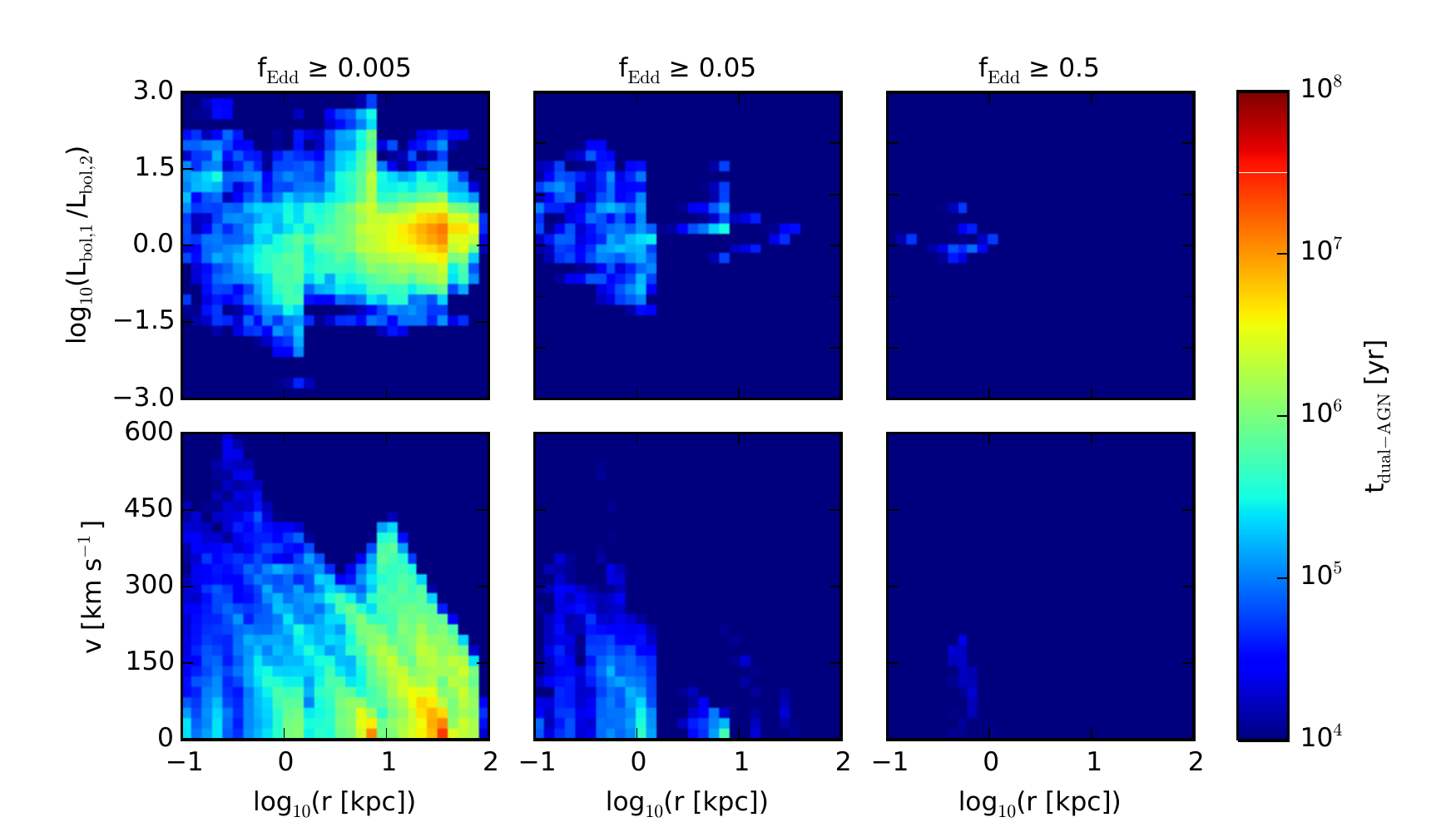}
\end{overpic}
\caption[]{Same as Fig.~\ref{dualagnpaper:fig:m2_hr_gf0_3_BHeff0_001_phi000000_dual_agn_map_proj_100los_stageall_loglog_no3b}, but for the 1:2 coplanar, prograde--prograde merger with 30 per cent gas fraction, standard BH feedback efficiency, and large BH mass.}
\label{dualagnpaper:fig:m2_hr_gf0_3_BHeff0_001_phi000000_largerBH_dual_agn_map_proj_100los_stageall_loglog_no3b}
\end{figure*}

\label{lastpage}
\end{document}